\documentclass[a4paper,12pt]{article}
\usepackage{authblk}
\usepackage{amsfonts}
\usepackage{amssymb}
\usepackage{amsthm}
\usepackage{amsmath}
\usepackage{array}
\usepackage{booktabs}
\usepackage{cases}
\usepackage{cite}
\usepackage{color}
\usepackage{dsfont}
\usepackage{enumerate}
\usepackage{epsfig}
\usepackage{graphicx}
\usepackage{geometry}%
\usepackage[colorlinks,linkcolor=red]{hyperref}
\usepackage{pdflscape}
\usepackage{longtable}
\usepackage{latexsym}
\usepackage{longtable}
\usepackage{lscape}
\usepackage{multirow}
\usepackage{mathrsfs}
\usepackage[numbers]{natbib}
\usepackage{rotating}
\usepackage{setspace}
\usepackage{pst-node}
\usepackage{pst-grad}
\usepackage{pst-xkey}
\usepackage{pstricks}
\usepackage{pst-gantt}
\usepackage[section]{placeins}
\usepackage{tipa}
\usepackage{titlesec}
\usepackage{titletoc}
\usepackage{times}
\usepackage{tikz}
\usepackage[T1]{fontenc}
\usepackage[utf8]{inputenc}
\usepackage{url}
\usepackage{subfigure}
\usepackage{diagbox}
\usepackage{color}
\usepackage{supertabular}
\usepackage{tabularx}
\usepackage{lscape}
\usepackage{amssymb}
\usepackage{bbding}
\usepackage{hyperref}
\title{Review and Assessment of Digital Twin--Oriented Social Network Simulators}

\author[1]{Jiaqi Wen, Bogdan Gabrys, Katarzyna Musial \\ Complex Adaptive Systems Lab, Data Science Institute, University of Technology Sydney \\ jiaqi.wen@student.uts.edu.au, bogdan.gabrys@uts.edu.au, katarzyna.musial-gabrys@uts.edu.au}

\begin{document}
\setlength{\baselineskip}{18pt}%

\date{~}
\maketitle
\vspace{-1cm}

\begin{abstract}
The ability to faithfully represent real social networks is critical from the perspective of testing various what-if scenarios which are not feasible to be implemented in a real system as the system's state would be irreversibly changed. High fidelity simulators allow one to investigate the consequences of different actions before introducing them to the real system. For example, in the context of social systems, an accurate social network simulator can be a powerful tool used to guide policy makers, help companies plan their advertising campaigns or authorities to analyse fake news spread.
In this study we explore different Social Network Simulators (SNSs) and assess to what extent they are able to mimic the real social networks. 
We conduct a critical review and assessment of existing Social Network Simulators under the Digital Twin-Oriented Modelling framework proposed in our previous study. We subsequently extend one of the most promising simulators from the evaluated ones, to facilitate generation of social networks of varied structural complexity levels. This extension brings us one step closer to a Digital Twin Oriented SNS (DT Oriented SNS). 
We also propose an approach to assess the similarity between real and simulated networks with the composite performance indexes based on both global and local structural measures, while taking runtime of the simulator as an indicator of its efficiency. We illustrate various characteristics of the proposed DT Oriented SNS using a well known Karate Club network as an example. While not considered to be of sufficient complexity, the simulator is intended as one of the first steps on a journey towards building a Digital Twin of a social network that perfectly mimics the reality.  

\vspace{1ex}
{\noindent{\bf Keywords:}
Social Networks; Network Dynamics; Machine learning methods}
\vspace{1ex}


\vspace{0.2cm}

\end{abstract}
\newpage
\tableofcontents
\newpage
\newpage
\section{Introduction}
Social network simulators (SNSs) aim at faithfully representing real networked systems and attempt to model the inner rules of network dynamics. They generate simulation-based networks (built with simulated data) or hybrid networks (built with both real and simulated data) to deal with the unobservability problems resulting from data sparsity, privacy concerns and lack of ground-truth \citep{IEEEexample:kavak2019location,IEEEexample:musial2013social}.
 
As proposed in our previous study \citet{IEEEexample:wen2022towards}, Digital Twin (DT) can be seen as a modelling paradigm that serves as an accurate reflection of reality, and this means that we can treat it as an ultimate goal of the representation and modelling of any Complex Networked System (CNS). Over the past decades, to achieve an accurate reflection of reality, variety of SNSs have been developed to embrace complexity of real world systems. Those efforts resulted in more and more realistic and complex social network simulations that, step by step, approach the desirable characteristics of a Digital Twin.

The complexity of the social network simulations results from the heterogeneity of the network components (nodes, edges and their attributes) as well as dynamically changing behaviour of those elements. Current studies have made good progress to attain a desired complexity level. They start from relatively simple simulations which are based on predetermined network statistics or connection principles about network topology, including such classical examples as Barabasi-Albert model for the scale-free network \citep{IEEEexample:barabasi1999emergence} or preferential attachment of nodes based on similarity, popularity or both of them \citep{IEEEexample:wahid2019predict}. Complexity of these networks increases in the structural dimension when node attributes (e.g. age, gender, geo-space, etc.) or edge attributes (e.g. direction, weight, relationship, etc.) are introduced. To incorporate the
impact of structural complexity on network simulation, social network simulators built on interaction rules have been employed due to their flexible definitions based on topology, attributes or both of them, like homophily \citep{IEEEexample:mcpherson2001birds,IEEEexample:asikainen2020cumulative}, triadic structure \citep{IEEEexample:asikainen2020cumulative,IEEEexample:musial2012triad} and geographic proximity \citep{IEEEexample:block2020social}. 
 
The social network simulations get even more complex when they enable the network structure to evolve using inner rules that define how components can change over time. Studies on SNSs focus mainly on the topology change (e.g. the instantaneous social contacts \citep{IEEEexample:fox2016modeling,IEEEexample:passino2021mutually}). Only a small number of studies account for both attribute change and topology change. For example, the SNS proposed by \citep{IEEEexample:ashraf2019simulation} generates social networks with node attributes including features (e.g. age, gender, etc.) and preference for each feature when interacting with others, termed as social DNA (sDNA). The network topology is formed based on sDNA and evolves as sDNA changes randomly over time. However, as an example of one its existing drawbacks, the mechanisms for attributes and preferences changes need to be further studied to allow more realistic modelling.

Studies on SNSs typically focus on the social intervention analysis and conduct an optimisation of inner rules for achieving the highest possible similarity with the target network \citep{IEEEexample:arora2017action,IEEEexample:asikainen2020cumulative}. The consistent network growth enabled by the iterative application of SNSs unfolds the desired network complexity at the cost of time and energy \citep{IEEEexample:hiesinger2021self}. The time step in these studies is often assumed to have the same meaning with the iteration, where the networks grow linearly with time (e.g. predetermined number of added nodes or rewired edges within one iteration between time step $t$ and $t+1$) \citep{IEEEexample:arora2017action,IEEEexample:asikainen2020cumulative}. However, the time truly spent on each iteration of SNS varies with the complexity of networks and computation resources \citep{IEEEexample:ashraf2019simulation} and the time that measures the duration of each SNS iteration can serve as an indicator of model efficiency. Current studies on SNSs take the similarity between simulated networks and the target networks as the objective of running an SNS \citep{IEEEexample:arora2017action,IEEEexample:asikainen2020cumulative,IEEEexample:ashraf2019simulation}, while none of them considers model efficiency.

In this study, we review and assess the current state-of-the-art SNSs under (i) the Digital Twin Oriented Modelling framework encompassing different complexity dimensions and (ii) the assessment framework based on similarity and efficiency, as proposed in our previous study \cite{IEEEexample:wen2022towards}. To provide a possible pathway of extending SNSs towards a Digital Twin Oriented SNS, we build an inner rule-based SNS that 
enables the social network simulation to take into account structural and dynamic dimensions of complexity that builds on \cite{IEEEexample:ashraf2019simulation}. 
Both the similarity and the model efficiency are used as assessment criteria of the proposed SNS. We evaluate the similarity between a simulated and a real network using the composite performance indexes based on both global and local measures and we consider runtime of SNS as an indicator of its efficiency. We use Karate Club network as an example to primarily illustrate but also examine to what extent the proposed SNS can model the structural complexity of a real social network. The diverse network patterns simulated by SNSs and their respective similarity levels to the real network assessed by the composite performance index, given different structural complexity levels, reveal the challenges of SNS performance evaluation for the future DT Oriented SNSs.


The rest of this paper is structured as follows. Section~\ref{section2} reviews the current state-of-the-art of social network simulations and offers comparison between different SNSs in the context of modelling framework proposed in~\cite{IEEEexample:wen2022towards}. Section~\ref{section3} presents the methodology of building an SNS to generate desired networks. Following this, section~\ref{section4} introduces the real data employed in the analysis and presents the results of the experiments. Finally, conclusions and future work are given in section~\ref{section5}.

\section{Current state-of-the-art of Social Network Simulators}
\label{section2}
We review here and discuss the current developments in the context of SNSs from the following four perspectives: (i) modelling prerequisite -- observability -- that determines the observable information to be modelled and the unobservable information to be simulated; (ii) modelling generations connected with the complexity of modelling SNSs; (iii) complexity dimensions that describe the complexity of social network simulations; and (iv) assessment of SNSs concerned with the distance between the social network simulations and the target networks. 

\paragraph{Modelling prerequiste} -- observability, in the context of SNSs, is concerned with the ability of reconstructing the desired complexity of networks from a limited set of observed network components in finite time with an understanding of their dynamically changing behaviour \cite{IEEEexample:wen2022towards}. To generate networks with desired topology and attributes, the existing SNSs simulate the unobservable (represented with a \CircleShadow) and partially observable network components (represented with a \HalfCircleRight) based on the observable network components (represented with a \CircleSolid) and an inner rule that directs the network growth (see Table~\ref{tab3}).

\begin{table}[h]
\centering
\caption{Current SNSs and the observability of network components
in their simulations}
\label{tab3}
\setlength{\tabcolsep}{3pt}
\renewcommand{\arraystretch}{1.5}
\begin{tabular}{|c|cc|cc|p{150pt}|}
\hline
\multirow{2}{*}{Type} & \multicolumn{2}{c|}{Topology} & \multicolumn{2}{c|}{Attributes}  & \multirow{2}{*}{Existing SNS}\\
& Nodes & Edges & Nodes & Edges &     \\
\hline
\multirow{4}{*}{Simulated}  & \multirow{4}{*}{\CircleShadow} & \multirow{4}{*}{\CircleShadow} & \multirow{4}{*}{\CircleShadow} & \multirow{4}{*}{\CircleShadow} & \cite{IEEEexample:barabasi1999emergence},\cite{IEEEexample:doye2002network},\cite{IEEEexample:watts1998collective},\cite{IEEEexample:fortunato2006scale},\cite{IEEEexample:amati2018social},\cite{IEEEexample:arora2017action},\cite{IEEEexample:seaton2004stations},\\
& & & & &\cite{IEEEexample:solomonoff1951connectivity},\cite{IEEEexample:petri2018simplicial},\cite{IEEEexample:elhesha2019co},\cite{IEEEexample:lou2020towards},\cite{IEEEexample:brinkmann1996fast},\cite{IEEEexample:brinkmann2017generation},\cite{IEEEexample:zamfirescu2019almost},\\
& & & & &\cite{IEEEexample:meringer1999fast},\cite{IEEEexample:gugisch2015molgen},\cite{IEEEexample:althofer2015alternating},\cite{IEEEexample:jones2018cograph},\cite{IEEEexample:goedgebeur2018exhaustive},\cite{IEEEexample:mckay2014practical},\cite{IEEEexample:brinkmann2007fast},\\
& & & & &\cite{IEEEexample:goedgebeur2015recursive},\cite{IEEEexample:fabrici2021non},\cite{IEEEexample:goedgebeur2020graphs}\\
\hline
\multirow{6}{*}{Hybrid} & \multirow{2}{*}{\CircleSolid} & \multirow{2}{*}{\CircleShadow}&\multirow{2}{*}{\CircleSolid} & &  \cite{IEEEexample:liu2021block},\cite{IEEEexample:hunter2008goodness},\cite{IEEEexample:asikainen2020cumulative},\cite{IEEEexample:verhoeven2020controlling},\cite{IEEEexample:ashraf2019simulation},\citep{IEEEexample:block2020social},\cite{IEEEexample:boda2020short},\\
& & & & &\cite{IEEEexample:shi2020evaluating},\cite{IEEEexample:lu2009similarity}\\ 
\cline{2-6}
& \CircleSolid & \CircleShadow&\CircleSolid & \CircleShadow & \cite{IEEEexample:lancichinetti2009benchmarks},\cite{IEEEexample:liu2020semi},\cite{IEEEexample:wang2019community},\cite{IEEEexample:lin2008facetnet}  \\ 
\cline{2-6}
& \CircleSolid & \HalfCircleRight& & &  \cite{IEEEexample:wahid2019predict},\cite{IEEEexample:gunecs2016link},\cite{IEEEexample:gao2016hybrid} \\ 
\cline{2-6}
& \CircleSolid & \HalfCircleRight&\CircleSolid & & \cite{IEEEexample:gao2017community},\cite{IEEEexample:wang2007local} \\ 
\cline{2-6}
& \CircleSolid & \HalfCircleRight&  & \HalfCircleRight&   \cite{IEEEexample:chen2018exploiting},\cite{IEEEexample:dai2017link}\\
\hline
\end{tabular}
\end{table}

As is shown in Table~\ref{tab3}, the simulated networks are built with purely simulated topology and attributes, dealing with the unobservability problem of all network components for the analysis of the topological features \citep{IEEEexample:solomonoff1951connectivity,IEEEexample:barabasi1999emergence,IEEEexample:doye2002network,IEEEexample:watts1998collective,IEEEexample:fortunato2006scale,IEEEexample:brinkmann1996fast,IEEEexample:brinkmann2017generation,IEEEexample:zamfirescu2019almost,IEEEexample:meringer1999fast,IEEEexample:gugisch2015molgen,IEEEexample:althofer2015alternating,IEEEexample:jones2018cograph,IEEEexample:goedgebeur2018exhaustive,IEEEexample:mckay2014practical,IEEEexample:brinkmann2007fast,IEEEexample:goedgebeur2015recursive,IEEEexample:fabrici2021non,IEEEexample:goedgebeur2020graphs} and mimicking of real networked systems \citep{IEEEexample:amati2018social,IEEEexample:arora2017action,IEEEexample:zhang2021modeling,IEEEexample:seaton2004stations,IEEEexample:zhang2021vulnerability}. 

The hybrid networks are built with the partially observable and partially simulated network components. Some SNSs incorporate the observable nodes and node attributes while simulating the unobservable edges for a desired states of networks \citep{IEEEexample:liu2021block,IEEEexample:hunter2008goodness,IEEEexample:asikainen2020cumulative,IEEEexample:verhoeven2020controlling,IEEEexample:ashraf2019simulation,IEEEexample:block2020social,IEEEexample:boda2020short}. Some SNSs consider the partial observable edges and conduct link prediction for the unobservable edges \citep{IEEEexample:wahid2019predict,IEEEexample:gunecs2016link,IEEEexample:gao2016hybrid,IEEEexample:gao2017community}, while some SNSs incorporate edge attributes in this process \cite{IEEEexample:wang2007local,IEEEexample:chen2018exploiting,IEEEexample:dai2017link}. In general, there is limited research that, in the simulation process, takes into account real-world attributes of both nodes and edges. There are some works that account for limited number of real node attributes \citep{IEEEexample:liu2021block,IEEEexample:hunter2008goodness,IEEEexample:asikainen2020cumulative,IEEEexample:verhoeven2020controlling,IEEEexample:ashraf2019simulation,IEEEexample:block2020social,IEEEexample:boda2020short,IEEEexample:shi2020evaluating,IEEEexample:lu2009similarity,IEEEexample:lancichinetti2009benchmarks,IEEEexample:liu2020semi,IEEEexample:wang2019community,IEEEexample:lin2008facetnet,IEEEexample:gao2017community,IEEEexample:wang2007local} and when it comes to edges only two simulators partially consider real-world edges \citep{IEEEexample:chen2018exploiting,IEEEexample:dai2017link}. As a result, most of the hybrid simulators take into account some of the topology information but not many consider attribute information and this points to the need for further work on SNSs, especially in the context of learning from continuously streaming data. Such advancement would bring current SNSs closer to the Digital Twin paradigm.

\paragraph{Modelling generations}, as proposed in \cite{IEEEexample:wen2022towards}, consider three key elements of modelling complex network dynamics, the simulation of (i) networks, (ii) processes over networks, and (iii) the interrelation between the two. There are five generations of modelling paradigms (G1, G2, G3, G4 and G5), where SNSs can be built with increasing dynamics complexity through generations and finally reach the goal of a DT in G5. 

We review and discuss the above mentioned three elements in the context of the modelling generations while considering two types of heterogeneity (i) the existence of a given element (represented with a \FiveStarOpen) and in addition to its existence, (ii) the capability of an element to change over time (represented with a \FiveStar) (See Table~\ref{tab1}). 

\begin{table}[h]
\centering
\caption{Current SNSs and their generations}
\label{tab1}
\setlength{\tabcolsep}{3pt}
\renewcommand{\arraystretch}{1.5}
\begin{tabular}{|c|c|c|c|p{120pt}|}
\hline
Stage & Network & Process & Interrelation & Existing SNS\\
\hline
\multirow{9}{*}{G1} & \multirow{7}{*}{\FiveStarOpen}&  &  &\cite{IEEEexample:barabasi1999emergence},\cite{IEEEexample:doye2002network},\cite{IEEEexample:watts1998collective},\cite{IEEEexample:fortunato2006scale},\cite{IEEEexample:amati2018social},\cite{IEEEexample:arora2017action}, \\
& & & &\cite{IEEEexample:lu2009similarity},\cite{IEEEexample:seaton2004stations},\cite{IEEEexample:solomonoff1951connectivity},\cite{IEEEexample:wahid2019predict},\cite{IEEEexample:liu2021block},\\
&&&& \cite{IEEEexample:hunter2008goodness},\cite{IEEEexample:asikainen2020cumulative},\cite{IEEEexample:dai2017link},\cite{IEEEexample:lou2020towards},\cite{IEEEexample:lancichinetti2009benchmarks}, \\
&&&& \cite{IEEEexample:verhoeven2020controlling},\cite{IEEEexample:wang2019community},\cite{IEEEexample:brinkmann1996fast},\cite{IEEEexample:brinkmann2017generation},\cite{IEEEexample:liu2020semi}, \\
&&&&\cite{IEEEexample:zamfirescu2019almost},\cite{IEEEexample:meringer1999fast},\cite{IEEEexample:gugisch2015molgen},\cite{IEEEexample:althofer2015alternating},\cite{IEEEexample:jones2018cograph},\\
&&&&\cite{IEEEexample:goedgebeur2018exhaustive},\cite{IEEEexample:mckay2014practical},\cite{IEEEexample:brinkmann2007fast},\cite{IEEEexample:goedgebeur2015recursive},\cite{IEEEexample:fabrici2021non},\\
&&&&\cite{IEEEexample:goedgebeur2020graphs}\\
\cline{2-5}
  & \multirow{2}{*}{\FiveStarOpen} & \multirow{2}{*}{\FiveStarOpen} &  &\cite{IEEEexample:zhang2021vulnerability},\cite{IEEEexample:zhang2018influence},\cite{IEEEexample:wang2019simulation},\cite{IEEEexample:wang2021multi},\cite{IEEEexample:pastor2001epidemic},\\
&&&&  \cite{IEEEexample:ganesh2005effect}  \\
\hline
G2a  & \FiveStarOpen & \FiveStar &  & \cite{IEEEexample:jovanovski2021modeling},\cite{IEEEexample:carchiolo2021mutual}\\
\hline
\multirow{4}{*}{G2b}  & \multirow{3}{*}{\FiveStar} &  &  & \cite{IEEEexample:gunecs2016link},\cite{IEEEexample:petri2018simplicial},\cite{IEEEexample:ashraf2019simulation},\cite{IEEEexample:block2020social},\cite{IEEEexample:boda2020short},\cite{IEEEexample:shi2020evaluating},\\
&&&&\cite{IEEEexample:gao2017community},\cite{IEEEexample:wang2007local},\cite{IEEEexample:kim2019advancing},\cite{IEEEexample:chen2018exploiting},\cite{IEEEexample:lin2008facetnet} \\
&&&&\cite{IEEEexample:gao2016hybrid},\cite{IEEEexample:kendrick2018change},\cite{IEEEexample:elhesha2019co},\cite{IEEEexample:budka2013molecular}\\
\cline{2-5}
   & \FiveStar &  \FiveStarOpen &  &\cite{IEEEexample:liu2020using},\cite{IEEEexample:kim2020location} \\
\hline
\multirow{2}{*}{G3} 
& \FiveStar &  \FiveStarOpen & \FiveStarOpen &\cite{IEEEexample:newman2005threshold}   \\
\cline{2-5}
& \FiveStarOpen &  \FiveStar & \FiveStarOpen &\cite{IEEEexample:koprulu2019battle},\cite{IEEEexample:fu2019analysis}, \cite{IEEEexample:pan2018effective}\\
\hline
\end{tabular}
\end{table}

\begin{figure}[t!]
\centering
\vspace{0.1cm}  
\setlength{\abovecaptionskip}{0.3cm}
\setlength{\belowcaptionskip}{-0.3cm}
  \includegraphics[width=3.5 in]{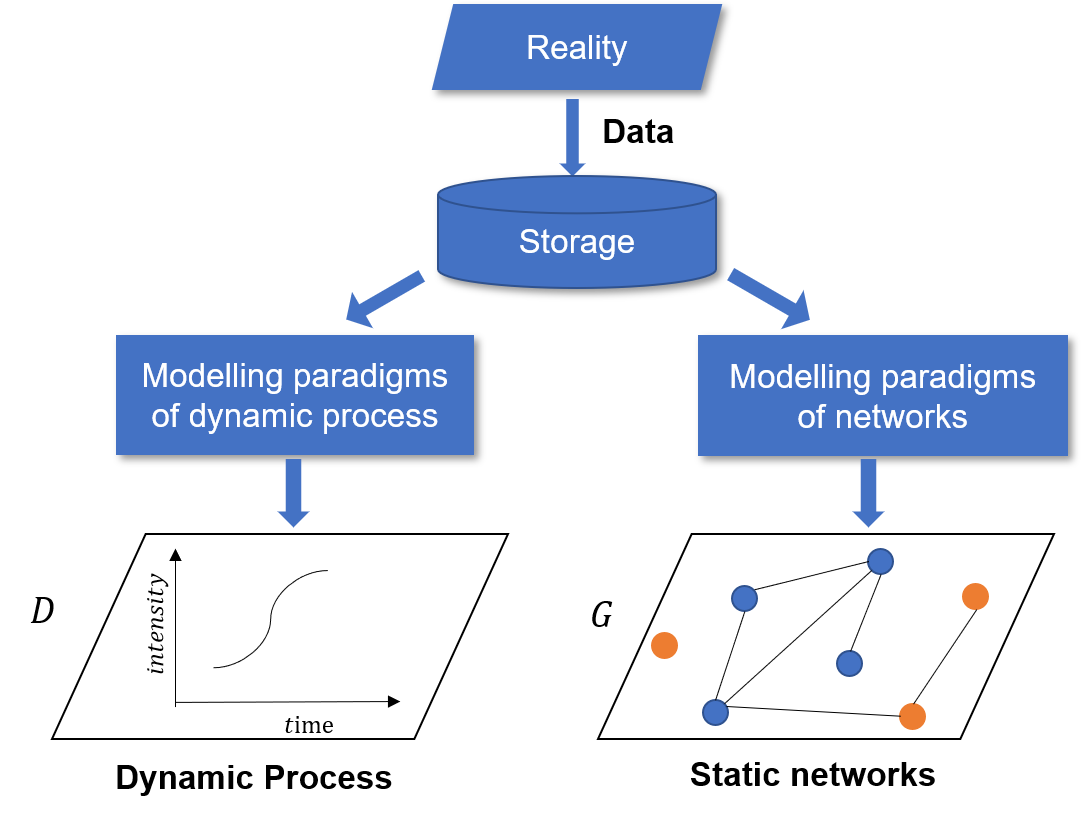}
\caption{G1.}
\label{gen1}      
\end{figure}

Generation 1 (G1) of models focuses on dynamic process on static networks (see Table~\ref{tab1} and Fig.~\ref{gen1}). They simulate networks that are "frozen" in time, with a dynamic process taking place on the networks where parameters of this process do not change during the simulation (e.g. epidemic spreading process on static social networks with a fixed infection rate \cite{IEEEexample:ganesh2005effect,krol2015propagation}). Most studies focus on SNSs in G1 \cite{IEEEexample:barabasi1999emergence,IEEEexample:doye2002network,IEEEexample:watts1998collective,IEEEexample:fortunato2006scale,IEEEexample:amati2018social,IEEEexample:arora2017action,IEEEexample:lu2009similarity,IEEEexample:seaton2004stations,IEEEexample:solomonoff1951connectivity,IEEEexample:wahid2019predict,IEEEexample:liu2021block,IEEEexample:hunter2008goodness,IEEEexample:asikainen2020cumulative,IEEEexample:dai2017link,IEEEexample:lou2020towards,IEEEexample:lancichinetti2009benchmarks,IEEEexample:liu2020semi,IEEEexample:verhoeven2020controlling,IEEEexample:wang2019community,IEEEexample:zhang2021vulnerability,IEEEexample:zhang2018influence,IEEEexample:wang2019simulation,IEEEexample:wang2021multi,IEEEexample:pastor2001epidemic,IEEEexample:ganesh2005effect,IEEEexample:brinkmann1996fast,IEEEexample:brinkmann2017generation,IEEEexample:zamfirescu2019almost,IEEEexample:meringer1999fast,IEEEexample:gugisch2015molgen,IEEEexample:althofer2015alternating,IEEEexample:jones2018cograph,IEEEexample:goedgebeur2018exhaustive,IEEEexample:mckay2014practical,IEEEexample:brinkmann2007fast,IEEEexample:goedgebeur2015recursive,IEEEexample:fabrici2021non,IEEEexample:goedgebeur2020graphs,IEEEexample:brodka2020interacting}. Also, many of the simulators generate static social networks without any consideration of dynamic processes \cite{IEEEexample:barabasi1999emergence,IEEEexample:doye2002network,IEEEexample:watts1998collective,IEEEexample:fortunato2006scale,IEEEexample:amati2018social,IEEEexample:arora2017action,IEEEexample:lu2009similarity,IEEEexample:seaton2004stations,IEEEexample:solomonoff1951connectivity,IEEEexample:wahid2019predict,IEEEexample:liu2021block,IEEEexample:hunter2008goodness,IEEEexample:asikainen2020cumulative,IEEEexample:dai2017link,IEEEexample:lou2020towards,IEEEexample:lancichinetti2009benchmarks,IEEEexample:liu2020semi,IEEEexample:verhoeven2020controlling,IEEEexample:wang2019community}. 

\begin{figure}[h]
\centering
\vspace{0.1cm}  
\setlength{\abovecaptionskip}{0.3cm}
\setlength{\belowcaptionskip}{-0.3cm}
  \includegraphics[width=3.5 in]{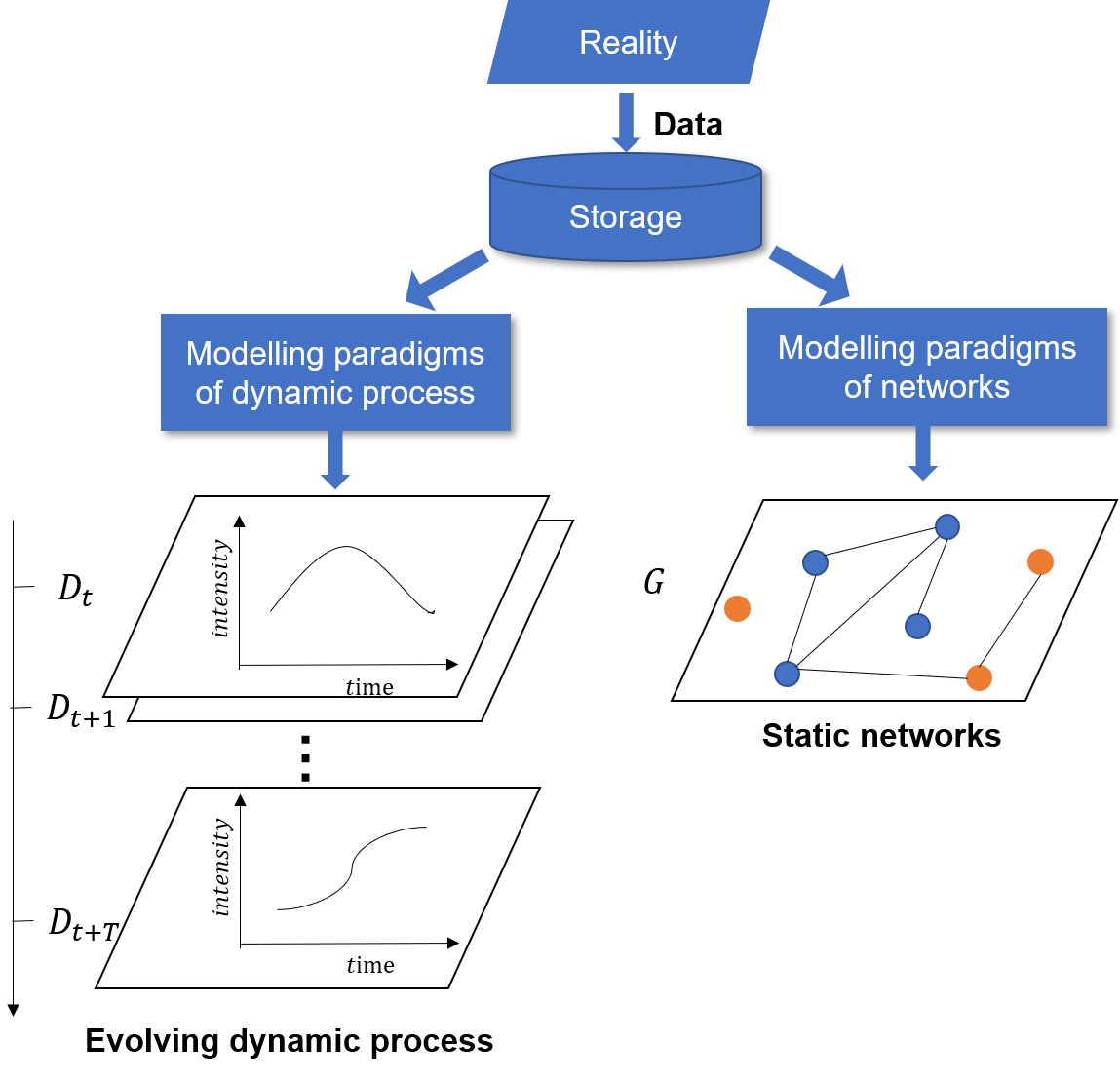}
\caption{G2a.}
\label{gen2a}      
\end{figure}

SNSs in Generstion 2 (G2) have two variations, one termed as G2a and the other as G2b. SNSs in G2a focus on evolving dynamic process on static networks (see Table~\ref{tab1} and Fig.~\ref{gen2a}). They simulate a static network where dynamic process changes its parameters over time and gets captured in snapshots (e.g. epidemic spreading processes on static social networks with a changeable infection rate \cite{IEEEexample:jovanovski2021modeling}). There are only few studies on SNSs in G2a \cite{IEEEexample:carchiolo2021mutual,IEEEexample:jovanovski2021modeling,IEEEexample:eletreby2020effects}.

\begin{figure}[h]
\centering
\vspace{0.1cm}  
\setlength{\abovecaptionskip}{0.3cm}
\setlength{\belowcaptionskip}{-0.3cm}
  \includegraphics[width=3.5 in]{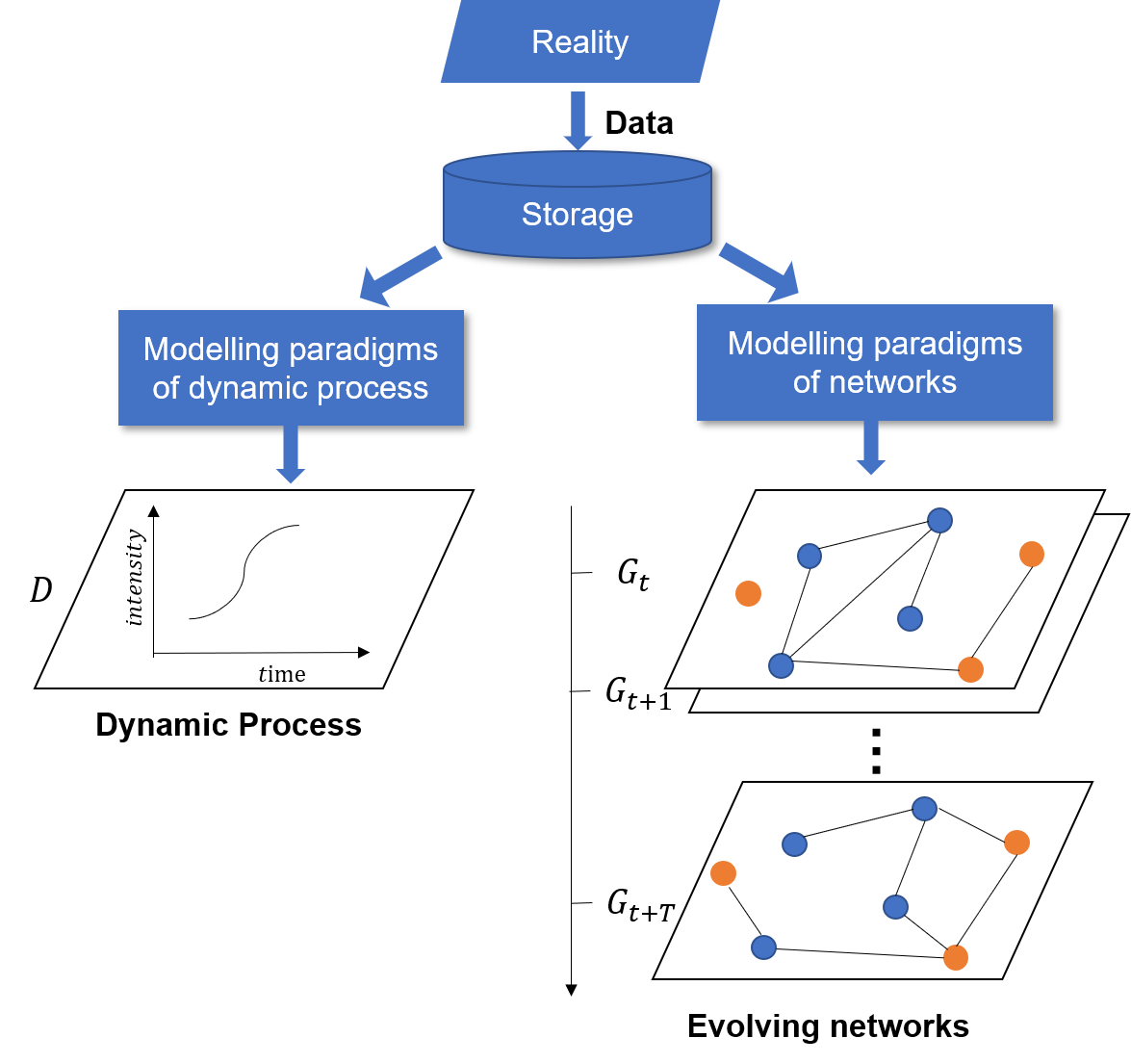}
\caption{G2b.}
\label{gen2b}      
\end{figure}

SNSs in G2b focus on dynamic process on evolving networks (see Table~\ref{tab1} and Fig.~\ref{gen2b}). They simulate network snapshots that describe the network topology changes over time, where the dynamic process takes place without changing its parameters (e.g. epidemic spreading process with a non-changeable infection rate on social network that evolves over time \cite{IEEEexample:kim2020location}). There are many studies on SNSs that model only the dynamics of network structure in this space \cite{IEEEexample:gunecs2016link,IEEEexample:petri2018simplicial,IEEEexample:ashraf2019simulation,IEEEexample:block2020social,IEEEexample:boda2020short,IEEEexample:shi2020evaluating,IEEEexample:gao2017community,IEEEexample:wang2007local,IEEEexample:kim2019advancing,IEEEexample:chen2018exploiting,IEEEexample:lin2008facetnet,IEEEexample:gao2016hybrid,IEEEexample:kendrick2018change,IEEEexample:elhesha2019co,IEEEexample:budka2013molecular,IEEEexample:liu2020using,IEEEexample:kim2020location,IEEEexample:verhoeven2020controlling,IEEEexample:musial2013creation} while few of them additionally consider the dynamic processes \cite{IEEEexample:liu2020using,IEEEexample:kim2020location}.

\begin{figure}[h]
\centering
\vspace{0.1cm}  
\setlength{\abovecaptionskip}{0.3cm}
\setlength{\belowcaptionskip}{-0.3cm}
  \includegraphics[width=3.5 in]{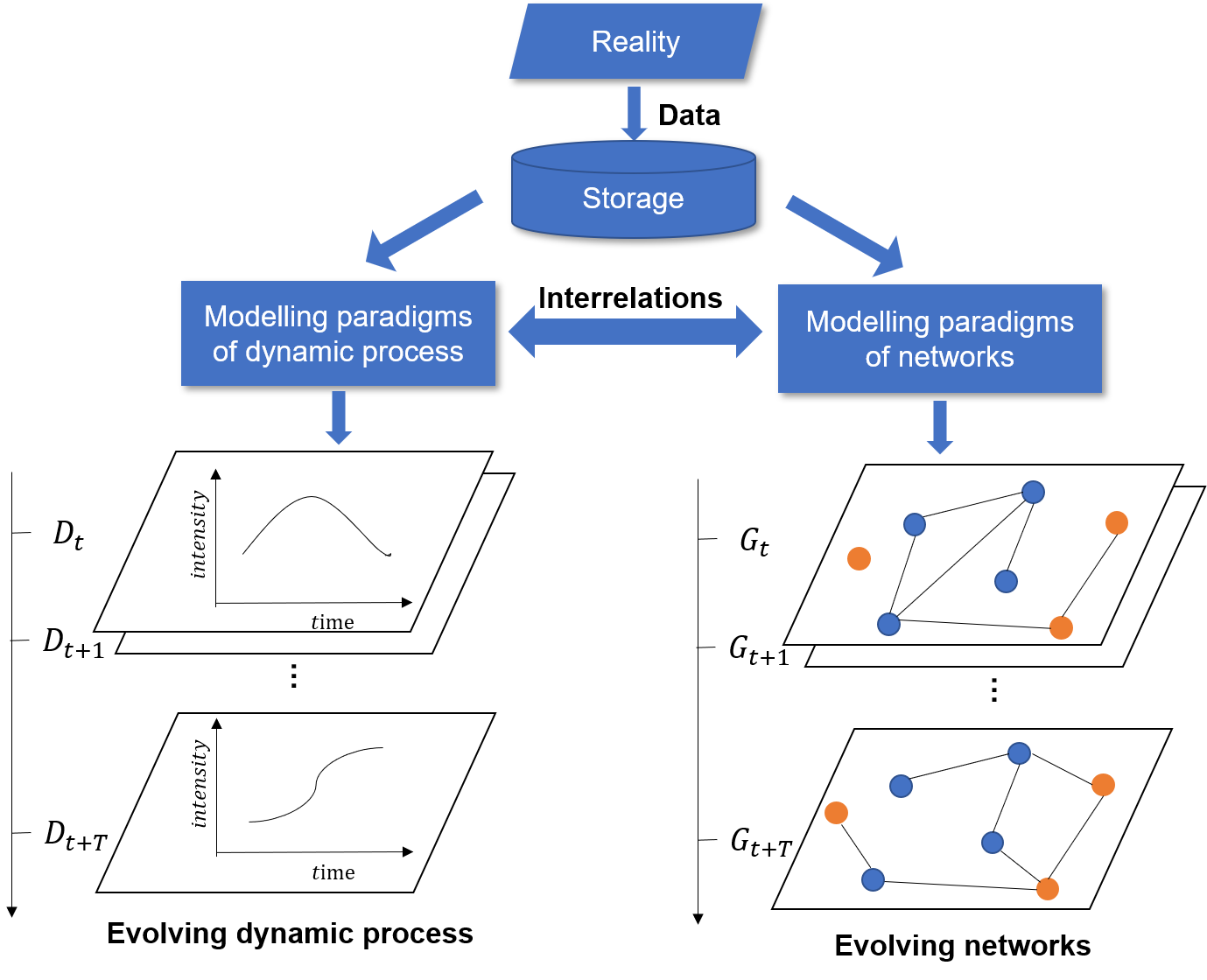}
\caption{G3.}
\label{gen3}      
\end{figure}

SNSs in G3 focus on evolving dynamic processes on evolving networks with interrelations between them (See Fig.~\ref{gen3}). They simulate networks, dynamic processes and their changes using snapshot approach (data are processed in batches), while incorporating the interactions between a network and a process. For example, the epidemic spreading across the network can leave some nodes dead and get them removed \cite{IEEEexample:newman2005threshold}, while the infection rate of epidemic spreading process can also vary depending on nodes' groups (structural patterns) \cite{IEEEexample:koprulu2019battle,IEEEexample:fu2019analysis}. Only few SNSs consider the interrelations between the networks and the dynamic processes, where they just study the dynamic processes with non-changeable parameters on evolving networks  \cite{IEEEexample:newman2005threshold} or evolving dynamic processes on static networks \cite{IEEEexample:koprulu2019battle,IEEEexample:fu2019analysis,IEEEexample:pan2018effective}

\begin{figure}[h]
\centering
\vspace{0.1cm}  
\setlength{\abovecaptionskip}{0.3cm}
\setlength{\belowcaptionskip}{-0.3cm}
  \includegraphics[width=3.5 in]{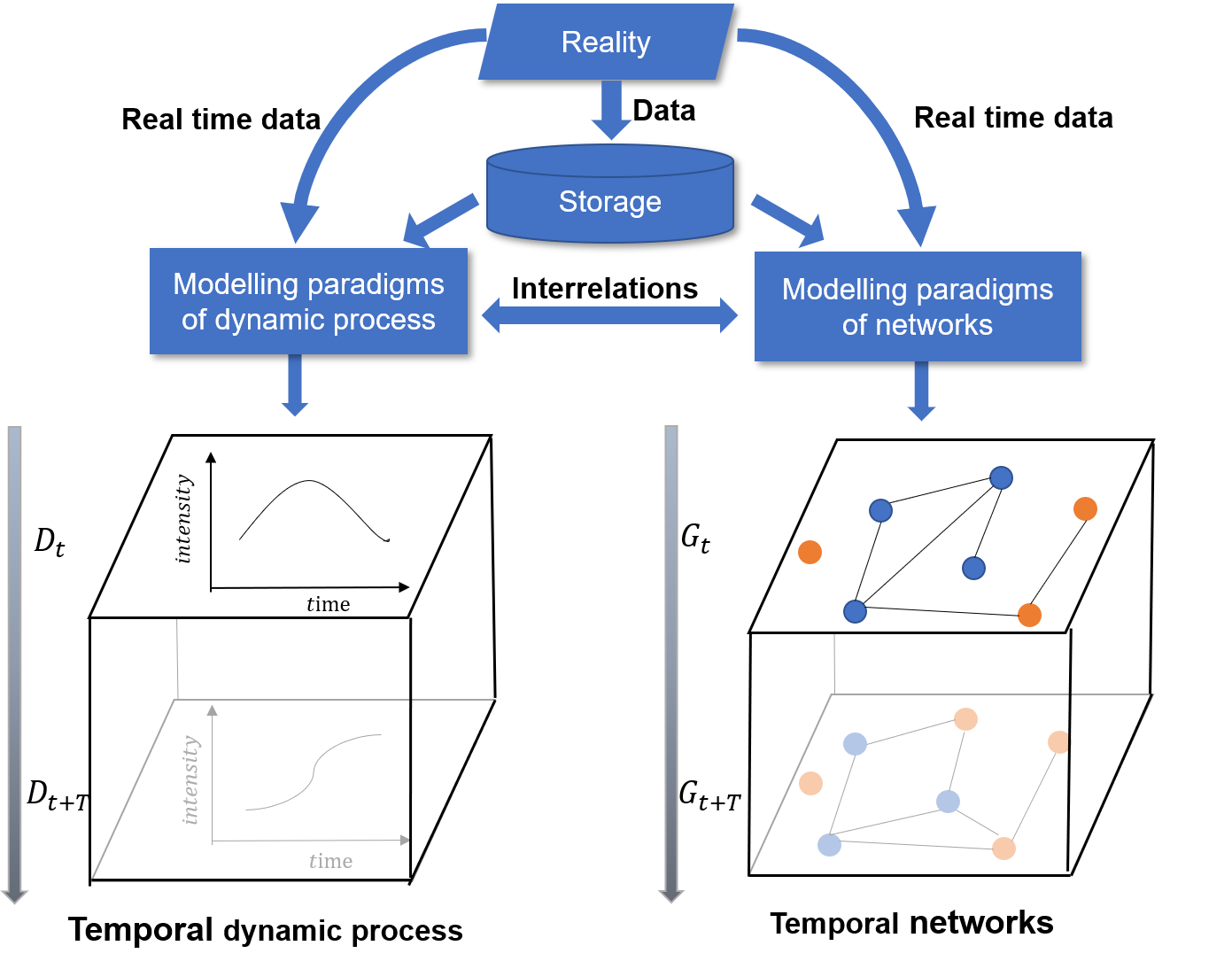}
\caption{G4.}
\label{gen4}      
\end{figure}

SNSs in G4 focus on temporal dynamic processes on temporal networks with interrelations between them and the continuous acquisition of real time information (see Fig.~\ref{gen4}). They simulate networks and the dynamic processes with instantaneous changes of network topology and parameters, while incorporating the interactions between a process and a network. SNSs in G5 further extend the SNSs in G4 by closing the feedback loop between the SNSs and the real system and can be identified as an idealised state that can be named as a Digital Twin. Currently, there are no studies on SNS in G4 and G5 and further studies are required to model such high complexity scenarios to approach a DT.

\paragraph{Complexity dimensions} describe the complexity of social networks resulting from the heterogeneity of their network components (topology and attributes) and temporal dynamics of those components and their attributes. 
As proposed in \cite{IEEEexample:wen2022towards}, there are four complexity dimensions (i.e. structural, spatial, temporal and dynamics) that need to be considered when representing and modelling a network. The structural and the spatial complexities are concerned with the existence of network attributes and the space where the topology can be embedded respectively. The temporal and the dynamics complexities are connected with the changeable network components and their dynamics, respectively. 

We review and discuss the network components and network dynamics of the social networks generated by the existing SNSs while considering two types of heterogeneity: (i) the existence of the components in a static time scale (represented with a \FiveStarOpen) and, in addition to its existence, (ii) the capability of the component to change over time (represented with a \FiveStar) (see Table~\ref{tab2}).

\begin{table}[h]
\centering
\caption{Current SNS and the components of their simulations}
\label{tab2}
\setlength{\tabcolsep}{3pt}
\renewcommand{\arraystretch}{1.5}
\begin{tabular}{|c|cc|cc|c|p{120pt}|}
\hline
\multirow{2}{*}{Stage} & \multicolumn{2}{c|}{Topology} & \multicolumn{2}{c|}{Attributes} & \multirow{2}{*}{Network dynamics} & \multirow{2}{*}{Existing SNS}\\
& Nodes & Edges & Nodes' & Edges' &    & \\
\hline
\multirow{9}{*}{G1/G2a}  & \multirow{6}{*}{\FiveStarOpen} & \multirow{6}{*}{\FiveStarOpen} &  &  &  \multirow{6}{*}{\FiveStarOpen} & \cite{IEEEexample:barabasi1999emergence},\cite{IEEEexample:doye2002network},\cite{IEEEexample:watts1998collective},\cite{IEEEexample:fortunato2006scale},\cite{IEEEexample:amati2018social},\cite{IEEEexample:arora2017action},\\
& & & & & &\cite{IEEEexample:zhang2021modeling},\cite{IEEEexample:lu2009similarity},\cite{IEEEexample:seaton2004stations},\cite{IEEEexample:solomonoff1951connectivity},\cite{IEEEexample:zhang2021vulnerability},\\
& & & & & &  \cite{IEEEexample:zhang2018influence},\cite{IEEEexample:wahid2019predict}\cite{IEEEexample:wang2019simulation},\cite{IEEEexample:wang2021multi},\cite{IEEEexample:goedgebeur2020graphs},\\
&&&&&& \cite{IEEEexample:jovanovski2021modeling},\cite{IEEEexample:ganesh2005effect},\cite{IEEEexample:pastor2001epidemic},\cite{IEEEexample:eletreby2020effects},\cite{IEEEexample:althofer2015alternating},\cite{IEEEexample:brinkmann1996fast},\\
&&&&&&\cite{IEEEexample:zamfirescu2019almost},\cite{IEEEexample:meringer1999fast},\cite{IEEEexample:gugisch2015molgen},\cite{IEEEexample:jones2018cograph},\cite{IEEEexample:brinkmann2017generation},\\
&&&&&&\cite{IEEEexample:goedgebeur2018exhaustive},\cite{IEEEexample:mckay2014practical},\cite{IEEEexample:brinkmann2007fast},\cite{IEEEexample:goedgebeur2015recursive},\cite{IEEEexample:fabrici2021non},\\
\cline{2-7}
   & \FiveStarOpen & \FiveStarOpen &\FiveStarOpen  &  &  \FiveStarOpen  & \cite{IEEEexample:liu2021block},\cite{IEEEexample:hunter2008goodness},\cite{IEEEexample:asikainen2020cumulative},\cite{IEEEexample:verhoeven2020controlling}  \\
\cline{2-7}
   &\FiveStarOpen  &\FiveStarOpen &   & \FiveStarOpen  &  \FiveStarOpen   & \cite{IEEEexample:dai2017link},\cite{IEEEexample:lou2020towards}\\
\cline{2-7}
    &\FiveStarOpen  &\FiveStarOpen & \FiveStarOpen  & \FiveStarOpen  &  \FiveStarOpen   &  \cite{IEEEexample:lancichinetti2009benchmarks},\cite{IEEEexample:liu2020semi},\cite{IEEEexample:wang2019community},\cite{IEEEexample:scata2016impact} \\
\hline
\multirow{7}{*}{G2b/G3}  & \FiveStarOpen  & \FiveStar & &  & \FiveStarOpen   & \cite{IEEEexample:gunecs2016link},\cite{IEEEexample:petri2018simplicial},\cite{IEEEexample:zhu2019investigation},\cite{IEEEexample:liu2020using}  \\
\cline{2-7}
 & \FiveStarOpen & \FiveStar &\FiveStarOpen  &  &  \FiveStarOpen  &  \cite{IEEEexample:ashraf2019simulation},\citep{IEEEexample:block2020social},\cite{IEEEexample:boda2020short},\cite{IEEEexample:shi2020evaluating},\cite{IEEEexample:gao2017community},\cite{IEEEexample:wang2007local} \\
 \cline{2-7}
&      \FiveStarOpen   &  \FiveStar      &\FiveStar &  & \FiveStarOpen  & \cite{IEEEexample:ashraf2019simulation},\cite{IEEEexample:kim2019advancing}  \\
\cline{2-7}
&      \FiveStarOpen   &  \FiveStar    & &\FiveStar  & \FiveStarOpen  &  \cite{IEEEexample:chen2018exploiting}\\
\cline{2-7}
&     \FiveStarOpen   &     \FiveStar    & \FiveStarOpen & \FiveStar  & \FiveStarOpen  & \cite{IEEEexample:budka2013molecular},\cite{IEEEexample:kim2020location}  \\
\cline{2-7}
&     \FiveStarOpen   &     \FiveStar    & \FiveStar & \FiveStar  & \FiveStarOpen  & \cite{IEEEexample:lin2008facetnet}   \\
\cline{2-7}
  & \FiveStarOpen  & \FiveStar & &  & \FiveStar   & \cite{IEEEexample:gao2016hybrid},\cite{IEEEexample:kendrick2018change},\cite{IEEEexample:elhesha2019co}\\
\hline
\end{tabular}
\end{table}

As is shown in Table~\ref{tab2}, most SNSs focus on static networks with no attributes, which are categorised as G1/G2a models and characterised with the lowest level of complexity in each dimension. They generate networks based on predetermined network statistics and connection principles about topology \citep{IEEEexample:barabasi1999emergence,IEEEexample:doye2002network,IEEEexample:watts1998collective,IEEEexample:fortunato2006scale,IEEEexample:amati2018social,IEEEexample:arora2017action,IEEEexample:zhang2021modeling,IEEEexample:lu2009similarity,IEEEexample:seaton2004stations,IEEEexample:solomonoff1951connectivity,IEEEexample:zhang2021vulnerability,IEEEexample:zhang2018influence,IEEEexample:wahid2019predict,IEEEexample:wang2019simulation,IEEEexample:wang2021multi,IEEEexample:brinkmann1996fast,IEEEexample:brinkmann2017generation,IEEEexample:zamfirescu2019almost,IEEEexample:meringer1999fast,IEEEexample:gugisch2015molgen,IEEEexample:althofer2015alternating,IEEEexample:jones2018cograph,IEEEexample:goedgebeur2018exhaustive,IEEEexample:mckay2014practical,IEEEexample:brinkmann2007fast,IEEEexample:goedgebeur2015recursive,IEEEexample:fabrici2021non,IEEEexample:goedgebeur2020graphs}. Some other SNSs, with higher level of structural complexity, incorporate node attributes \citep{IEEEexample:liu2021block,IEEEexample:hunter2008goodness,IEEEexample:asikainen2020cumulative,IEEEexample:verhoeven2020controlling}, edge attributes \citep{IEEEexample:dai2017link,IEEEexample:lou2020towards} or both of them \citep{IEEEexample:lancichinetti2009benchmarks,IEEEexample:liu2020semi,IEEEexample:wang2019community} into the generation process of static networks. The SNSs in G2b/G3 have higher level of temporal complexity as network components are allowed to change over time, and they generally just consider the topology change \citep{IEEEexample:gunecs2016link,IEEEexample:petri2018simplicial,IEEEexample:ashraf2019simulation,IEEEexample:block2020social,IEEEexample:boda2020short,IEEEexample:shi2020evaluating,IEEEexample:gao2017community,IEEEexample:wang2007local}. Few SNSs focus on changeable network dynamics, 
where the existing ones just consider topology and its changes over time, with a higher level of dynamics complexity but the lowest level of structural complexity \cite{IEEEexample:gao2016hybrid,IEEEexample:kendrick2018change,IEEEexample:elhesha2019co}. To the best of our knowledge, there are no other SNSs in G4/G5 that would enable to model the desired level of complexity of network structure and its dynamics.

\paragraph{Assessment of SNSs} is concerned with investigating how close the generated networks are to the real systems that they attempt to model. 

For SNSs, which generate simulated or hybrid networks to deal with the unobservable information, their assessment involves network statistics like density, degree distribution, shortest path, assortativity, modularity, clustering coefficient and betweenness \citep{IEEEexample:musial2013kind}. Most SNSs employ network statistics concerned with the global network structures like degree centrality, betweenness centrality, and PageRank coefficient \cite{IEEEexample:arora2017action} to measure the similarity between the social network simulations and the target network.  As an example, SNS proposed by~\citep{IEEEexample:arora2017action} simulates the unobservable edges to achieve the desired betweenness centrality, PageRank, local clustering coefficient and degree distribution of the networks. There are also SNSs that simulate the missing edges for networks with partially observable edges, which can be treated as a link prediction task and evaluated considering the prediction performance with precision~\citep{IEEEexample:lu2011link,IEEEexample:gao2017community}, Micro Precision ~\citep{IEEEexample:chen2020multi}, Area Under the ROC (AUC)~\citep{IEEEexample:lu2011link}, Error Rate~\citep{IEEEexample:chen2018gc}, etc. However, the above mentioned measures just focus on the states of static networks, whereas further studies are required for the measures considering network changes over time. 

As an indicator of the SNSs' efficiency, the runtime of the simulation is considered by few studies \citep{IEEEexample:kim2019advancing,IEEEexample:ashraf2019simulation}, which is proved to be influenced by the network components such as the number of simulated features and the network size. 

To summarise, SNSs in the current studies can not capture enough complexity from the real world for an idealised DT modelling due to observability reasons. Though some SNSs consider real information, such as node attributes, there remains an undeveloped research space concerning how to gradually craft the increase of complexity for an improved SNS performance in one or several aspects. By increasing structural complexity, we aim to provide one of the possible pathways for extending the existing SNSs towards a DT Oriented SNS discussed in Section \ref{section3}. Through experiments in Section \ref{section4}, we also reveal the challenges of DT Oriented modelling by presenting the complex network patterns and, thus, the difficulties they pose to the SNS performance evaluation and SNS extension towards a DT. 

\section{Towards Digital Twin Oriented Social Network Simulator}
\label{section3}

To generate social networks with the desired level of complexity, we build an inner rule-based SNS referring to and extending work in \cite{IEEEexample:ashraf2019simulation}, while enabling it to simulate social networks of increased complexity level and iterate with an optimised performance in efficiency and similarity. The two steps of building such an SNS that mimics the target network include: (i) the proposal of an inner rule-based SNS with extensible complexity in section \ref{section31} and (ii) an optimised iterative application of this SNS for similarity and efficiency in section \ref{section32} considering the observability and complexity. We have included some SNSs in the \nameref{appendix}, while employing the extensible SNS initially proposed by \cite{IEEEexample:ashraf2019simulation} as an example in the following part of this study.

\subsection{An Extensible Social Network Simulator}
\label{section31}
The inner rule-based SNS framework proposed by \citep{IEEEexample:ashraf2019simulation} models people as nodes, with the node attributes including features (i.e. such as age, gender, etc.) and individual preferences towards particular features termed as social-DNA (sDNA). The edges between these nodes represent relationships between people, which are connected according to preferential attachment rule based on topological and non-topological (attributes) characteristics. The edges can be directed or undirected and  they are time-stamped with the iterative application of the SNS. 

Based on this framework, we treat each node attribute as a tuple defined with a feature and sDNA (preference for this feature and weight of preference), while allowing their variations among individuals and changes over time. This enables the social network simulations with an extensible complexity in structural, spatial, temporal and dynamics dimensions. To implement this complexity, the detailed settings about network components and network dynamics are proposed and presented below.

\paragraph{Network components} include: (i) nodes (e.g. people), (ii) edges (e.g. social contact or relations), (iii) attributes including node attributes (e.g. age, gender, geo-space, etc.) and edge attributes (e.g. direction, weight, relationship, etc.). They vary in diversity and their capability to change over time. 

We assume that the SNS simulates an undirected network $\hat{G}$ with $N$ attributed nodes within $M$ iterations, the network formed in the $k_{th}$ iteration is represented as
\begin{equation}
G_k=(V, E_k,A_k)\quad k\in[1,M]
\end{equation}
which is composed of a fixed node set $V=\{v_1,\cdots, v_N\}$, an edge set that grows through iterations: $E_k=\{(v_{i},v_{j})_k|,v_i,v_j\in V, i\neq j\}$ and a node attribute set $A_k=\{a_k(v_i),\cdots,a_k(v_N)\}$ that can vary with nodes and change for each iteration.

Especially, referring to \citep{IEEEexample:ashraf2019simulation}, the set of attributes $a_k(v_i)$ for node $v_i$ is defined as
\begin{equation}
    a_k(v_i) = \{f_k(v_i), p_k(v_i), w_k(v_i)\}  \quad v_i\in V
\end{equation}
which includes a feature vector $f_k(v_i)$ and the social-DNA (sDNA) defined with two vectors of the same length as $f(v_i)$. Any attribute, in this context, can be represented with a three-value tuple that is composed of feature, preference for this feature and weight of preference. 

$p_k(v_i)$ determines whether to prefer similar feature with a binary value of $1$ (prefer) or $-1$ (not prefer), and $w_k(v_i)$ represents the weight of preference for the feature with the range of $[0,1]$. 

The sDNA can be set at the level of individual nodes, groups or the whole populations in a static manner or can mutate over time. This variability of node attributes forms another source of structural complexity and requires further study. 

\paragraph{Network dynamics} drives the network growth based on the current network topology and attributes. Referring to the previous study by \citep{IEEEexample:ashraf2019simulation}, the edges of the undirected network are created without removal according to the ranking scores of each node pair based on the two-way evaluation between node $v_i$ and $v_j$:
\begin{equation}
    s_k(v_i,v_j) = q_k\Phi_k(v_i,v_j) + r_k\Delta_k(v_i,v_j)+c_k^\tau\Pi_k(v_i,v_j) \quad v_i,v_j\in V, i\neq j
\end{equation}
where $\Phi_k(v_i,v_j)$, $\Delta_k(v_i,v_j)$ and $\Pi_k(v_i,v_j)$ each represents the feature-based, popularity-based and shortest-path score with the weights of $q_k$, $r_k$ and $c_k$ respectively.

The feature-based score is based on the nodes' preferences $sDNA(v_{i})$ and $sDNA(v_{j})$ for their features $f(v_i)$ and $f(v_j)$:
\begin{equation}
\begin{aligned}
    \Phi_k(v_i,v_j) = |f(v_i)-f(v_j)|^\tau (w_k(v_i) \odot l_k(v_i))+|f(v_i)-f(v_j)|^\tau (w_k(v_j) \odot l_k(v_j))  
\end{aligned}
\end{equation}

The popularity-based score incorporates the preference for nodes with higher degrees, with $m_k$ representing the preferential attachment parameter and $d_k$ represents the calculation of node degrees in the $k_{th}$ iteration:
\begin{equation}
    \Delta_k(v_i,v_j) = m_k d_k(v_i) +m_k d_k(v_j) 
\end{equation}

The shortest path-based score is calculated based on the preference for the shortest path-length $l$ within the range $[2,q]$ between the node pair:
\begin{equation}
    \Pi_{k}(v_i,v_j) = \sum\limits_{l=2}^{q} \gamma(v_i)^lI^l(v_i,v_j)+\sum\limits_{l=2}^{q} \gamma(v_j)^lI^l(v_i,v_j) \quad l\in[2,q] 
\end{equation}
where $\gamma(v_i)^l$ represents node $v_i$'s weight of preference for the path-length $l$ and $I^l(v_i,v_j)$ represents the existence of the path-length $l$ between node pair $v_i$ and $v_j$ with a binary value of $0$ or $1$.
\begin{equation}
    s_k(v_i,v_j) = q\Phi_k(v_i,v_j) + r\Delta_k(v_i,v_j)+c\Pi_k(v_i,v_j) 
\end{equation}
where $\Phi_k(v_i,v_j)$, $\Delta_k(v_i,v_j)$ and $\Pi_k(v_i,v_j)$ each represents the feature-based, popularity-based and shortest-path score with the weights of $q$, $r$ and $c$.


\subsection{Optimisation towards a Digital Twin}
\label{section32}
A high quality SNS, in order to approach an idealised Digital Twin status, needs the optimised simulation of network components and network dynamics considering the three elements reviewed in section \ref{section2}: (i) observability, (ii) complexity, and (iii) assessment of social network simulators. 

\paragraph{Development path} of the SNS towards a DT can be divided into small steps, where with each step the complexity of a simulated social network gradually increases. Under the assumption of a fixed node set and non-attributed edges, we define a development of SNS towards a DT with increasing structural complexity and temporal complexity (see Table~\ref{tab4}).

\begin{table}[h]
\centering
\caption{Small steps of building up the complexity.}
\label{tab4}
\setlength{\tabcolsep}{3pt}
\renewcommand{\arraystretch}{1.5}
\begin{tabular}{|c|cc|cc|c|}
\hline
 \multirow{2}{*}{Stage} & \multicolumn{2}{c|}{Topology} & \multicolumn{2}{c|}{Attributes} & \multirow{2}{*}{Dynamics}\\
& Nodes & Edges & Nodes' & Edges' &  \\
\hline
\multirow{2}{*}{G1}  &  \FiveStarOpen & \FiveStarOpen &  &  &  \FiveStarOpen \\
   & \FiveStarOpen & \FiveStarOpen &\FiveStarOpen  &  &  \FiveStarOpen \\
\hline
\multirow{2}{*}{G2}  & \FiveStarOpen  & \FiveStar &\FiveStarOpen &  & \FiveStarOpen  \\
&      \FiveStarOpen   &  \FiveStar      &\FiveStar &  & \FiveStarOpen \\
\hline
\multirow{1}{*}{G3}  
       &  \FiveStarOpen &\FiveStar &\FiveStar &  & \FiveStar \\
\hline
\end{tabular}
\end{table}

In G1, the structural complexity increases when more features are considered with a longer feature vector $f_k(v_i)$ as part of a node attribute $a_k(v_i)$. The sDNA, $p_k$ and $w_k$, represent the corresponding preference and weight of preference for respective similar features in the feature vector, composing another parts of node attribute $a_k(v_i)$. The sDNA also increases the structural complexity when more variability of the nodes' preferences (ranging from population-based, group-based to individual-based) is incorporated. Feature selection and the sDNA simulation is the necessary step required for an appropriate structural complexity level.

In G2, the temporal complexity increases when edges $E_k$ and the node attributes $A_k$ start to change with every simulation step, creating snapshots of evolving networks (with overall number of iterations $k$). This requires further study on what changes, how it changes, which network state the change leads to. In G3, the network dynamics can change with the changeable weighting factor $q_k$, $r_k$ and $c_k$ for the ranking scores of node pairs, as well as the changeable mechanism that generates sDNA through various stochastic distributions.


We can simulate the unobservable networked information and remove the unnecessary steps if the increased complexity depreciates the performance of an SNS measured using criteria of similarity and efficiency. For each step ahead, we optimise the unobservable sDNA, $p_k$ and $w_k$, with the SNS proposed in section \ref{section31} to achieve the maximised similarity between the $G_k$, the desired $\hat{G}$ and the minimised runtime for its simulation with efficiency (See equation \ref{equ}).

\begin{equation}
\label{equ}
\begin{aligned}
    \max\limits_{p_k,w_k} ||G_k-\hat{G}||\\
   s.t. \min \sum \limits_{z=1}^{k}t_z \\
\end{aligned}
\end{equation}

The efficiency of SNS is measured with the runtime $\sum \limits_{z=1}^{k}t_z$ spent on the $k$ iterations of running the SNS. We propose the composite performance indexes to measure the similarity of social network simulations and the target network. 

\paragraph{composite performance indexes} measure the distance $||G_k-\hat{G}||$ between the simulated social network $G_k$ and the target network $\hat{G}$ based on their similarity from both the global and the local perspectives \cite{IEEEexample:musial2013kind}. As shown in equation \ref{summaryequ}, the composite performance index is calculated as the average distance between the simulated and the real networks in terms of each network measure $\eta_i$. 
\begin{equation}
\label{summaryequ}
    ||G_k-\hat{G}|| = \frac{1}{n}\sum\limits_{i=1}^{n} ||\eta_i(G_k)-\eta_i(\hat{G})|| 
\end{equation}

The global perspective, concerned with the high level outcomes of interactions between the number of actors, includes measures such as node degree distribution, shortest path length distribution and the values of modularity and assortativity \cite{IEEEexample:musial2013kind}. The local perspective, connected with interpersonal relations within subgraphs, includes measures such as the clustering coefficient distribution and the triad significance profile \cite{IEEEexample:musial2013kind,IEEEexample:milo2004superfamilies,IEEEexample:jia2021measuring,IEEEexample:jia2021directed}. 

We employ the 2-sample Kullback-Leibler divergence (KL divergence) to quantify their differences in distribution and the Manhattan distance to measure the differences of normalised values. Their missing values are replaced with $1$ to indicate the large gaps between the network simulations and the target network. Overall, the composite performance index of each network simulation, with values ranging from $0$ to $1$, is calculated as the weighted average of these values. A lower value of this quality indicator means a more similar simulation compared with the target network.   

\paragraph{Multi-objective optimisation,} towards an optimal similarity and efficiency, is required for the iterative application of DT Oriented SNS. To figure out an optimised sDNA in this process, we employ the fast elitist Non-dominated Sorting Genetic Algorithm (NSGA-2) proposed by \cite{IEEEexample:deb2002fast} for three reasons.

First, as it is impossible to have a single solution which simultaneously optimises all objectives, NSGA-2, as a multi-objective evolutionary algorithm, enables a Pareto-optimal solution set with alternative solutions in or on Pareto-optimal front \cite{IEEEexample:deb2002fast}. Second, NSGA-2 is characterised with low computational requirements, elitist approach, and parameter-less sharing approach \cite{IEEEexample:deb2002fast}. Third and also the most important, NSGA-2, as a type of genetic algorithm, works best in the search of multiple parameters \cite{IEEEexample:lambora2019genetic} while enabling the optimisation of sDNA based on genetics selection principles.  




\section{Results and Assessment}
\label{section4}
The social networks generated using the extended SNS vary with the node attributes, including the feature $f_k(v_i)$ and the feature's preference: $p_k(v_i)$ and $w_k(v_i)$. Specifically, we assume a population-based preference, where all nodes share the same preference vector $p_k(v_i)\equiv p$ and $w_k(v_i)\equiv w$, without any change across iterations, which are calculated for an optimal level of similarity and efficiency based on a target network. The Zachary's Karate Club network \citep{IEEEexample:zachary1977information}, composed of 34 nodes with one binary attribute, is used as an illustratinve example in this experiment. For each iteration, $4\%$ of edges are added without removal.
Ten network statistics are used in the assessment stage, including the density, assortativity, modularity, shortest path length and degree distribution in the global perspective, and the clustering coefficient and the significance profile of four types of triadic closures considering the binary node attributes.

Under the above assumptions and settings, the impact of feature selection for feature vector $f(v_i)$, concerned with the structural complexity, over the iterations is studied. This involves the benchmark SNS for networks without attributes (zero feature--based SNS), SNS for networks with one real node attribute (real feature--based SNS), SNS for networks with one unobservable node attribute that is randomly simulated as $0$ or $1$ (simulated feature--based SNS) and SNS for networks that are built with both the real and the simulated attributes (hybrid feature--based SNS). The complexity of the generated social network and the performance of the social network simulator are each presented and analysed in section~\ref{complexity} and section~\ref{performance} respectively. Please note that all of these analyses are intended to provide insights into a discussion of how feasible and challenging the possible pathways to the development of more realistic DT oriented SNSs are likely to be and what one can expect along such a journey.

\subsection{Performance of the proposed social network simulator}
\label{performance}
\begin{figure*}[h!] 
	\centering
	\subfigure[The comprehensive composite performance index]{
		\begin{minipage}[b]{0.48\linewidth}
		\includegraphics[width=1\linewidth]{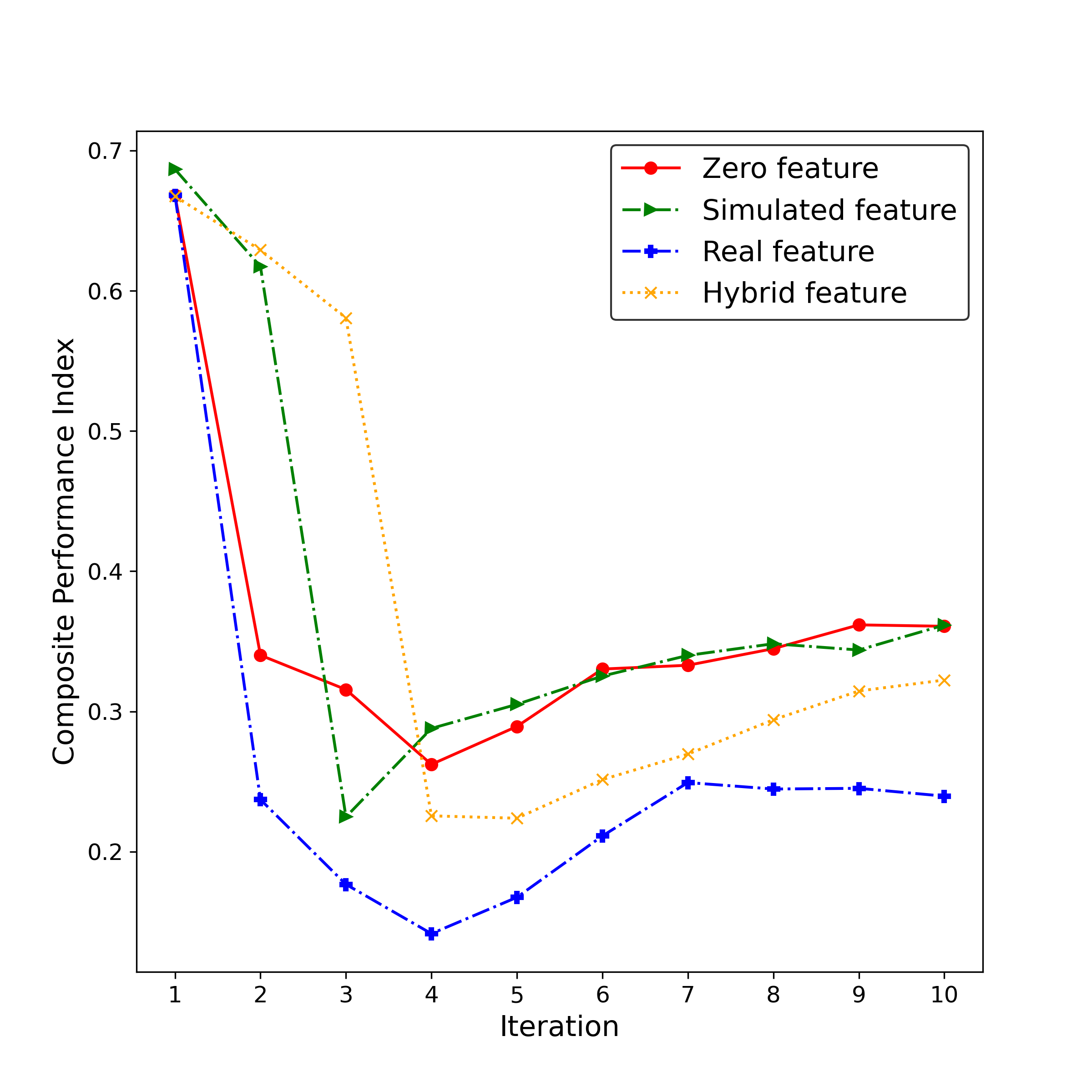}
	\end{minipage}}
	\subfigure[The runtime]{
		\begin{minipage}[b]{0.48\linewidth}
			\includegraphics[width=1\linewidth]{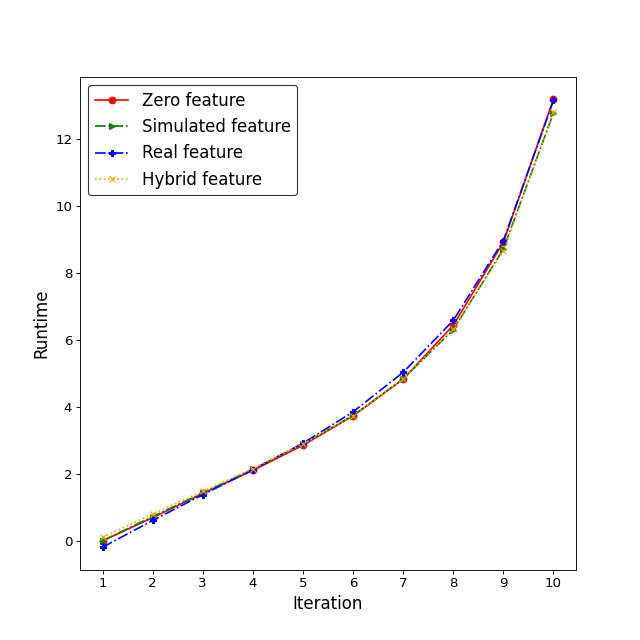}
	\end{minipage}}\\
	\caption{The performance of SNS based on the comprehensive composite performance index (a) and the efficiency (b)}
\label{fig1}
\end{figure*}

We measure the similarity between the simulated social networks and the target network with the composite performance indexes based on a variety of measures from both the global and the local perspectives and take the runtime of each simulation as an indicator for efficiency. Rather than proposing a performance measure or reaching final conclusions to be universally employed for SNSs, we aim to reveal the challenges of SNS performance evaluation for the future DT Oriented SNSs by discussing the diverse network patterns and the respective varying similarity levels that contribute to the composite performance index. The performance obtained based on the similarity and efficiency is shown in Fig. \ref{fig1}.

As we can see in Fig.~\ref{fig1}(a), the distance between the simulated social networks and the target network topology fluctuates over the iterations. All generated networks share a similar trend of composite performance index,
which firstly decreases in the first three iterations and then gradually increases. 
Generally all the SNSs achieve the lower values of composite performance index around the fourth iteration. The real feature--based SNS, with the introduction of one feature, achieves the lowest composite performance index among all the SNSs since the second iteration, within the shortest time. 
The hybrid feature-based SNS introduces a simulated feature to the real feature-based SNS, which, compared with the zero feature-based and the simulated feature-based SNSs, produces a higher composite performance index in the first three iterations and a lower composite performance index afterwards.

For each SNS, the time spent on each iteration increases as more edges are added
and more topological information is considered in the network growth, which, given the similar computation load of feature-based score with the non-changeable feature set, can be attributed to the calculation of popularity-based and shortest path score. There is little difference in runtime among the benchmark: zero feature--based SNS and other SNSs considering the small feature set that varies little. 

Overall speaking, the introduction of the real feature to the SNS, with an increased structural complexity, improves the model performance in terms of similarity. In addition to the real feature, the consideration of the unobservable feature and its random simulation with the hybrid feature--based SNS do not bridge its distance with the target. Based on the SNS modelled with one real feature and an optimised population--based preference (the real feature-based SNS), we are able to approach the state of the target network while having the network growth captured in snapshots.

\begin{figure*}[h] 
	\centering
	\subfigure[The composite performance index from a global perspective]{
		\begin{minipage}[b]{0.48\linewidth}
			\includegraphics[width=1\linewidth]{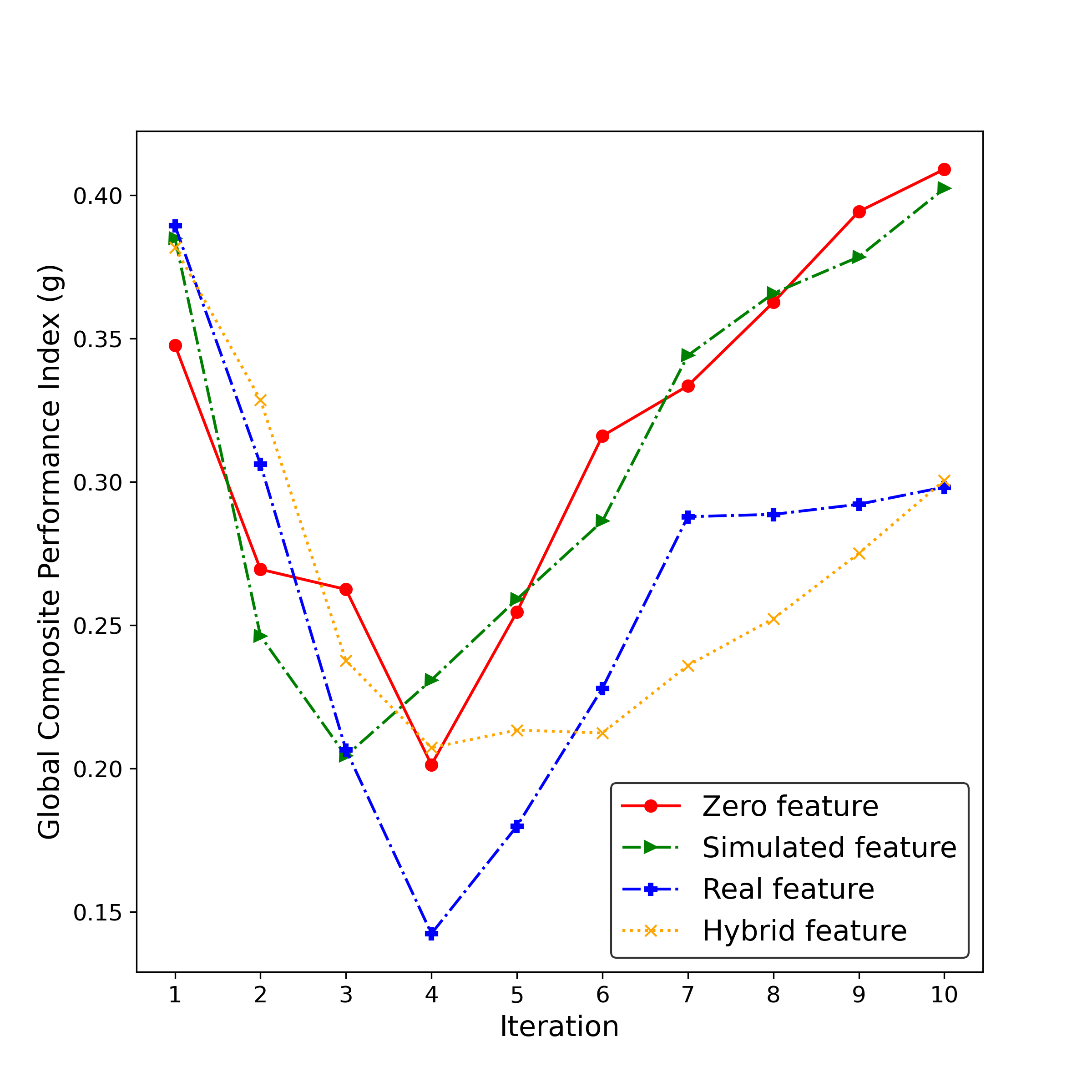}
	\end{minipage}}
	\subfigure[The composite performance index from a local perspective]{
		\begin{minipage}[b]{0.48\linewidth}
			\includegraphics[width=1\linewidth]{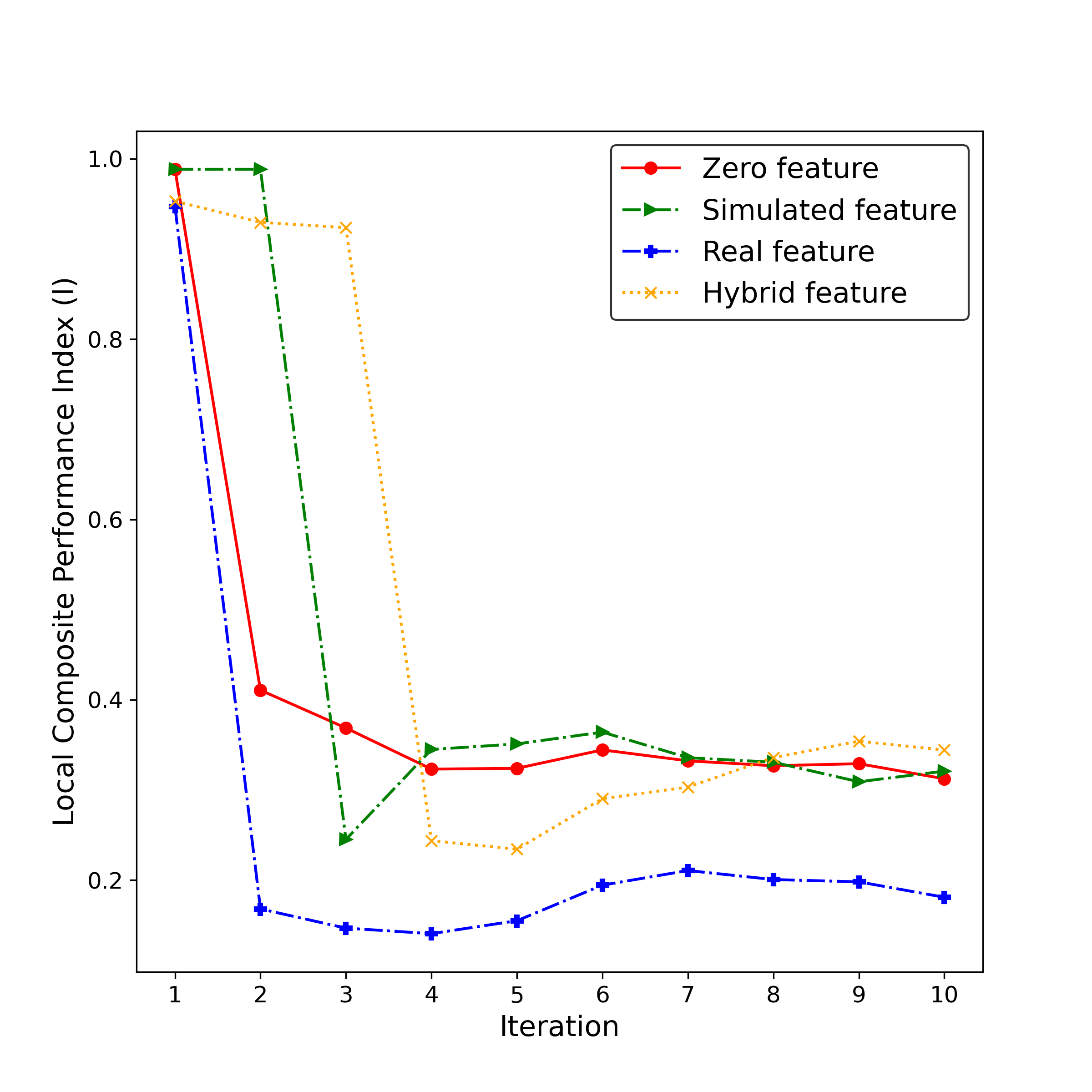}
	\end{minipage}}\\
	\caption{The performance of SNS respectively based on the global and the local composite performance indexes.}
\label{figKLlocalglobal}
\end{figure*}

To better understand the contributions of the global measures and the local measures to the composite performance index, we respectively calculate the global composite performance index (Fig.~\ref{figKLlocalglobal}(a)) and the local composite performance index (Fig.~\ref{figKLlocalglobal}(b)), which each covers the five global measures and the five local measures of networks that respectively contribute to the $50\%$ of the composite performance index employed in the optimisation process.

Fig.~\ref{figKLlocalglobal}(a) shows the values of composite performance index from the global perspective (composite performance index (g)) based on the average differences between simulations and target networks in density, modularity, assortativity, shortest path length distribution and degree distribution. For value-based measures, including density, modularity and assortativity, we use Manhattan distance to calculate their respective difference from the target. In contrast, the differences of distribution-based measures including the shortest path length and the degree distributions are measured with the KL divergence.
Their values range from $0.14$ to $0.41$ (both for real feature SNS), and share a similar decreasing trend in the first three iterations with the composite performance index calculated from both the global and the local perspective (see Fig.~\ref{fig1}). Generally all the SNS reach the lowest global composite performance index (g) around the fourth iteration and then deviate from the target with an increasing global composite performance index (g). The real feature-based SNS, compared with the other SNSs, achieves a lower value in the third, fourth and fifth iterations.

Fig.~\ref{figKLlocalglobal}(b) shows the composite performance index from the local perspective (composite performance index (l)) based on the differences between simulations and target network in the significance profiles and the clustering coefficient. Similarly, the differences in the value-based measures including the significance profiles, and the distribution-based measures including the clustering coefficient, are respectively measured with the Manhattan distance and the KL divergence.
The trend of local composite performance index (l) for each SNS is similar, with a larger value range from $0.11$ (for real feature SNS) to $1.0$ (for simulated feature SNS and hybrid feature SNS) , which decreases sharply in the first two iterations and then fluctuates around the lowest values between $0.12$ and $0.99$. Generally all the SNSs reach their lowest composite performance index (l) around $0.2$, except for the real feature-based SNS, which achieves the lowest value of $0.12$. The real feature-based SNS keeps a lower composite performance index (l) than the other SNSs starting from the second iteration, similar to the trend of composite performance index calculated from both the global and the local perspective (see Fig.~\ref{fig1} and Fig.~\ref{figKLlocalglobal}(b)). 

For each SNS, with its iterative application, the global composite performance index (g) is smaller than the local composite performance index (l) in the first iteration. This indicates that the similarity between real and simulated networks, with small number of edges added, is higher when comparing networks from the global perspective than from the local perspective. Additionally, the real feature-based SNS achieves the smallest composite performance index (best performance) with the highest level of efficiency and similarity when looking at the local and global perspectives separately as well as when they are combined. 

\subsection{Complexity of the social network simulation}
\label{complexity}
We conduct a comparative analysis of the structural complexity built up through the iterative application of zero feature--based SNS, real feature--based SNS, simulated feature--based SNS and the hybrid feature--based SNS. The network states of social network simulations and their changes across iterations are assessed based on the global and the local measures involved in the composite performance index of the optimisation process in section \ref{section32}.

\subsubsection{Global perspective}
To have a better understanding of the network states and their changes across iterations from the global perspective connected with interactions over large number of actors, we conduct a comparative analysis of social network simulations from each SNS based on density, modularity, assortativity, degree distribution and shortest path length distribution.

\begin{figure*}[h!] 
	\centering
	\subfigure[The number of edges]{
		\begin{minipage}[b]{0.48\linewidth}
			\includegraphics[width=1\linewidth]{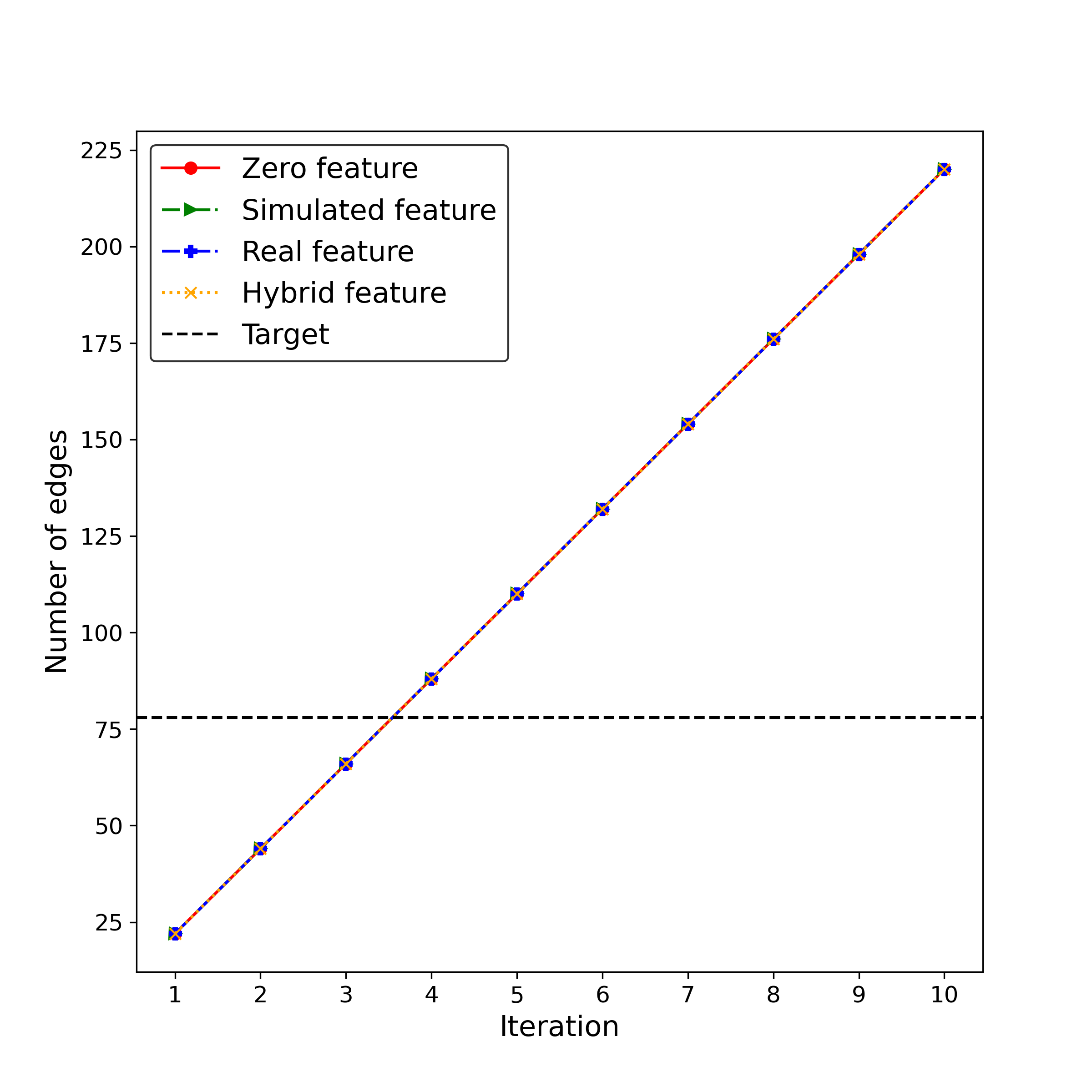}
	\end{minipage}}
	\subfigure[Density of networks]{
		\begin{minipage}[b]{0.48\linewidth}
			\includegraphics[width=1\linewidth]{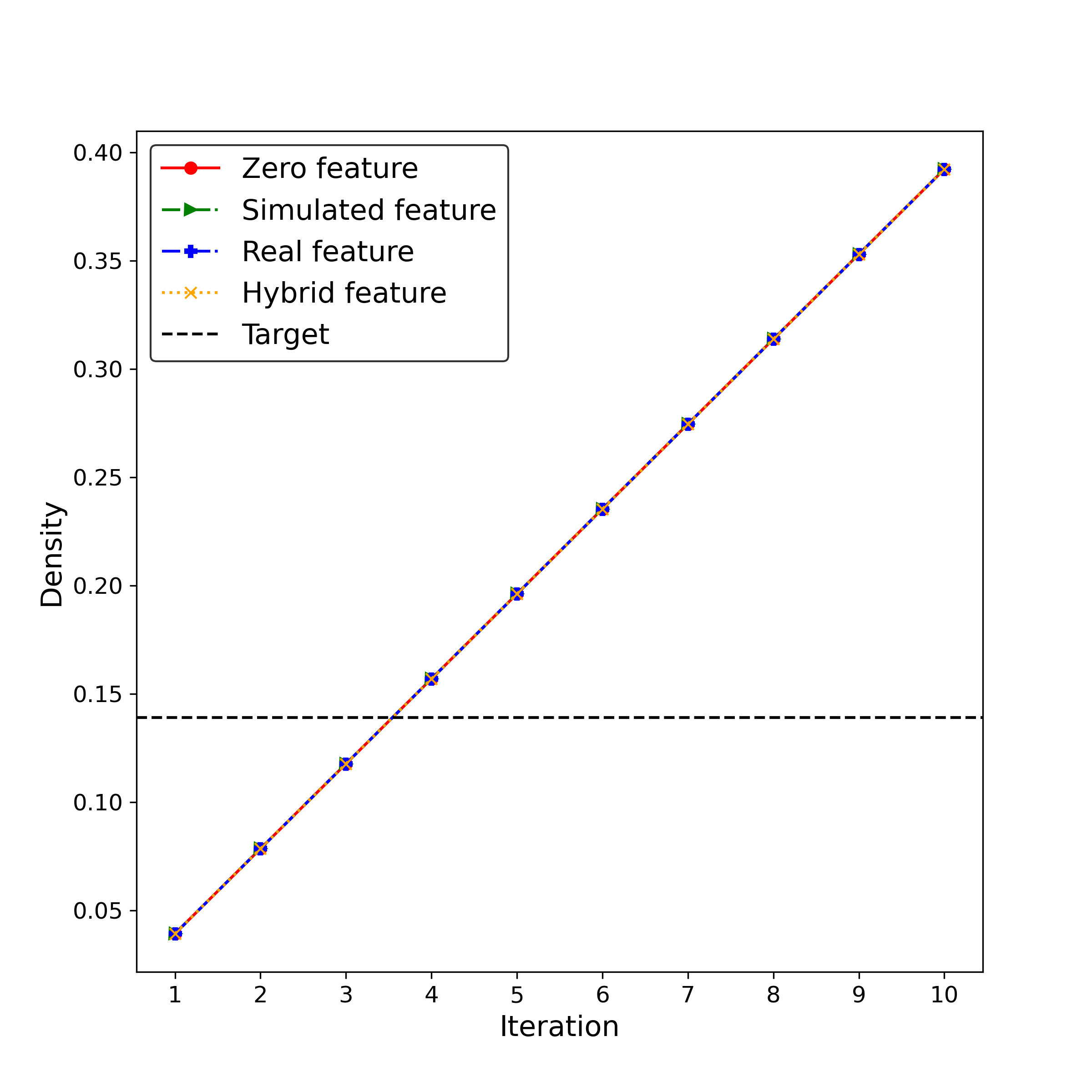}
	\end{minipage}}\\
	\caption{The number of edges and the density of social network simulations over the iterations}
\label{fig3}
\end{figure*}

\paragraph{Density} of networks is calculated as a fraction of existing edges to all possible connections between all node pairs \citep{IEEEexample:musial2013kind}. The density of networks increases with more edges added over the iterations. As is shown in Table~\ref{densityTarget}, there are $78$ edges in the target network, with a network density of $0.1390$. In Fig.~\ref{fig3}, all the simulated social networks share the same values of the edge number and density within each iteration under the assumed 4\% edges (22 edges) to be rewired.
The number of edges and the density of networks approach the values of the target network in the $4_{th}$ iteration.

\begin{table}[h!]
\centering
\caption{The number of nodes and edges, as well as the density of the target network.}
\label{densityTarget}
\setlength{\tabcolsep}{3pt}
\renewcommand{\arraystretch}{1.5}
\begin{tabular}{|c|c|c|c|c|c|}
\hline
  & No. nodes & No. edges & Density &  Modularity & Assortativity \\
\hline
Target & 34 & 78 & 0.14 &0.42 & -0.48\\
\hline
\end{tabular}
\end{table}

\begin{figure*}[h!] 
	\centering
		\subfigure[Modularity of networks]{
		\begin{minipage}[b]{0.48\linewidth}
			\includegraphics[width=1\linewidth]{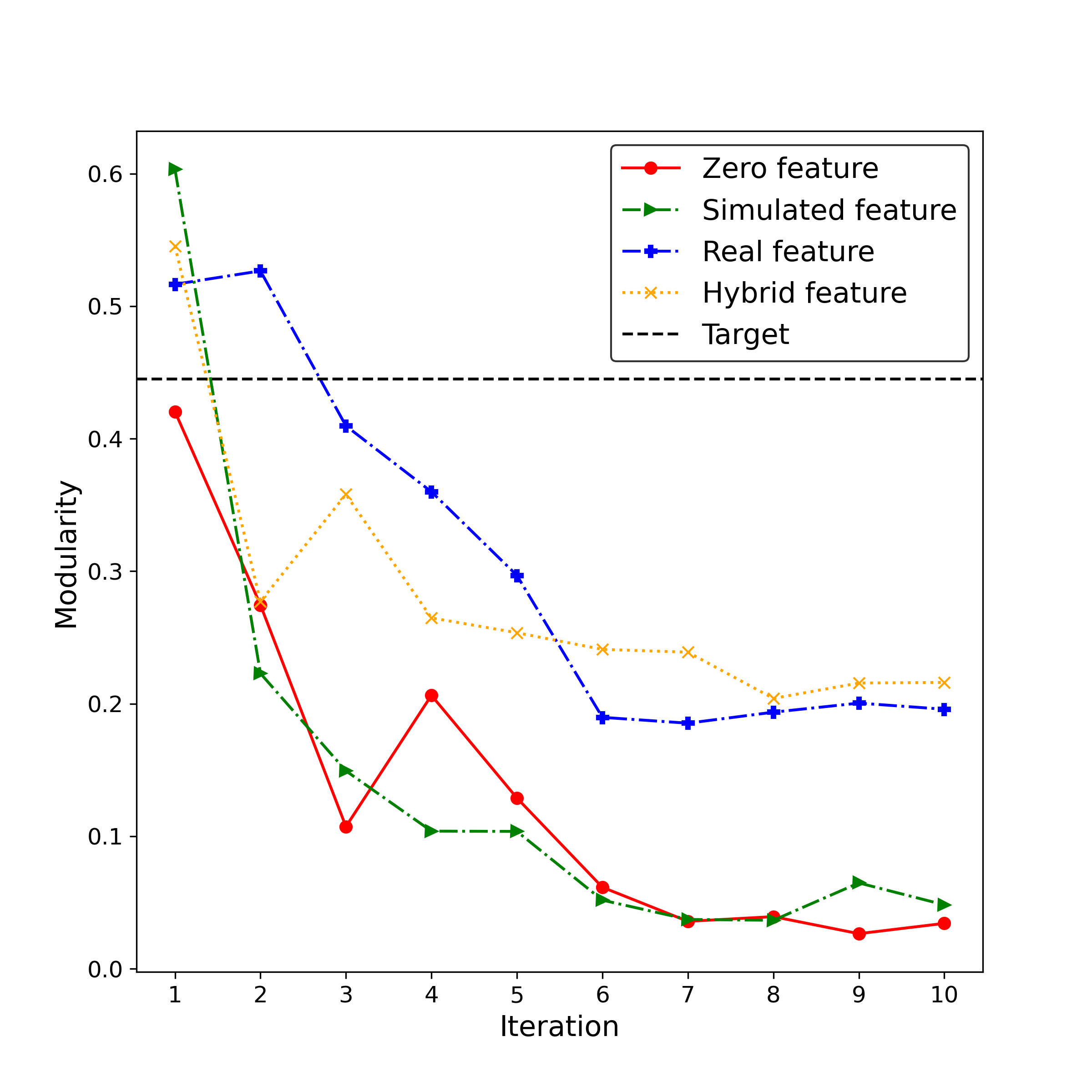}
	\end{minipage}}
	\subfigure[Assortivity of networks]{
		\begin{minipage}[b]{0.48\linewidth}
			\includegraphics[width=1\linewidth]{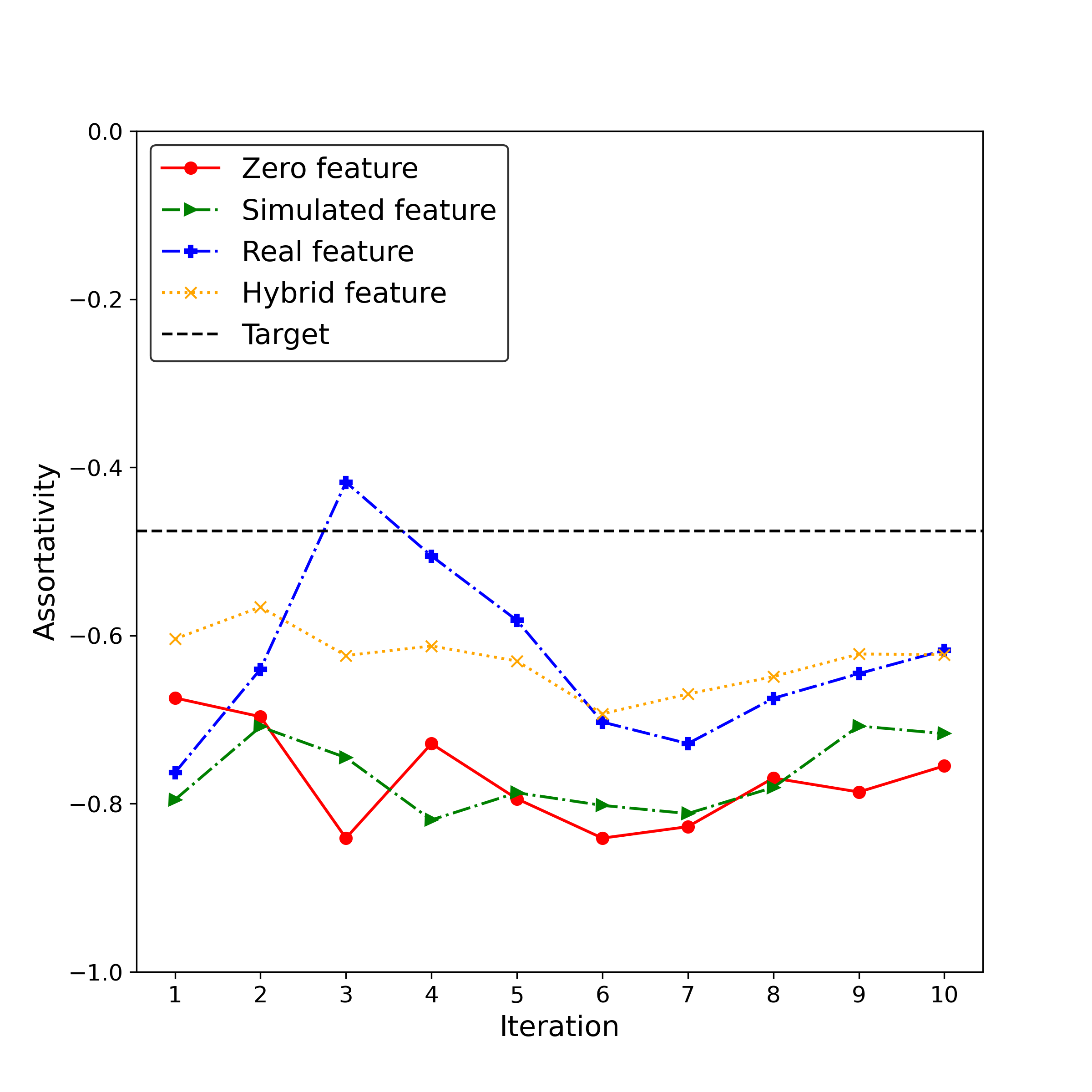}
	\end{minipage}}\\
	\caption{The modularity and the assortivity of social network simulations}
\label{MA}
\end{figure*}

\paragraph{Modularity} measures the extent to which the network is clustered and how strong those clusters are. Its value is between $0$ and $1$, where a larger value represents a strong community structure. As is shown in Fig.~\ref{MA}(a), the modularity of each simulated network decreases over iterations, with a convergence to the target modularity $0.4156$ (See Table~\ref{densityTarget}) in the second iteration. The real feature--based and the hybrid feature--based SNSs generally have the largest modularity over the iterations, indicating the strongest community structure for all the simulated networks. This can be caused by the binary feature, as nodes with the same feature and the same preference tend to have similar behaviours, increasing the number of connections within a community. The social network simulated by the zero feature-based SNS gets nearest to the target modularity in the second iteration.

\paragraph{Assortativity} measures whether the linked nodes have a similar degree (number of connections). Its value ranges between $-1$ and $1$. With a larger positive value of assortativity, nodes tend to link to other nodes with the same or similar degree, indicating a stronger assortative mixing pattern \citep{IEEEexample:musial2013kind}. As is shown in Fig.~\ref{MA}(b), throughout the iterations, the assortativity of simulated networks is negative and fluctuates around $-0.7$. This can be caused by the weak preference for higher degrees and the strong preference for dissimilar features and shortest path lengths.
The real feature-based SNS is closest to the target assortativity of $-0.4756$ (see Table~\ref{densityTarget}) in the third iteration, while the other simulated networks approach the target in the third or fourth iteration.

\begin{figure*}[h!] 
	\centering
	\subfigure[Zero feature--based SNS]{
		\begin{minipage}[b]{0.48\linewidth}
			\includegraphics[width=1\linewidth]{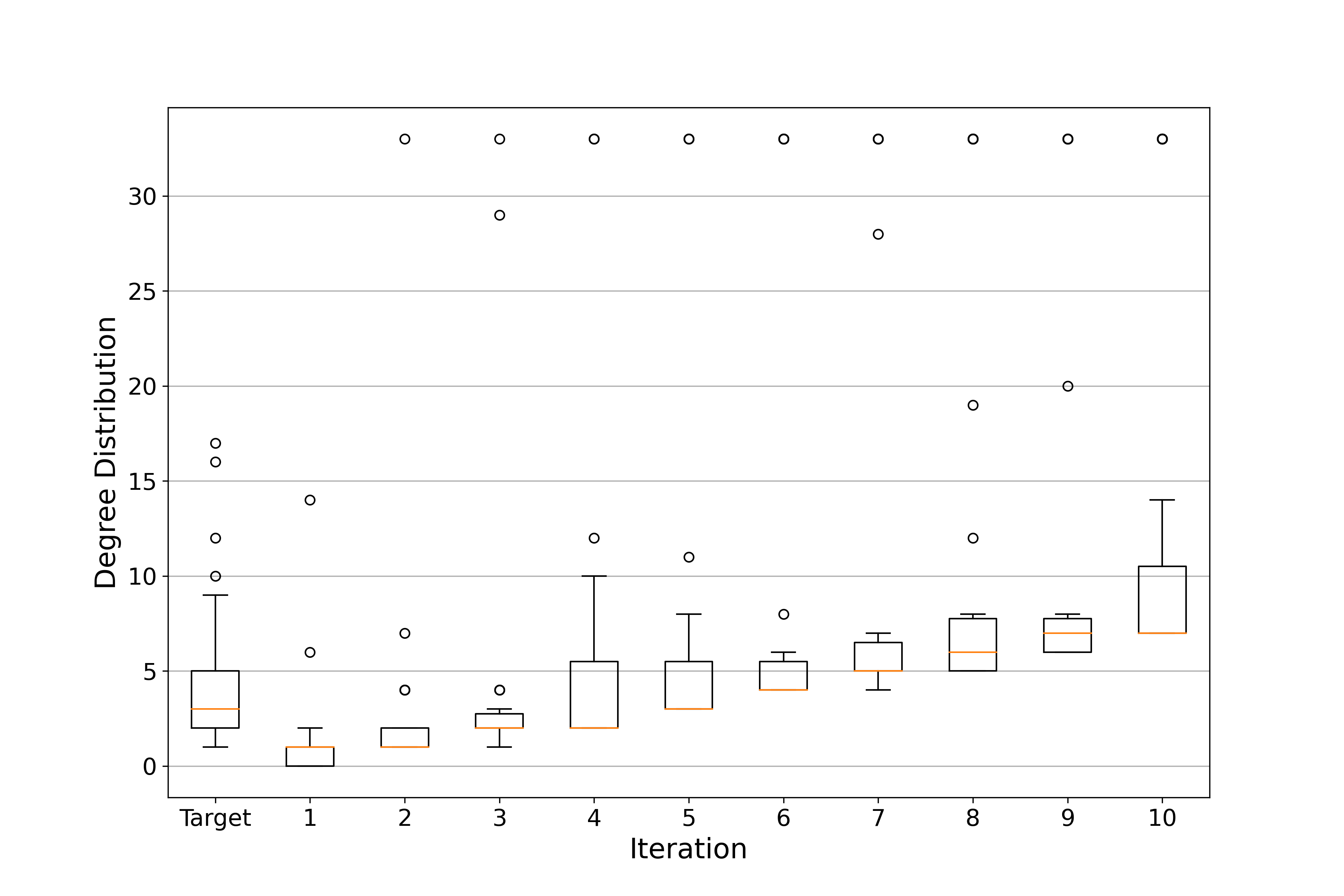}
	\end{minipage}}
	\subfigure[Real feature--based SNS]{
		\begin{minipage}[b]{0.48\linewidth}
			\includegraphics[width=1\linewidth]{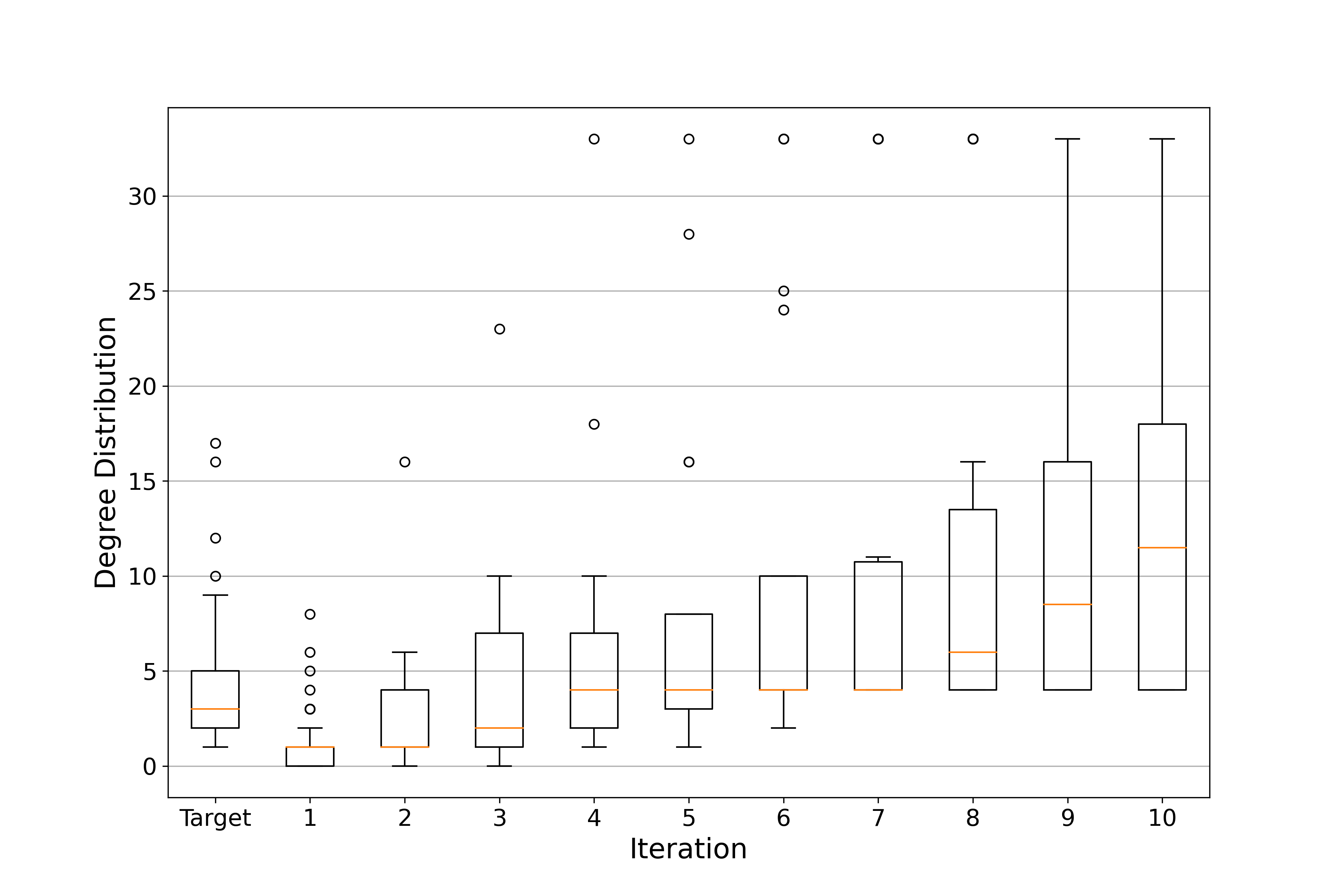}
	\end{minipage}}\\
	\subfigure[Simulated feature--based SNS]{
		\begin{minipage}[b]{0.48\linewidth}
			\includegraphics[width=1\linewidth]{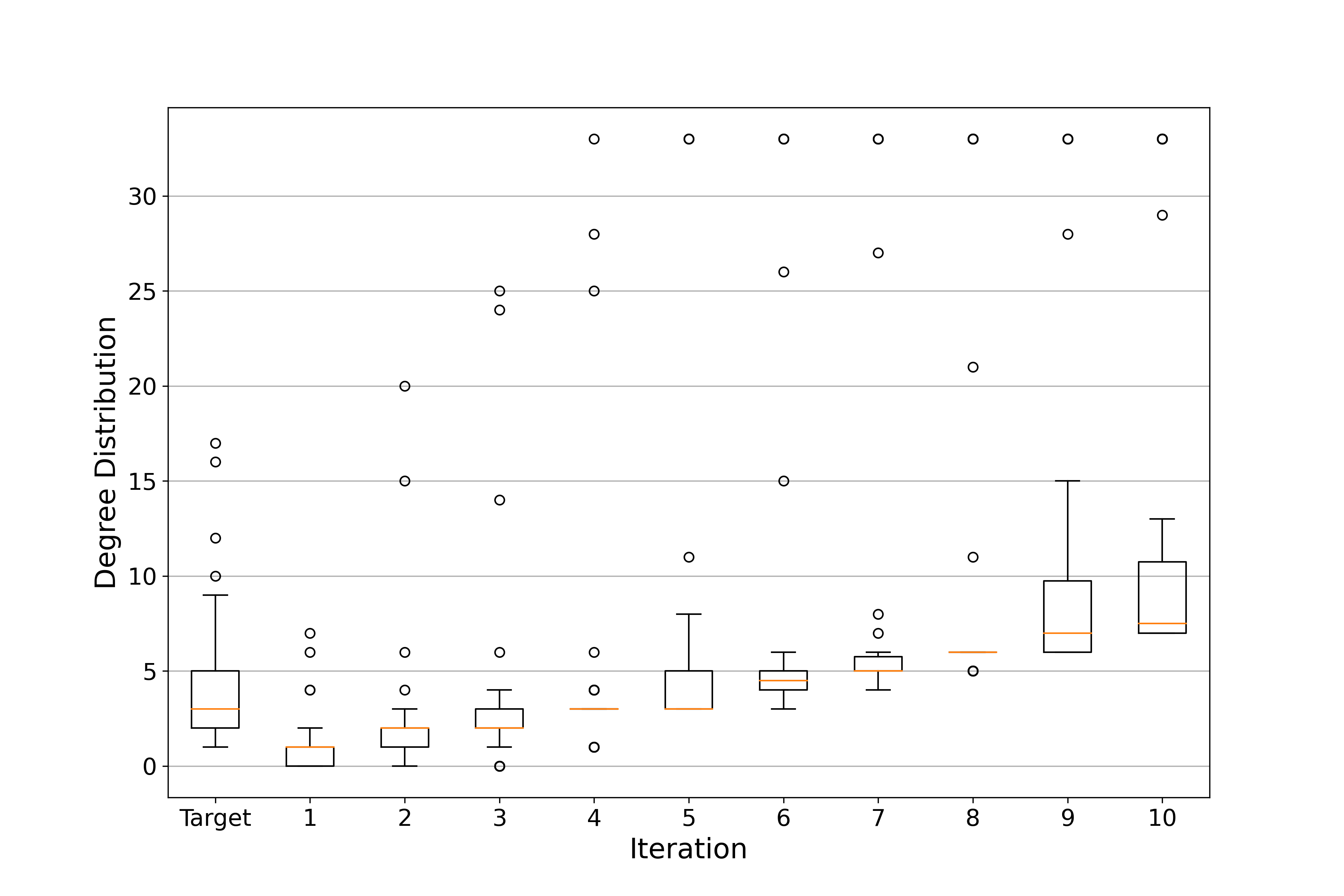}
	\end{minipage}}
	\subfigure[Hybrid feature--based SNS]{
		\begin{minipage}[b]{0.48\linewidth}
			\includegraphics[width=1\linewidth]{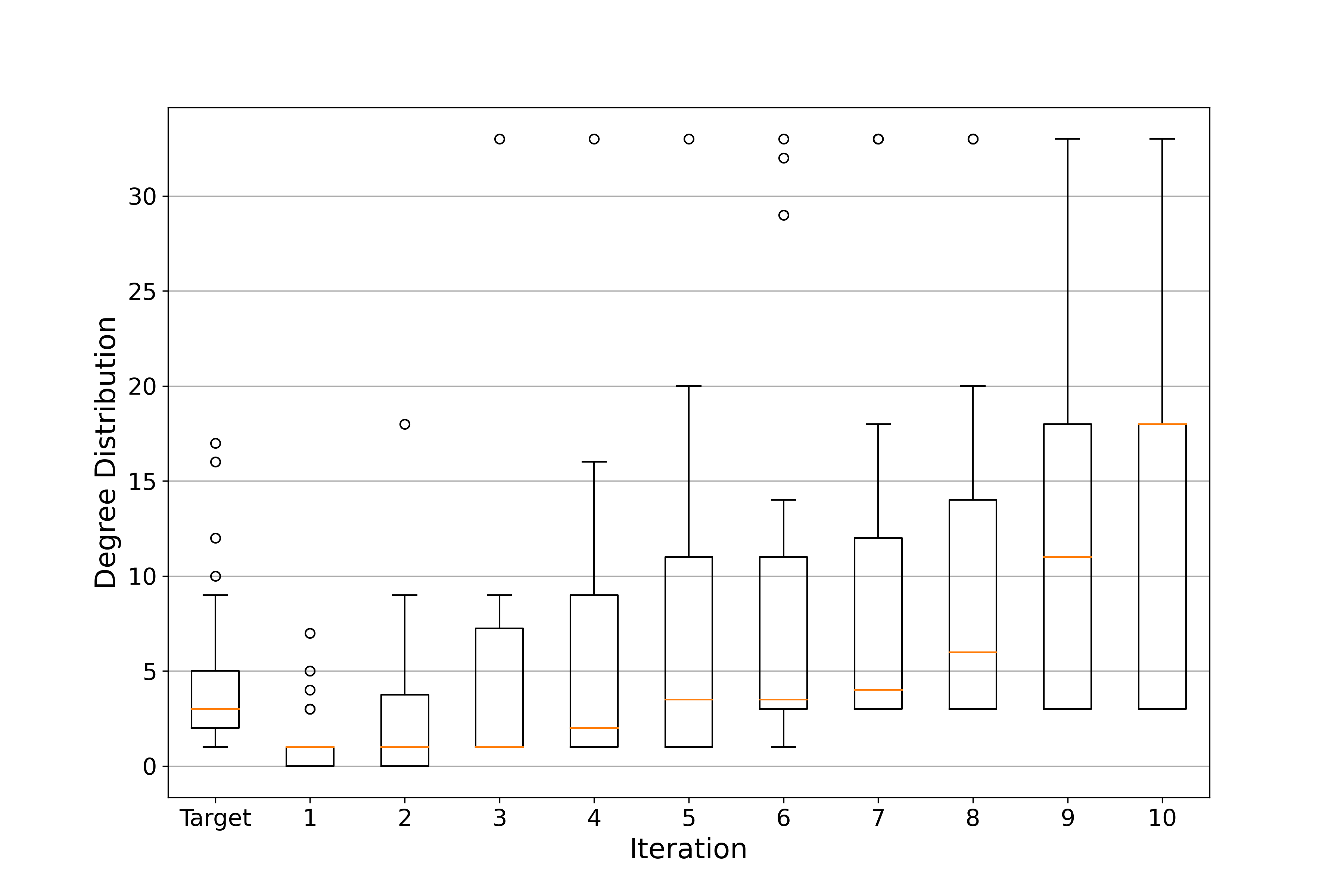}
	\end{minipage}}\\
	\caption{The degree distribution of networks}
\label{degD}
\end{figure*}

\begin{figure*}[h!] 
	\centering
		\subfigure[Average value]{
		\begin{minipage}[b]{0.48\linewidth}
			\includegraphics[width=1\linewidth]{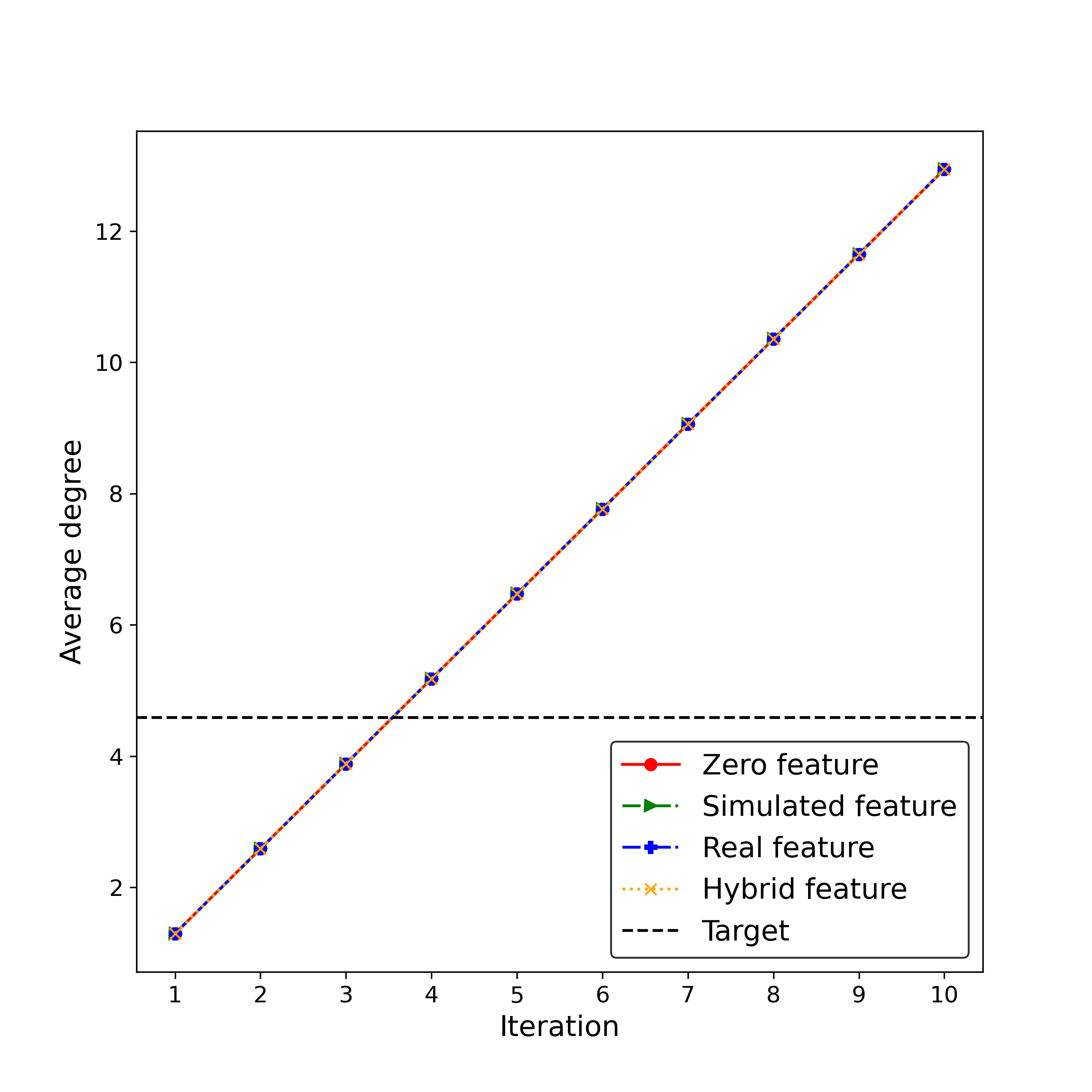}
	\end{minipage}}
	\subfigure[KL divergence]{
		\begin{minipage}[b]{0.48\linewidth}
			\includegraphics[width=1\linewidth]{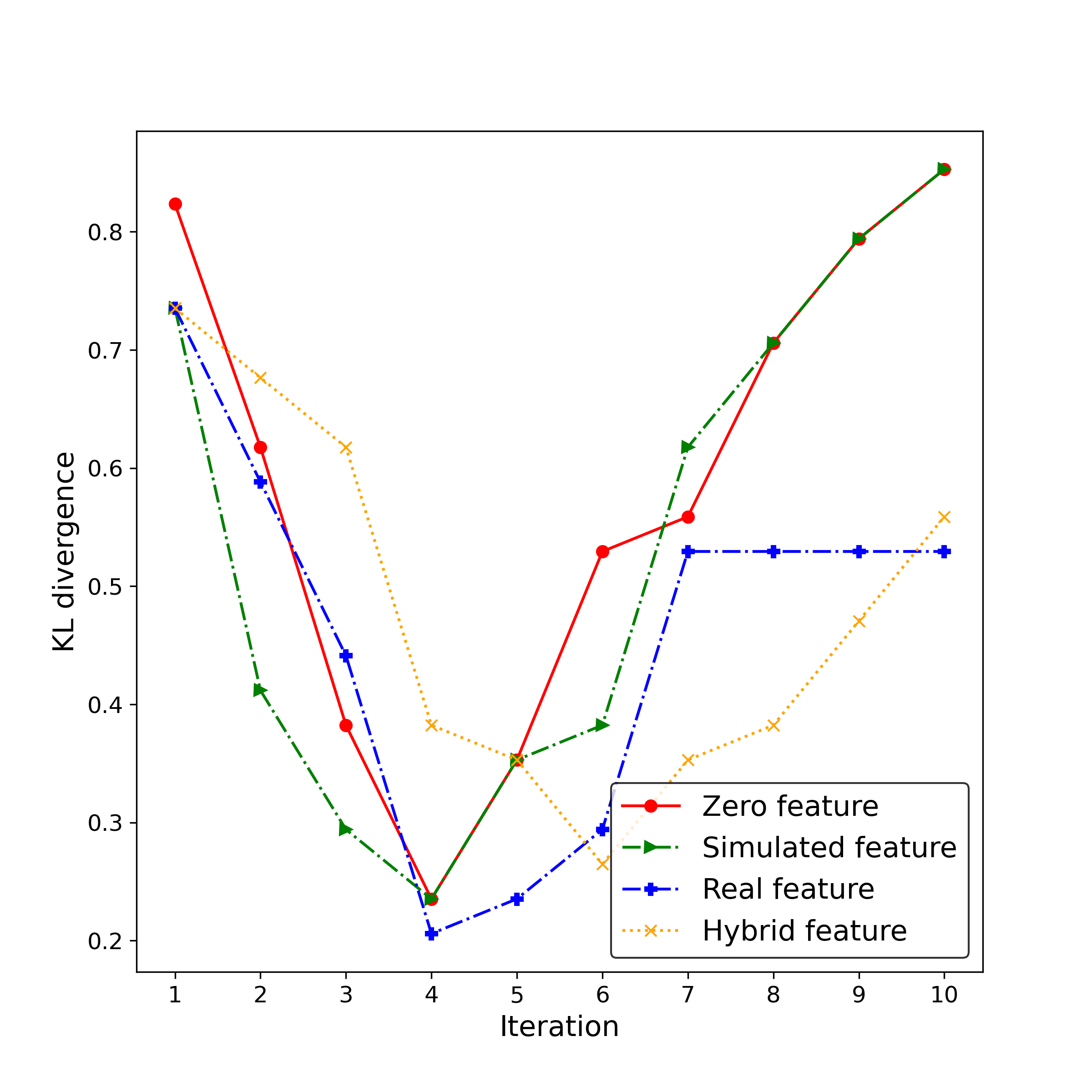}
	\end{minipage}}\\
	\caption{The average value and the KL divergence of degree distribution.}
\label{degKL}
\end{figure*}

\paragraph{Degree distribution} is a distribution of node degrees in a given network
~\citep{IEEEexample:musial2013kind}. As shown in Table~\ref{degTarget}, the node degree of the target network fluctuates around the average value of $4.59$ with a standard deviation of $3.82$, ranging from $1.00$ to $17.00$.  Fig.~\ref{degD} shows node degree distributions of the target network and the simulated social networks over iterations. For each SNS, the range of the degree distribution generally gets larger with more edges added over the iterations. 
This indicates that the node degrees vary within the network and change over time, implying the diversity of network topology. The average values of the degree distribution fluctuates around $5$ with a small increase over time, which firstly approach the average node degree of the network and then deviates through iterations. The degree distributions of zero feature-based SNS (see Fig.~\ref{degD}(a)) and the real feature-based SNS (see Fig.~\ref{degD}(b)) have patterns more similar to the target network than the simulated feature-based SNS (see Fig.~\ref{degD}(c)) and the hybrid feature-based SNS (see Fig.~\ref{degD}(d)). And to quantify the level of similarity, we calculate the KL divergence based on the target degree distribution (see Fig.~\ref{degKL}(a)).

\begin{table}[h!]
\centering
\caption{The degree distribution of the target network.}
\label{degTarget}
\setlength{\tabcolsep}{3pt}
\renewcommand{\arraystretch}{1.5}
\begin{tabular}{|c|c|c|c|c|c|c|}
\hline
  & Average & Standard deviation &  75\% quantile & 25\% quantile &Maximum & Minimum\\
\hline
Target & 4.59 &3.82  &5.00 &  2.00 & 17.00 & 1.00 \\
\hline
\end{tabular}
\end{table}

As is shown in Fig.~\ref{degKL}(a), all the simulated social networks have the same average node degree value, which increases systematically from $1$ to $13$ with the same number of edges added over the iterations. The average degree approaches the average degree of the target network in the third iteration and then deviates to higher values. 

In Fig.~\ref{degKL}(b), all the simulated networks have similar trend of KL divergence in terms of degree distribution, which firstly decreases to the lowest point at around fourth iteration and then increases. 
The real feature-based SNS, in the sixth iteration, reaches the lowest KL divergence among all the SNS and across all iterations. Correspondingly, the degree distribution of the simulated social network obtained using the real feature-based SNS in the fourth iteration (see Fig.~\ref{degD}(c)) has similar average values and quantiles as the target network, with few nodes having larger degree than nodes in the target network.

\begin{figure*}[h!] 
	\centering
	\subfigure[Zero feature--based SNS]{
		\begin{minipage}[b]{0.48\linewidth}
			\includegraphics[width=1\linewidth]{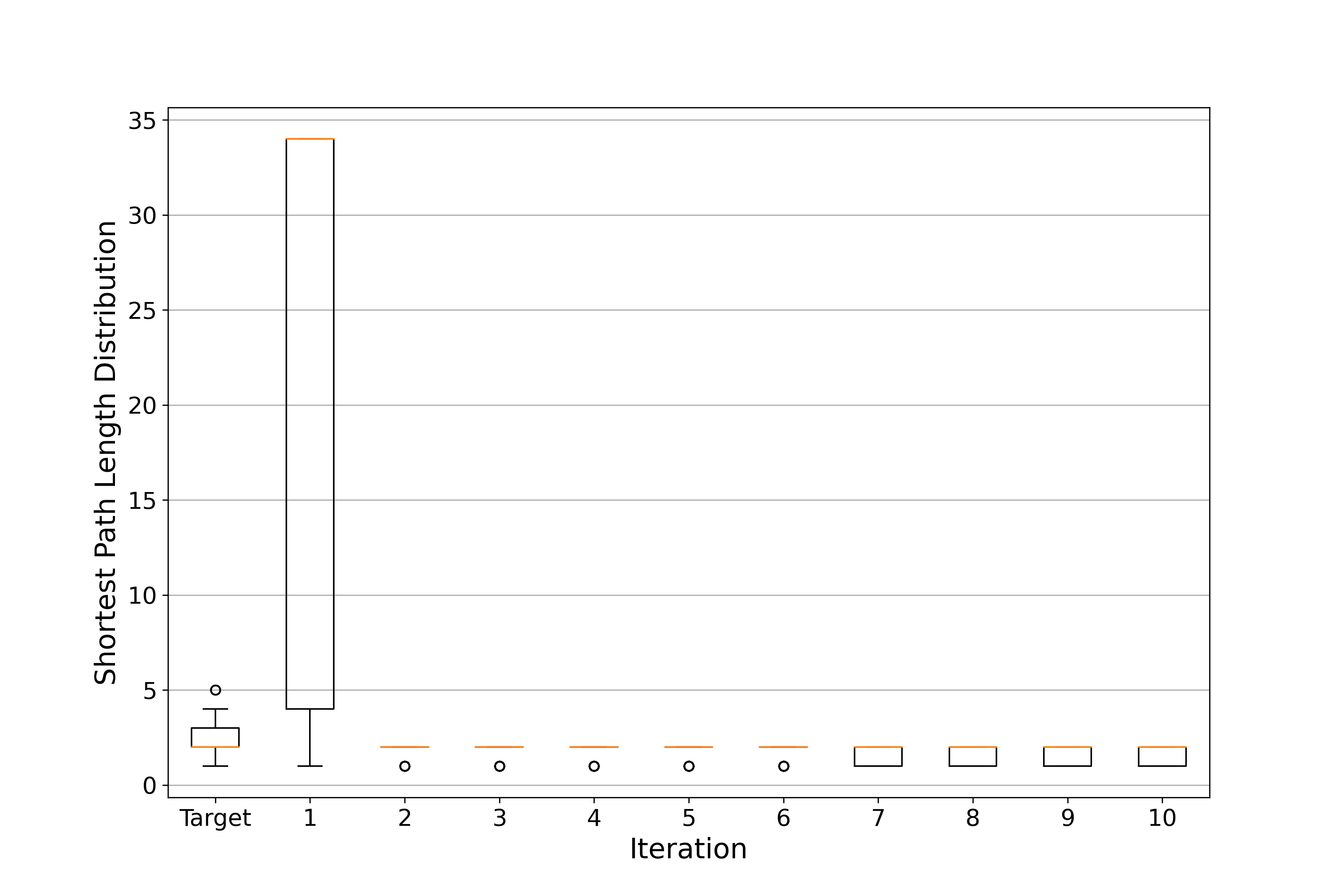}
	\end{minipage}}
	\subfigure[Real feature--based SNS]{
		\begin{minipage}[b]{0.48\linewidth}
			\includegraphics[width=1\linewidth]{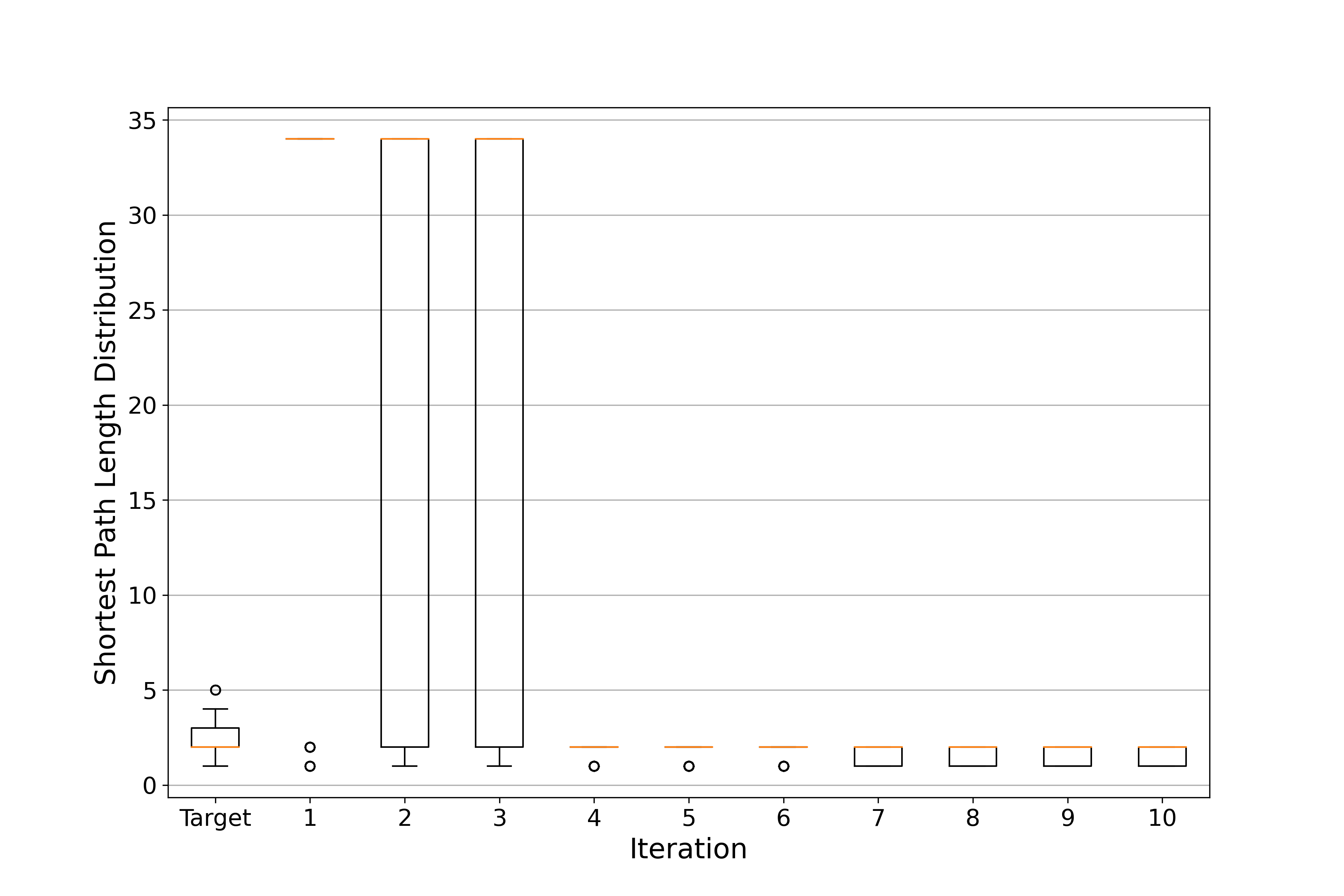}
	\end{minipage}}\\
	\subfigure[Simulated feature--based SNS]{
		\begin{minipage}[b]{0.48\linewidth}
			\includegraphics[width=1\linewidth]{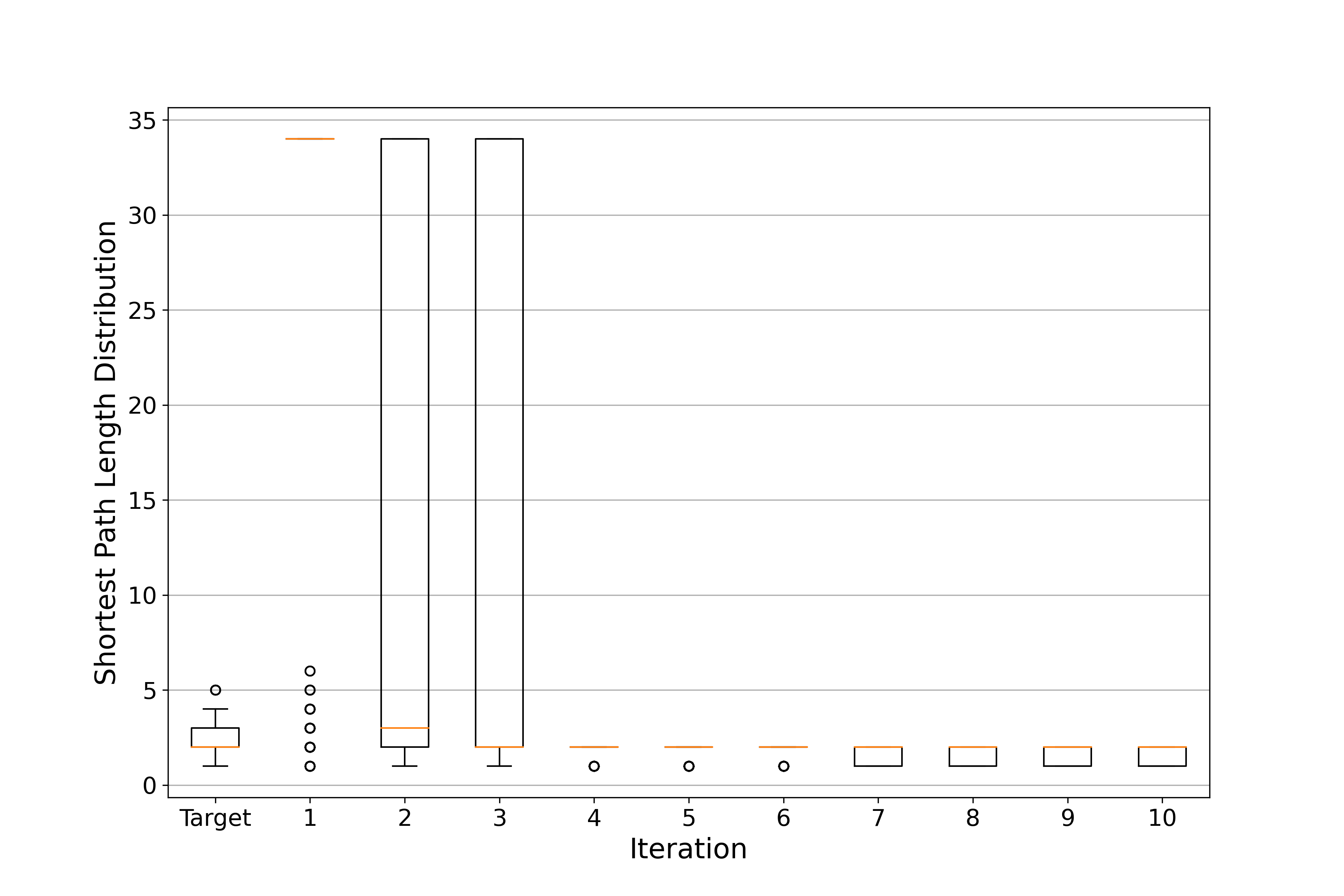}
	\end{minipage}}
	\subfigure[Hybrid feature--based SNS]{
		\begin{minipage}[b]{0.48\linewidth}
			\includegraphics[width=1\linewidth]{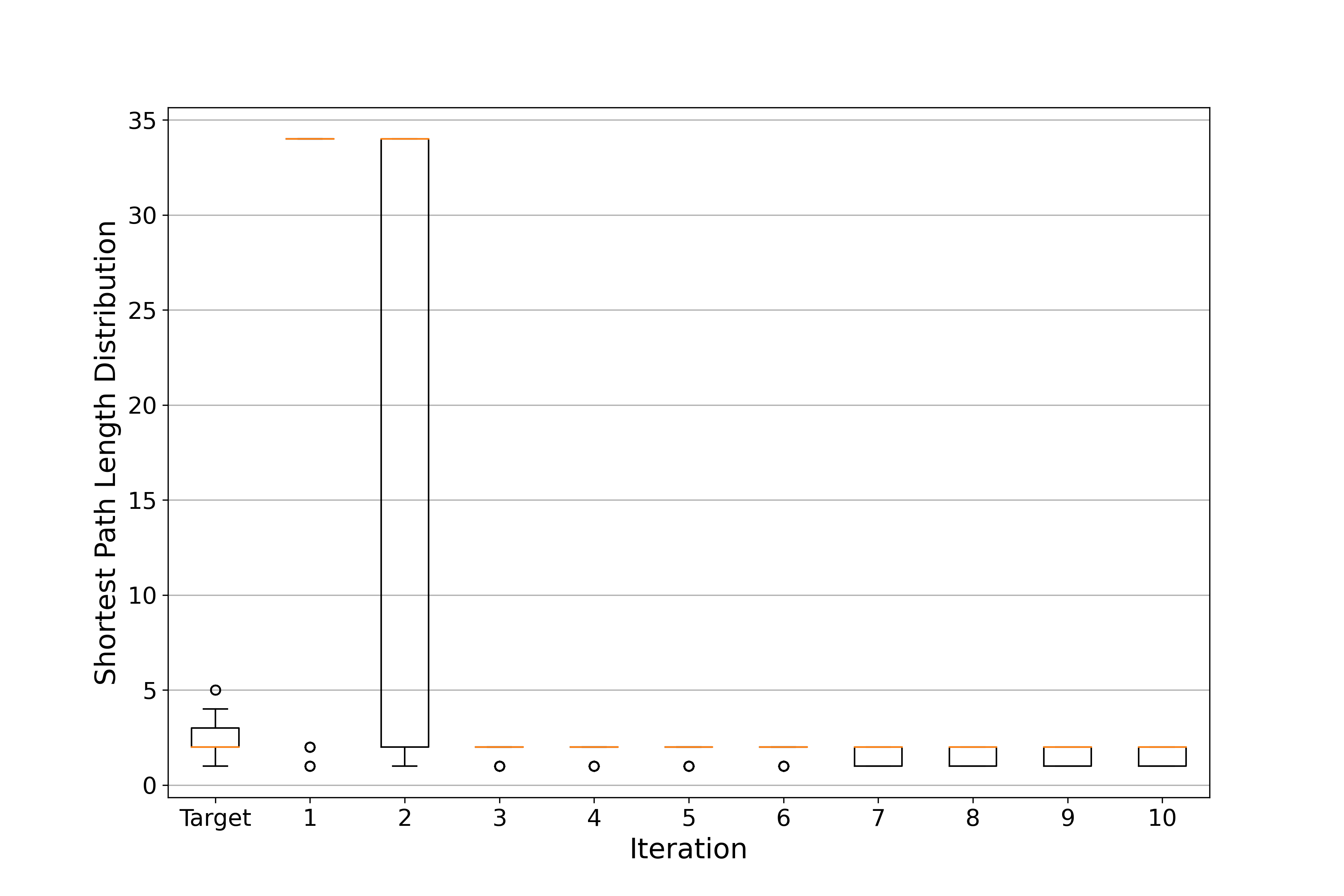}
	\end{minipage}}\\
	\caption{The shortest path length distribution of networks, where the path lengths value between $1$ and $34$ as we assume no self-links and replace the unavailable (infinite) paths with the lengths of $34$.}
\label{pathpath}
\end{figure*}

\paragraph{Shortest path length} between two nodes is defined as the number of edges along the shortest path between a pair of nodes \citep{IEEEexample:musial2013kind}. As shown in Table~\ref{pathTarget}, the shortest path length in the target network fluctuates around an average value of $2.41$ with a standard deviation of $1$, ranging from $1.00$ to $5.00$. Fig.~\ref{pathpath} shows the shortest path length distributions of the target network and the simulated social networks over iterations when different SNSs were used. The shortest path lengths of simulated networks value between $1$ and $33$ as we assume no self-links exist. The shortest path length between connected nodes can not approach an upper limit at $34$ given $33$ nodes. In contrast, the shortest path length between the unconnected nodes is infinite, which is hard to be measured and visualised in a distribution. Therefore, to describe the unavailable paths between unconnected node pairs, we additionally assign these node pairs with the upper limit for shortest path length $34$. For all SNSs, the shortest path length is distributed around the average value over $20$ in the first iteration and then converges to $2$. This can be caused by the unavailability of paths between node pairs given a small number of edges in the first iteration. 

\begin{table}[h!]
\centering
\caption{The shortest path length distribution of the target network.}
\label{pathTarget}
\setlength{\tabcolsep}{3pt}
\renewcommand{\arraystretch}{1.5}
\begin{tabular}{|c|c|c|c|c|c|c|}
\hline
  & Average & Standard deviation &  75\% quantile & 25\% quantile &Maximum & Minimum\\
\hline
Target & 2.41 &0.93  &3.00 & 2.00 &5.00 & 1.00\\
\hline
\end{tabular}
\end{table}

As is shown in Fig.~\ref{pathpath}(a), the average values of the shortest path length  for all for all the social network simulations decrease with addition of edges, approach the target $2.4$ and then fluctuate around $2$, indicating small-world properties \cite{IEEEexample:musial2013kind}. The zero feature--based SNS hits the target in the second iteration within the shortest time. 

\begin{figure*}[h!] 
	\centering
		\subfigure[Average value]{
		\begin{minipage}[b]{0.48\linewidth}
			\includegraphics[width=1\linewidth]{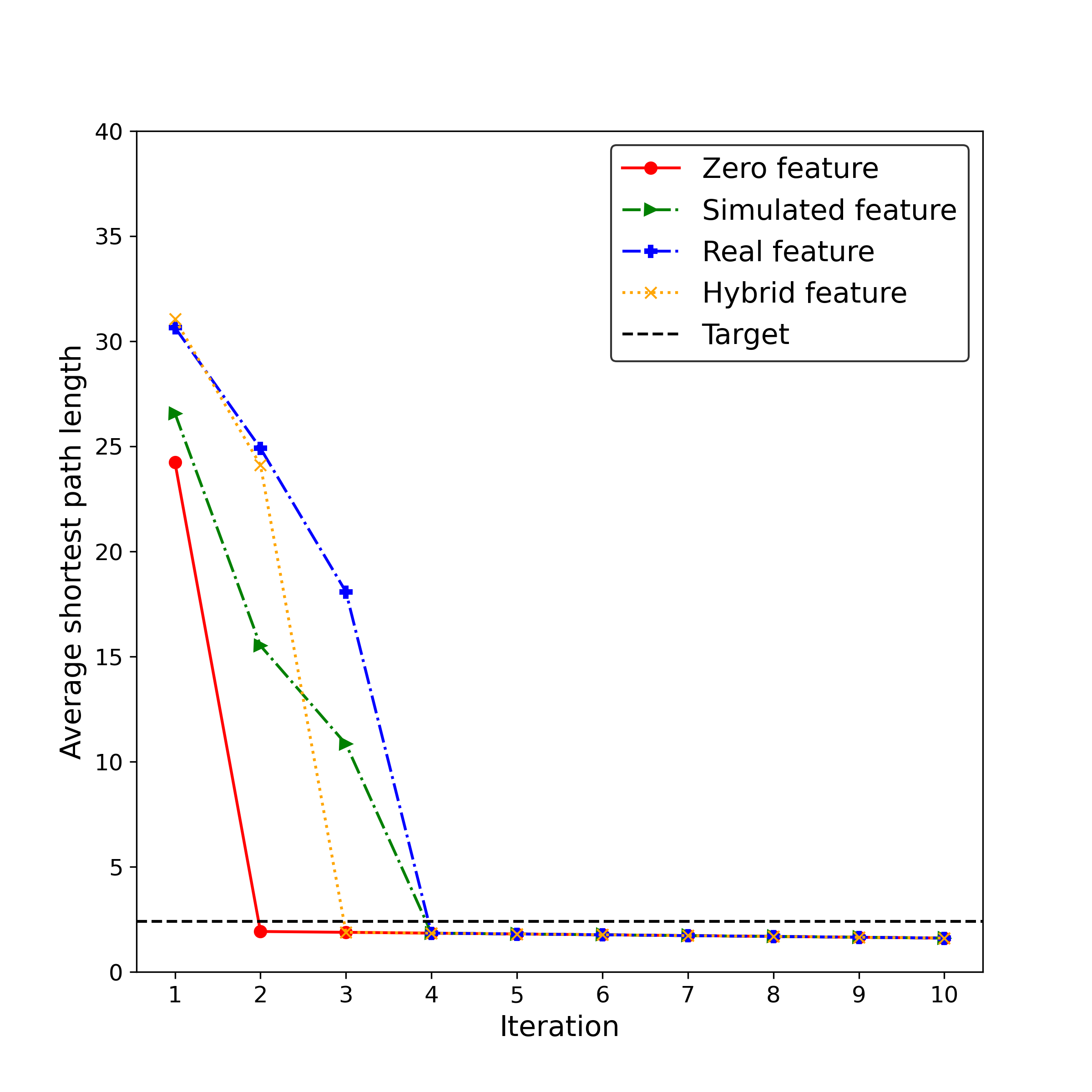}
	\end{minipage}}
	\subfigure[KL divergence]{
		\begin{minipage}[b]{0.48\linewidth}
			\includegraphics[width=1\linewidth]{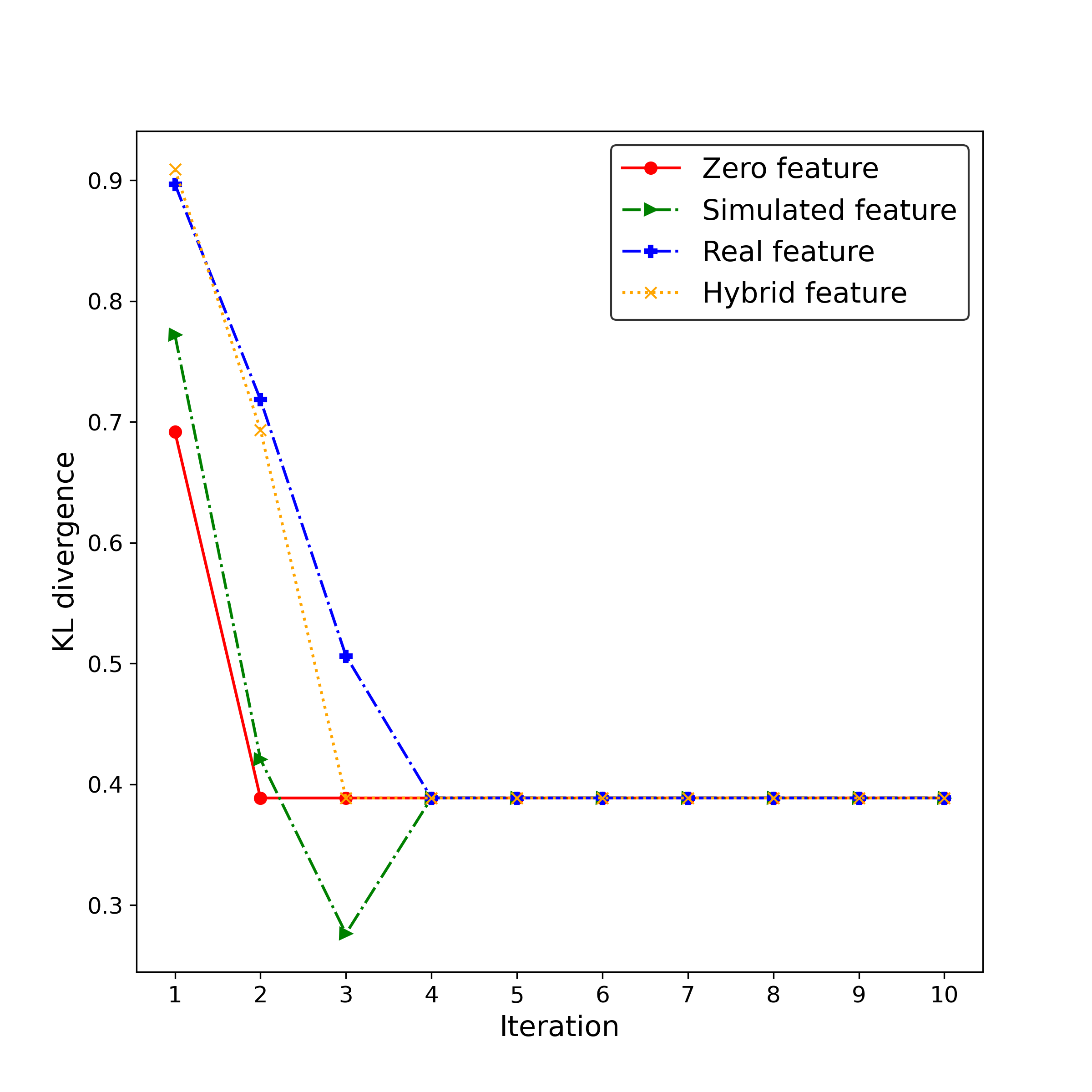}
	\end{minipage}}\\
	\caption{The average value and the KL divergence of shortest path length distribution.}
\label{pathKL}
\end{figure*}

In Fig.~\ref{pathKL}(b), the KL divergence of shortest path length distribution firstly decreases to the lowest values in the third iteration and then keeps static at $0.39$. The SNSs gradually get all the nodes connected and finally result in the same shortest path length distributions, between the value $1$ and $2$, given the same number of edges (see Fig.~\ref{pathpath}). The simulated feature-based SNS achieves the lowest KL divergence in the third iteration, compared with the other SNSs across iterations. 

\subsubsection{Local perspective}
To have a better understanding of the network states and their changes in between, we conduct a comparative analysis of the social networks obtained from running each SNS focusing on the local perspective. We focus here on the clustering coefficient distribution and the significance profiles of the four types of triads.

\begin{figure*}[h!] 
	\centering
	\subfigure[Zero feature--based SNS]{
		\begin{minipage}[b]{0.48\linewidth}
			\includegraphics[width=1\linewidth]{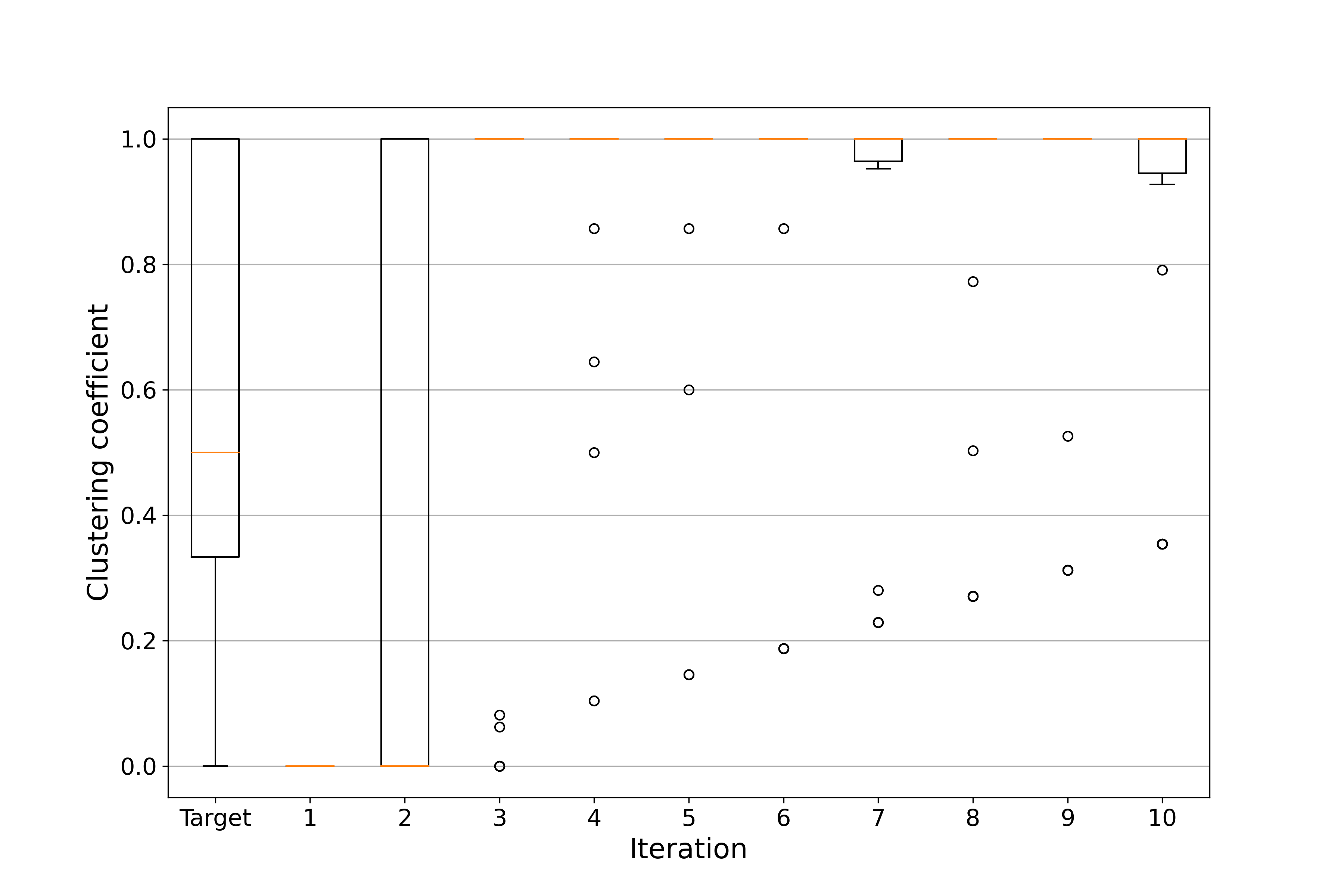}
	\end{minipage}}
	\subfigure[Real feature--based SNS]{
		\begin{minipage}[b]{0.48\linewidth}
			\includegraphics[width=1\linewidth]{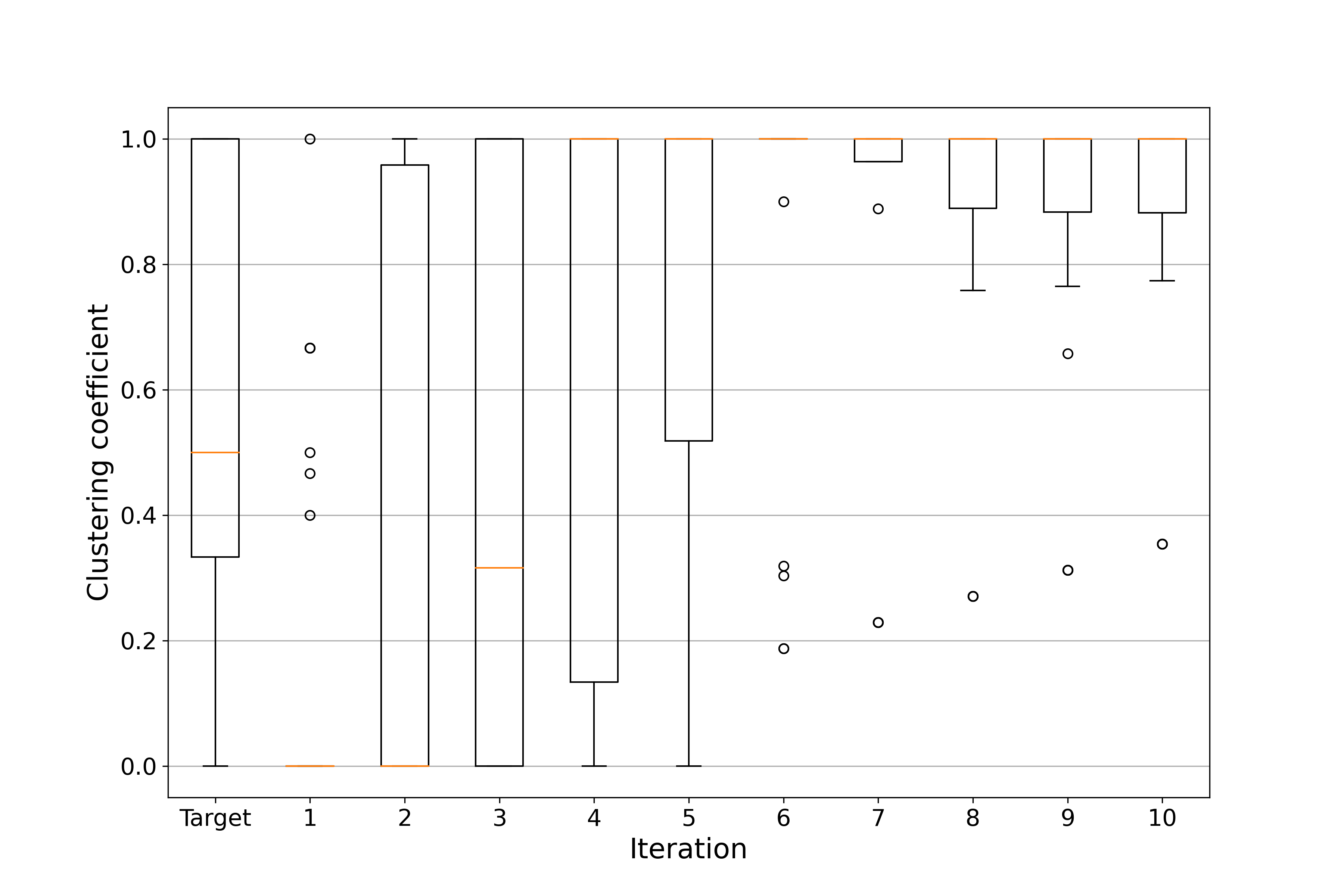}
	\end{minipage}}\\
	\subfigure[Simulated feature--based SNS]{
		\begin{minipage}[b]{0.48\linewidth}
			\includegraphics[width=1\linewidth]{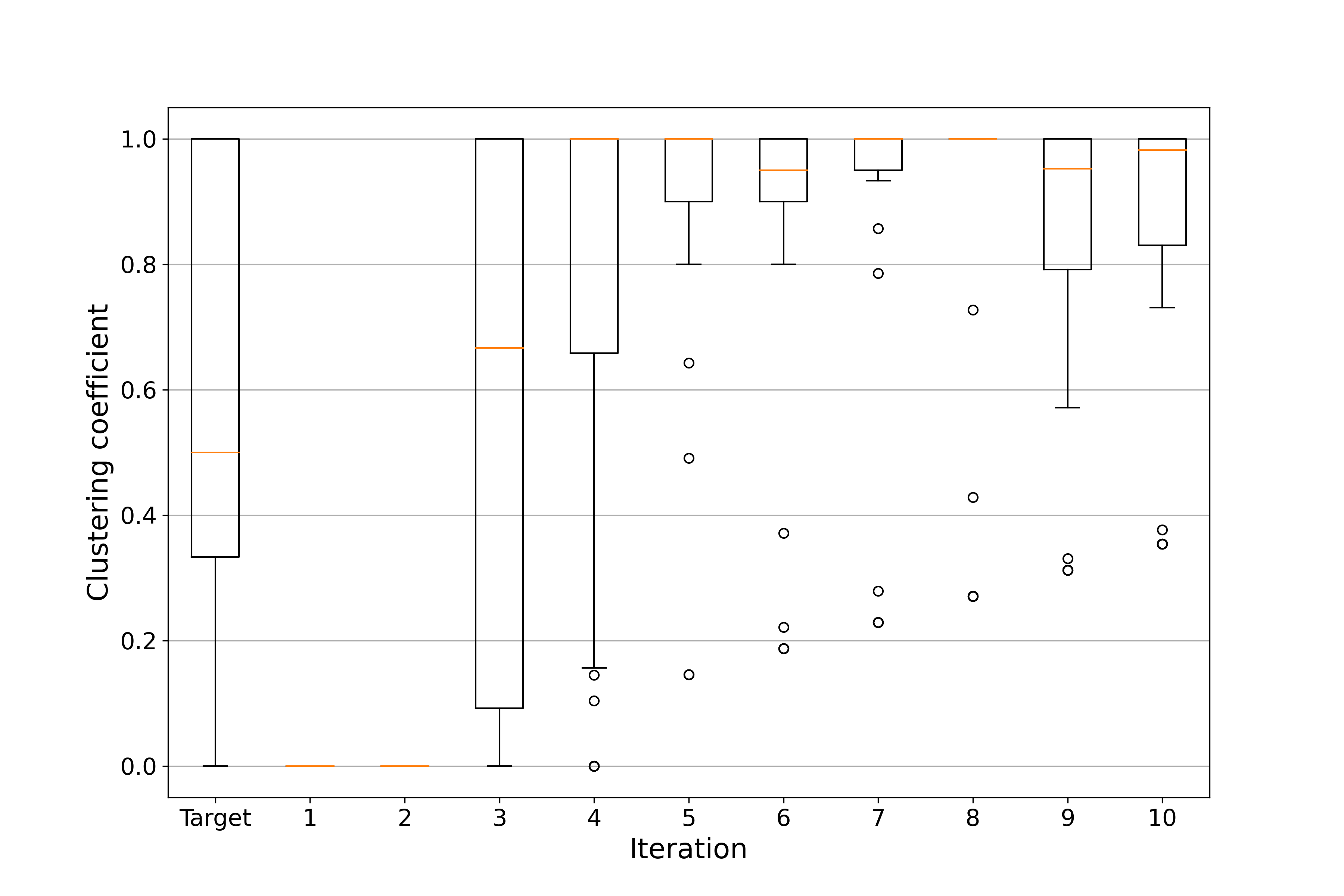}
	\end{minipage}}
	\subfigure[Hybrid feature--based SNS]{
		\begin{minipage}[b]{0.48\linewidth}
			\includegraphics[width=1\linewidth]{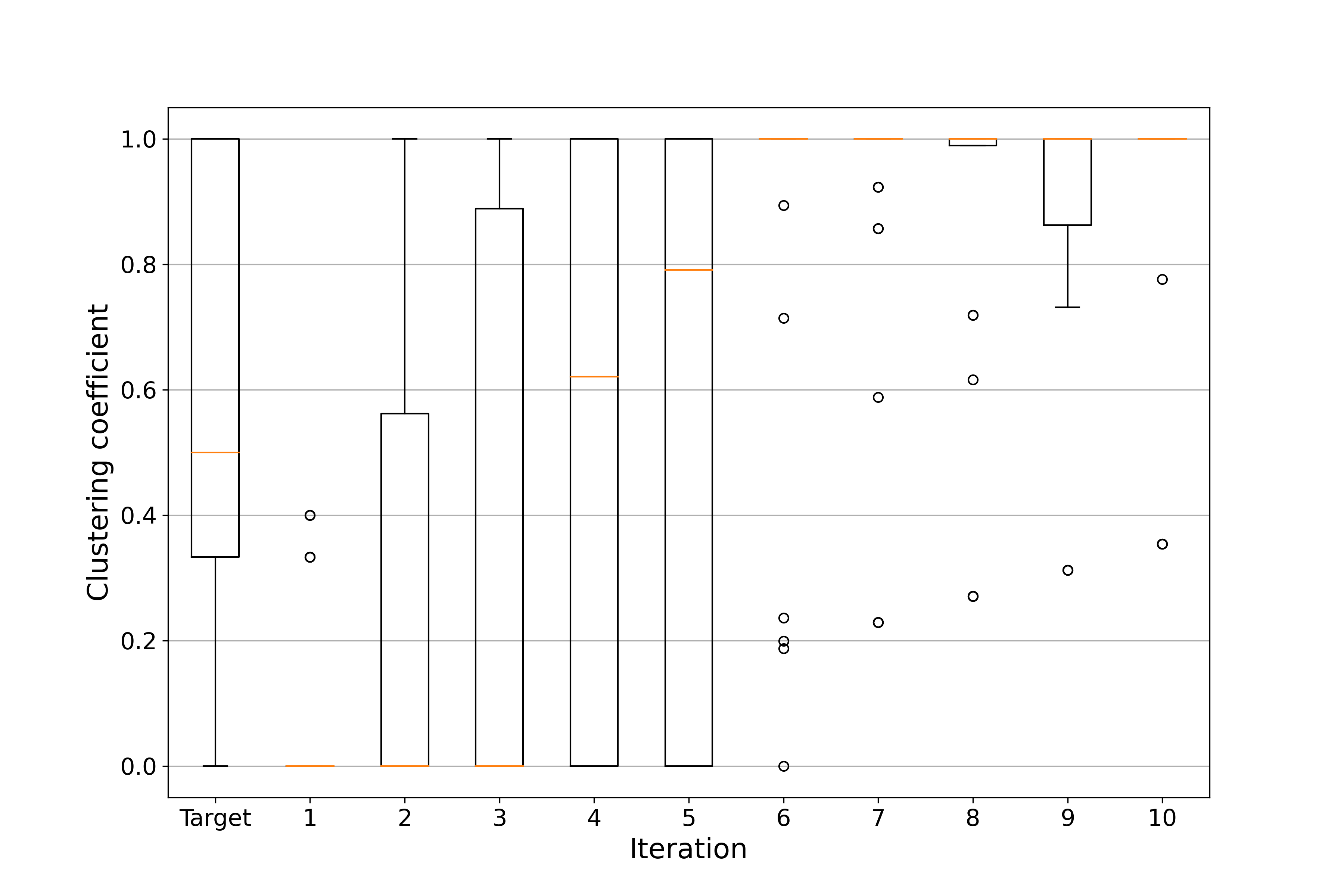}
	\end{minipage}}\\
	\caption{The clustering coefficient distribution of network simulations}
\label{clusD}
\end{figure*}

\paragraph{Clustering coefficient} describes the probability of a node's neighbours to be connected. Its value is between $0$ and $1$~\cite{IEEEexample:musial2013kind}.  As shown in Table~\ref{clusTarget}, in the target network, clustering coefficient fluctuates around an average value of $0.57$ with a standard deviation of $0.34$, ranging from $0.00$ to $1.00$. Fig.~\ref{clusD} shows, for each SNS and each iteration, the clustering coefficient distribution of the simulated social networks. For all the networks, the values of the clustering coefficient start with $0$ in the first iteration and then increase over iterations, with its distribution getting closer to that of the target and then converging to the value $1$. The clustering coefficient values of the feature-based SNSs converge slower
than that of the zero feature-based SNS. The simulated feature--based SNS gets closest to the target in the third iteration within the shortest time. 

\begin{table}[h!]
\centering
\caption{The clustering coefficient distribution of the target network.}
\label{clusTarget}
\setlength{\tabcolsep}{3pt}
\renewcommand{\arraystretch}{1.5}
\begin{tabular}{|c|c|c|c|c|c|c|}
\hline
  & Average & Standard deviation &  75\% quantile & 25\% quantile &Maximum & Minmum\\
\hline
Target & 0.57 &0.34  &1.00 & 0.33 &1.00 & 0.00\\
\hline
\end{tabular}
\end{table}

\begin{figure*}[h!] 
	\centering
		\subfigure[Average value]{
		\begin{minipage}[b]{0.48\linewidth}
			\includegraphics[width=1\linewidth]{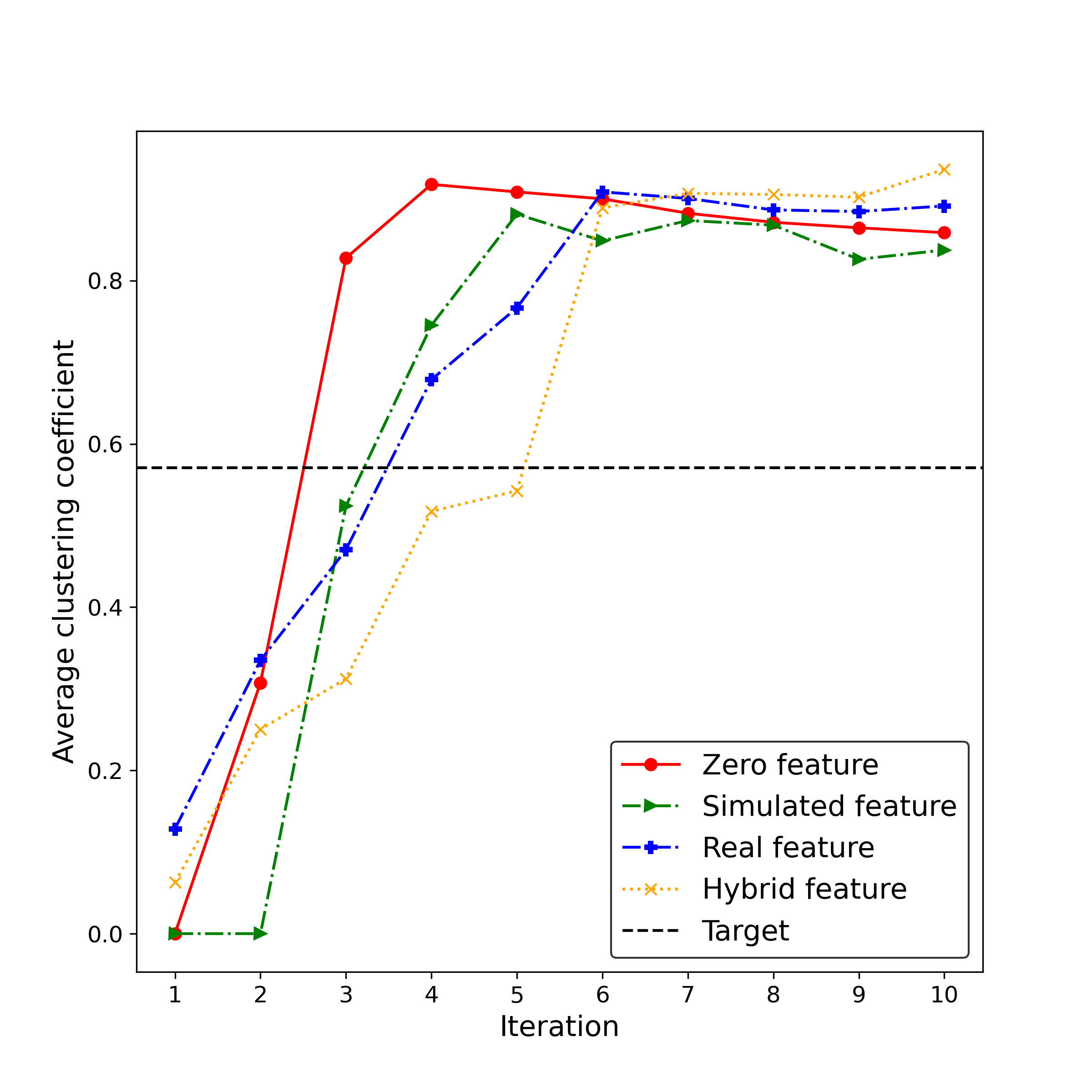}
	\end{minipage}}
	\subfigure[KL divergence]{
		\begin{minipage}[b]{0.48\linewidth}
			\includegraphics[width=1\linewidth]{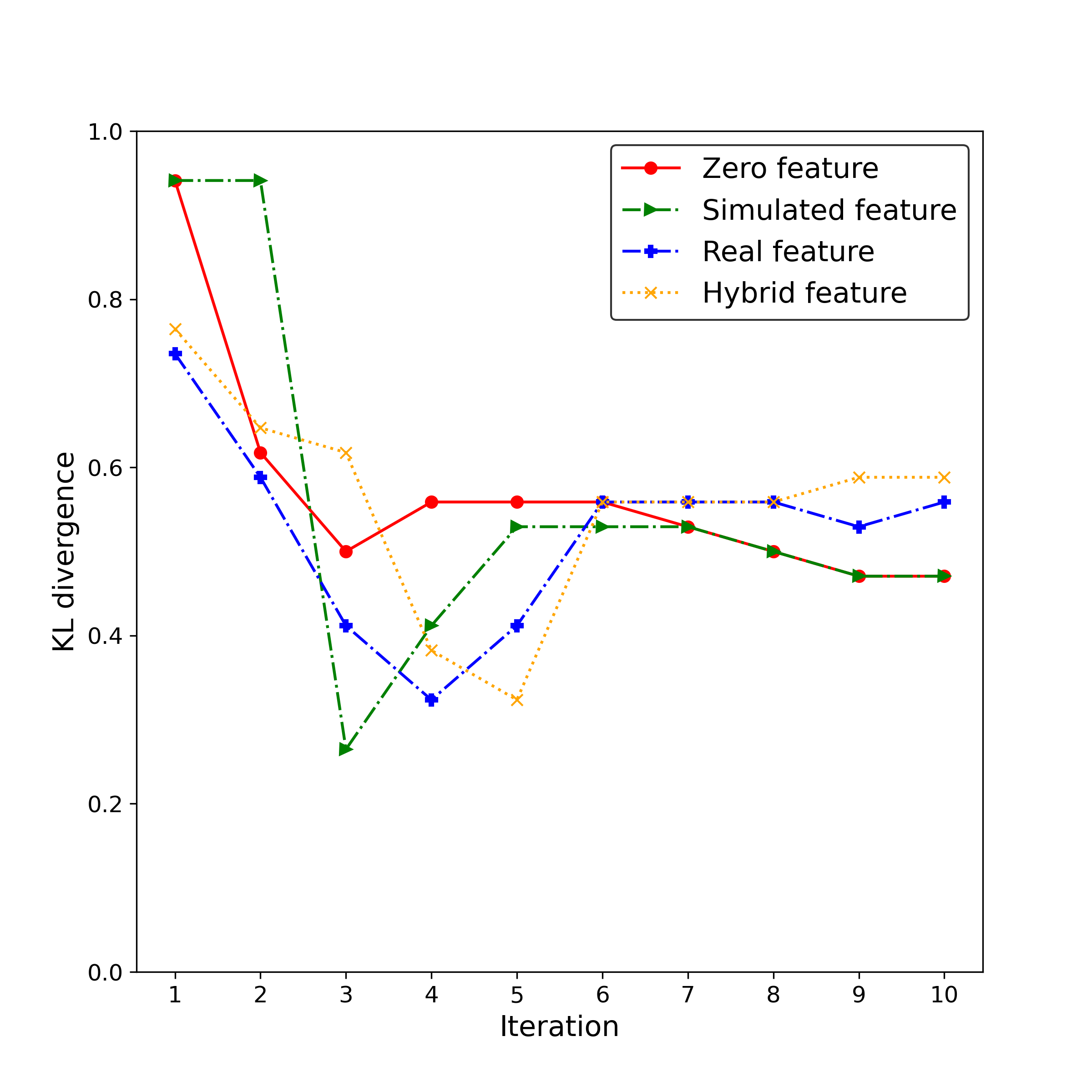}
	\end{minipage}}\\
	\caption{The average value and the KL divergence of clustering coefficient distribution.}
\label{clusKL}
\end{figure*}

As is shown in Fig.~\ref{clusKL}(a), the average values of the clustering coefficient increase over iterations, indicating a larger probability of neighbours of
one node to be connected \cite{IEEEexample:musial2013kind}. The SNSs generally approach the target network in the third iteration, except for the hybrid feature-based SNS, which reaches the target in the fifth iteration. The average clustering coefficient of the zero feature-based SNS converges to the higher bound $1$ faster than in other SNSs and keeps steady at around $0.9$ starting from the sixth iteration. 

In Fig.~\ref{clusKL}(b), the KL divergences of the feature-based SNSs firstly decrease to the lowest values at around $0.26$ and then gradually increase to around $0.5$. The simulated feature-based SNS achieves the lowest KL divergence in the third generation, getting closest to the target network within the shortest time.

\paragraph{Subgraphs significance profile} describes the similarity of subgraphs and their numbers in a given network when compared to random networks of the same size and number of edges~\cite{IEEEexample:milo2004superfamilies}. Assuming $M$ types of subgraphs, the statistical significance of subgraph $i$ ($i\in \{1,cdots,M\}$) is defined by its Z-score \cite{IEEEexample:milo2004superfamilies,IEEEexample:juszczyszyn2008temporal}:
\begin{equation}
\label{zscore}
    Z_i = \frac{n_i-<n_i^{rand}>}{\sigma_i^{rand}} \quad i \in \{1,\cdots,M\}
\end{equation}
where $n_i$, $<n_i^{rand}>$ and $\sigma_i^{rand}$ represent the frequency of subgraph $i$ in the studied network, the mean value of its occurrences in the random network ensemble and the corresponding standard deviation respectively. The significance profile $SP_i$ after normalisation is defined in equation \ref{sp}, where a positive value indicates a more often occurrence of subgraph $i$ in a network than that in a set of random network ensembles.
\begin{equation}
\label{sp}
    SP_i =\frac{Z_i}{(\sum Z_i^2)^{1/2}} 
\end{equation}

\begin{figure}[h!] 
	\centering
			\includegraphics[width=1\linewidth]{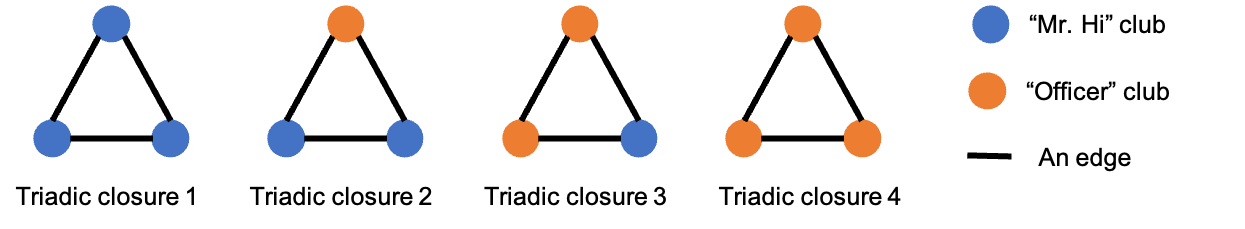}
	\caption{The four types of triadic closure in the network simulations}
	\label{trdplot}
\end{figure}

We focus on the significance profile of a social structure, triadic closure, which is an interconnected three-node subgraph and can be categorised into four types considering the binary node attributes of the target network (See Fig.~\ref{trdplot} and Table~\ref{triad}). The binary attribute determines whether a node (individual) belongs to the "Mr. Hi" club or the "Officer" club, while the edges represent the relations between the nodes. We respectively identify the triadic closures as triadic closure 1,  triadic closure 2, triadic closure 3 and triadic closure 4 based on their node diversity considering the binary attribute (See Fig.~\ref{trdplot}).

\begin{table}[h!]
\centering
\caption{The definition, numbers and significance profiles of triadic closure of the target network.}
\label{triad}
\setlength{\tabcolsep}{3pt}
\renewcommand{\arraystretch}{1.5}
\begin{tabular}{|c|c|c|c|c|}
\hline
\multirow{2}{*}{Type} & \multicolumn{2}{c|}{No. nodes} & \multirow{2}{*}{No. subgraphs} &  \multirow{2}{*}{Z--score} \\
 \cline{2-3}
 & "Mr. Hi" club &"Officer" club 
 &  & \\
\hline
 Triadic closure 1 & 0 & 3 & 15 & 0.59  \\
\hline
 Triadic closure 2 & 1 & 2 & 3 & -0.06\\
\hline
 Triadic closure 3 & 2 & 1 & 1 & -0.12\\
\hline
 Triadic closure 4 & 3 & 0 & 26 & 0.80\\
\hline
\end{tabular}
\end{table}

In Table~\ref{triad}, four types of triadic closures are identified with $i\in\{1,2,3,4\}$ based on a binary node attribute that decides the number of nodes attributed with the "Mr Hi" club or the "Officer" club. There are $15$ observations for tragic closure $1$ and $26$ observations for triadic closure $4$, more than the numbers of the other two triadic closures, which means that nodes within the same club are more likely to be connected and form a triadic closure. The significance profile of triadic closure $1$ and triadic closure $4$ is positive, contrasting with the negative values of triadic closure $2$ and triadic closure $3$.  The triadic closure $1$ and triadic $4$ appear in the target network more often than in a set of random network ensembles.

\begin{figure*}[h!] 
	\centering
	\subfigure[Triadic closure 1]{
		\begin{minipage}[b]{0.48\linewidth}
			\includegraphics[width=1\linewidth]{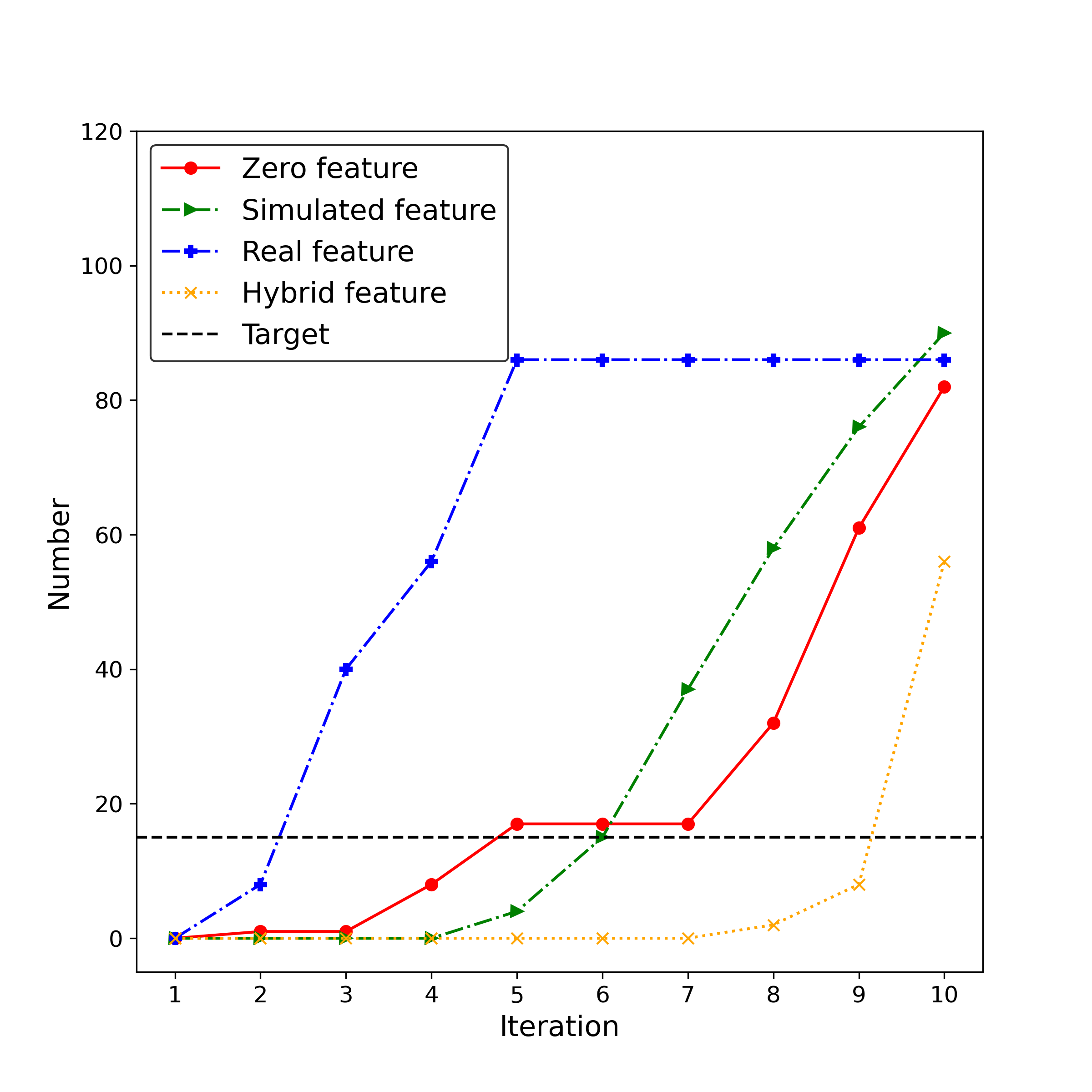}
	\end{minipage}}
	\subfigure[Triadic closure 2]{
		\begin{minipage}[b]{0.48\linewidth}
			\includegraphics[width=1\linewidth]{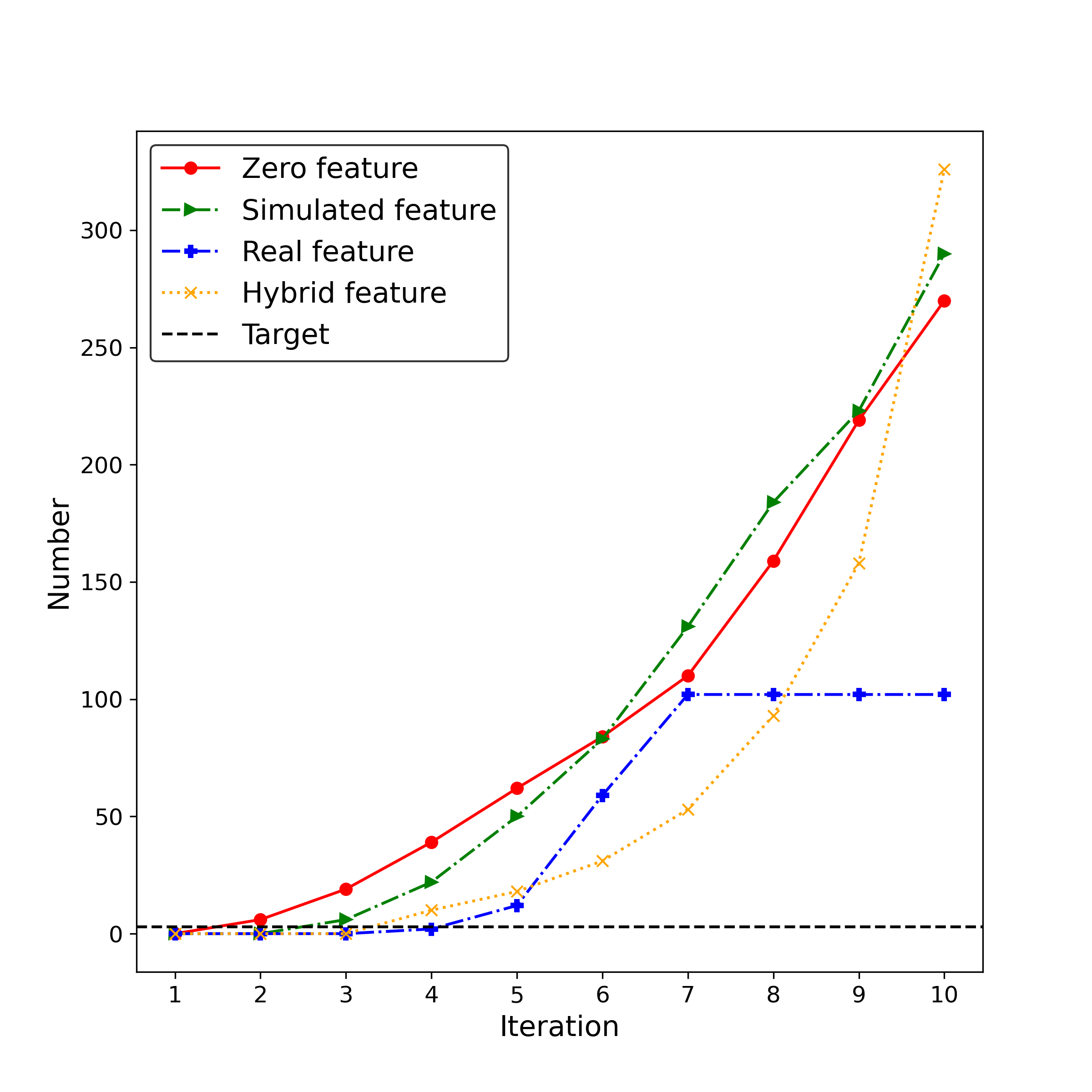}
	\end{minipage}}\\
	\subfigure[Triadic closure 3]{
		\begin{minipage}[b]{0.48\linewidth}
			\includegraphics[width=1\linewidth]{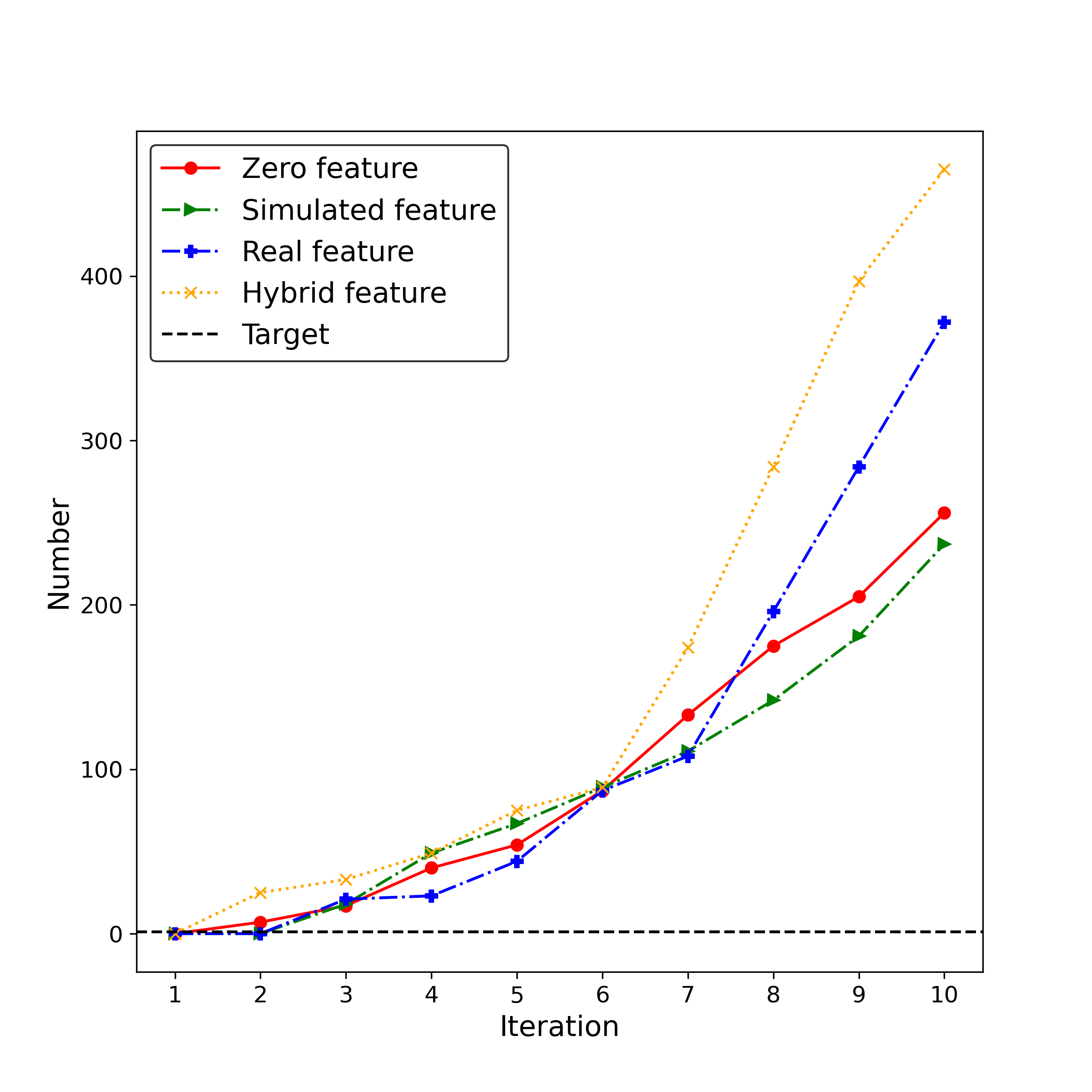}
	\end{minipage}}
	\subfigure[Triadic closure 4]{
		\begin{minipage}[b]{0.48\linewidth}
			\includegraphics[width=1\linewidth]{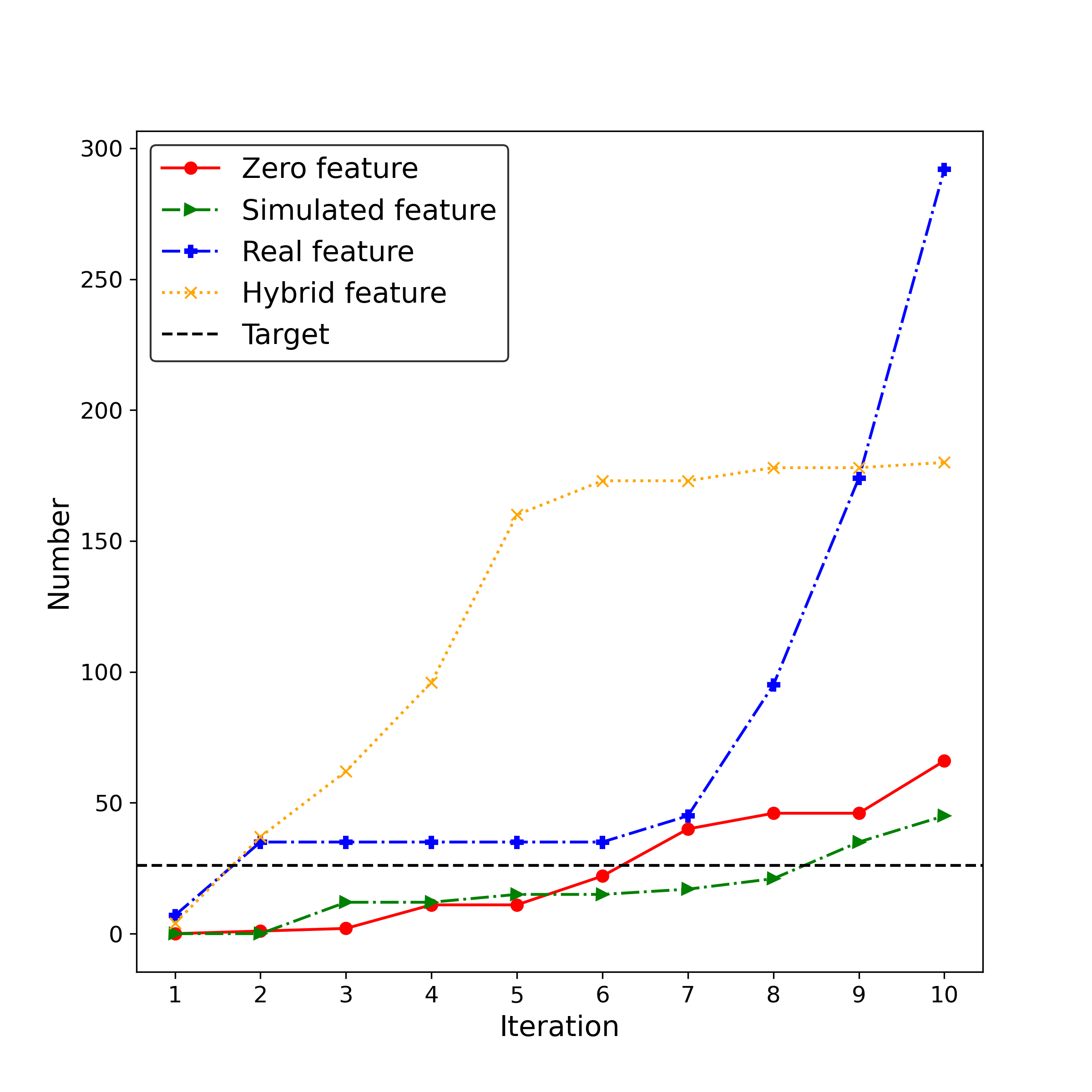}
	\end{minipage}}\\
	\caption{The number of the four types of triadic closure in the network simulations}
\label{trdNum}
\end{figure*}

\begin{figure*}[h] 
\centering
	\subfigure[Triadic closure 1]{
		\begin{minipage}[b]{0.48\linewidth}
			\includegraphics[width=1\linewidth]{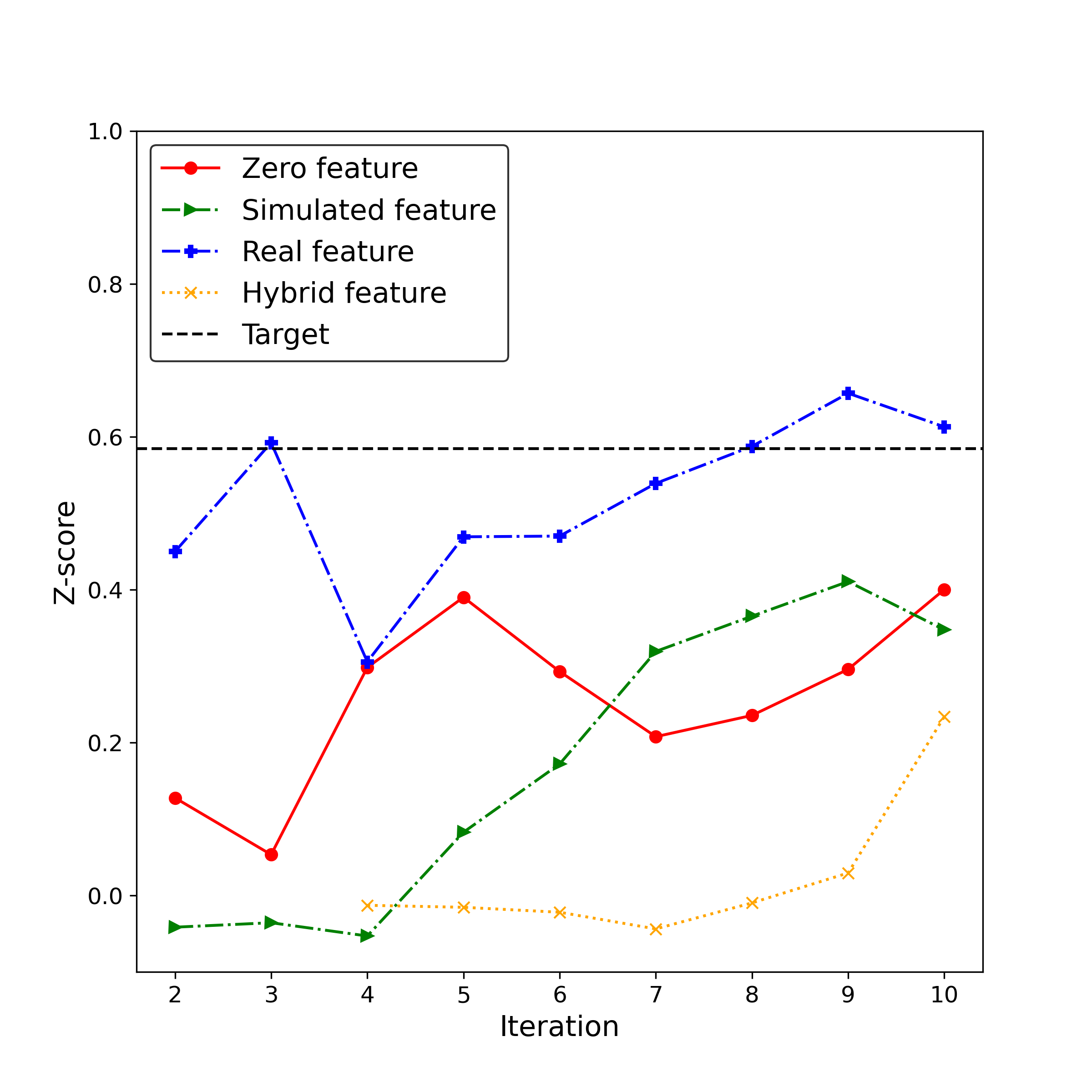}
	\end{minipage}}
	\subfigure[Triadic closure 2]{
		\begin{minipage}[b]{0.48\linewidth}
			\includegraphics[width=1\linewidth]{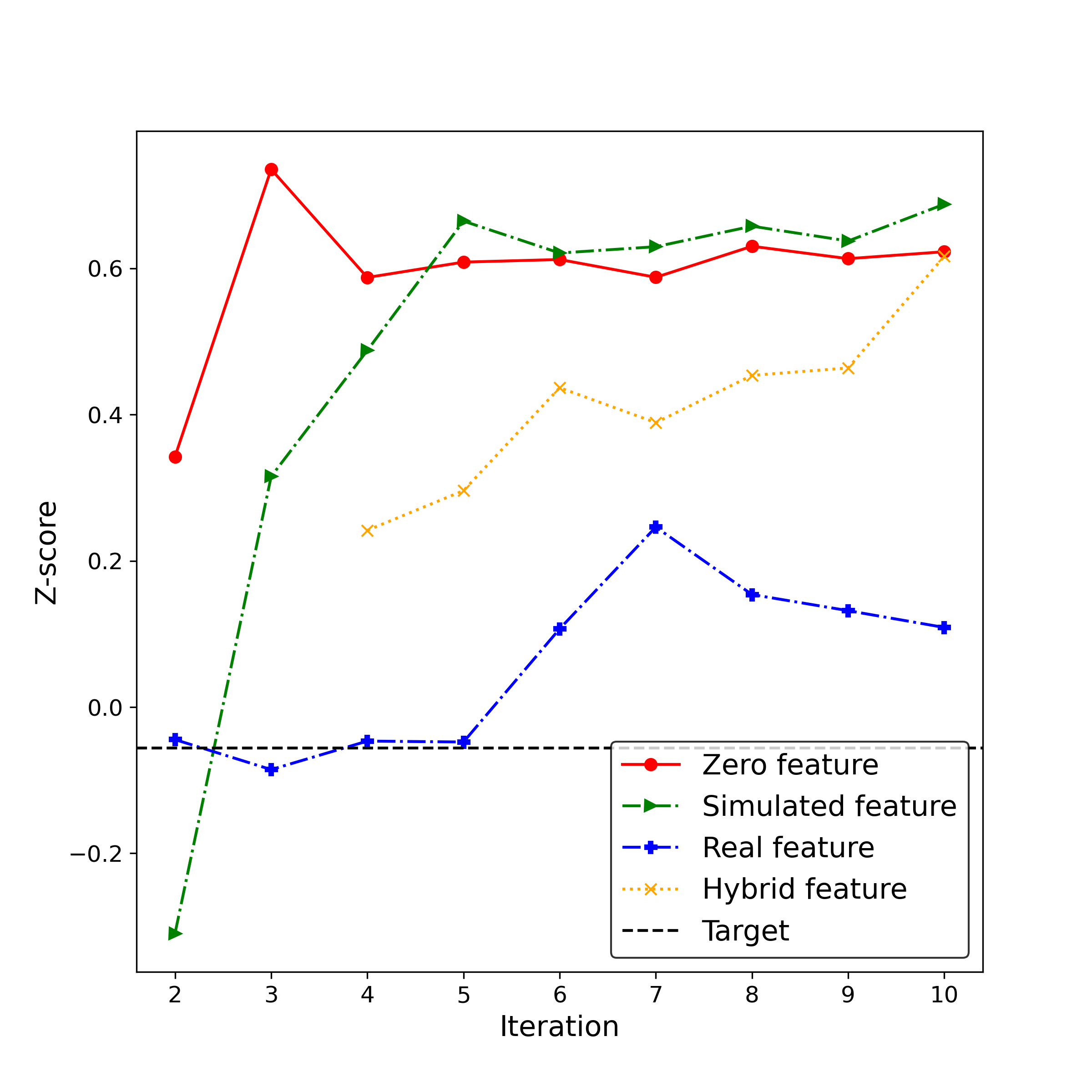}
	\end{minipage}}\\
	\subfigure[Triadic closure 3]{
		\begin{minipage}[b]{0.48\linewidth}
			\includegraphics[width=1\linewidth]{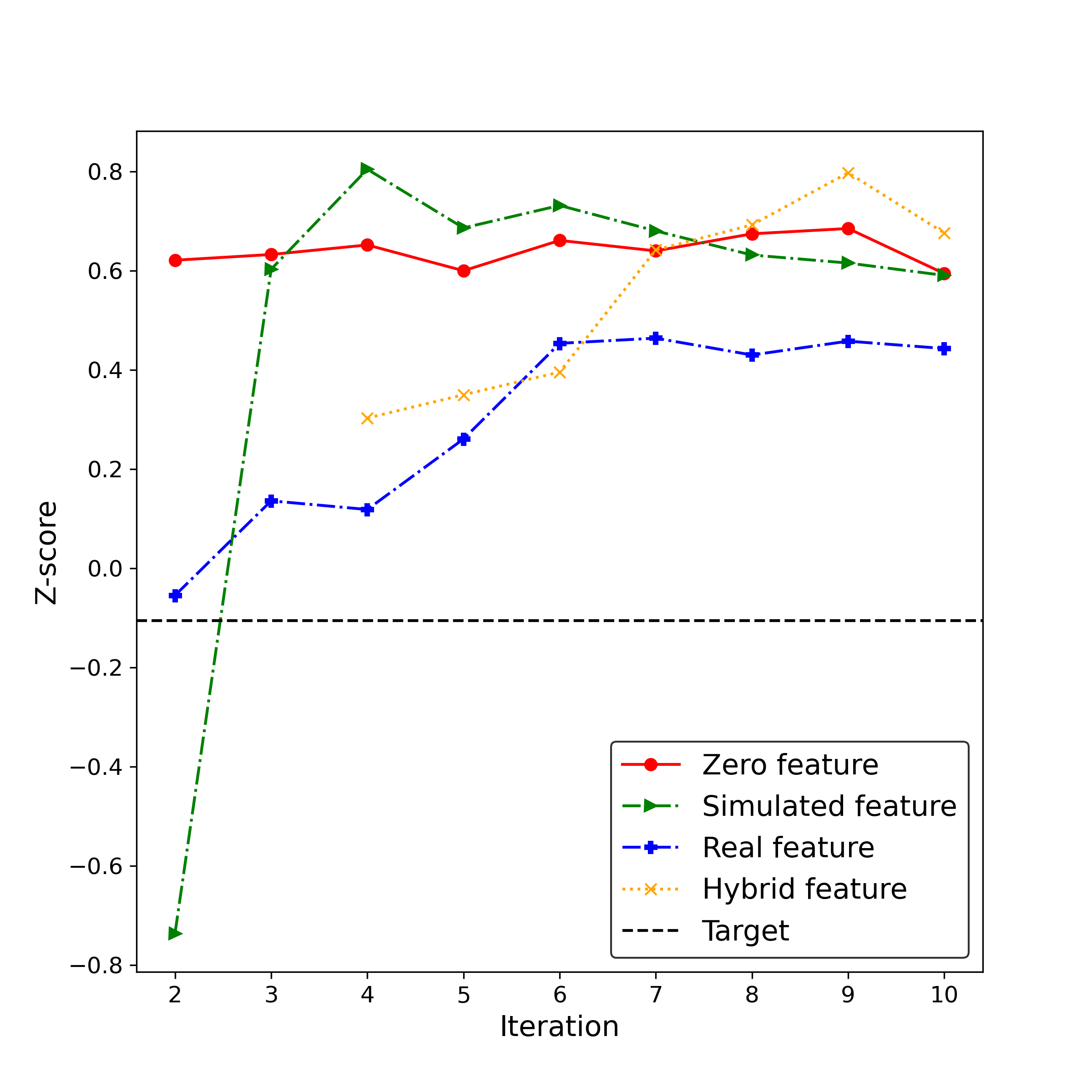}
	\end{minipage}}
	\subfigure[Triadic closure 4]{
		\begin{minipage}[b]{0.48\linewidth}
			\includegraphics[width=1\linewidth]{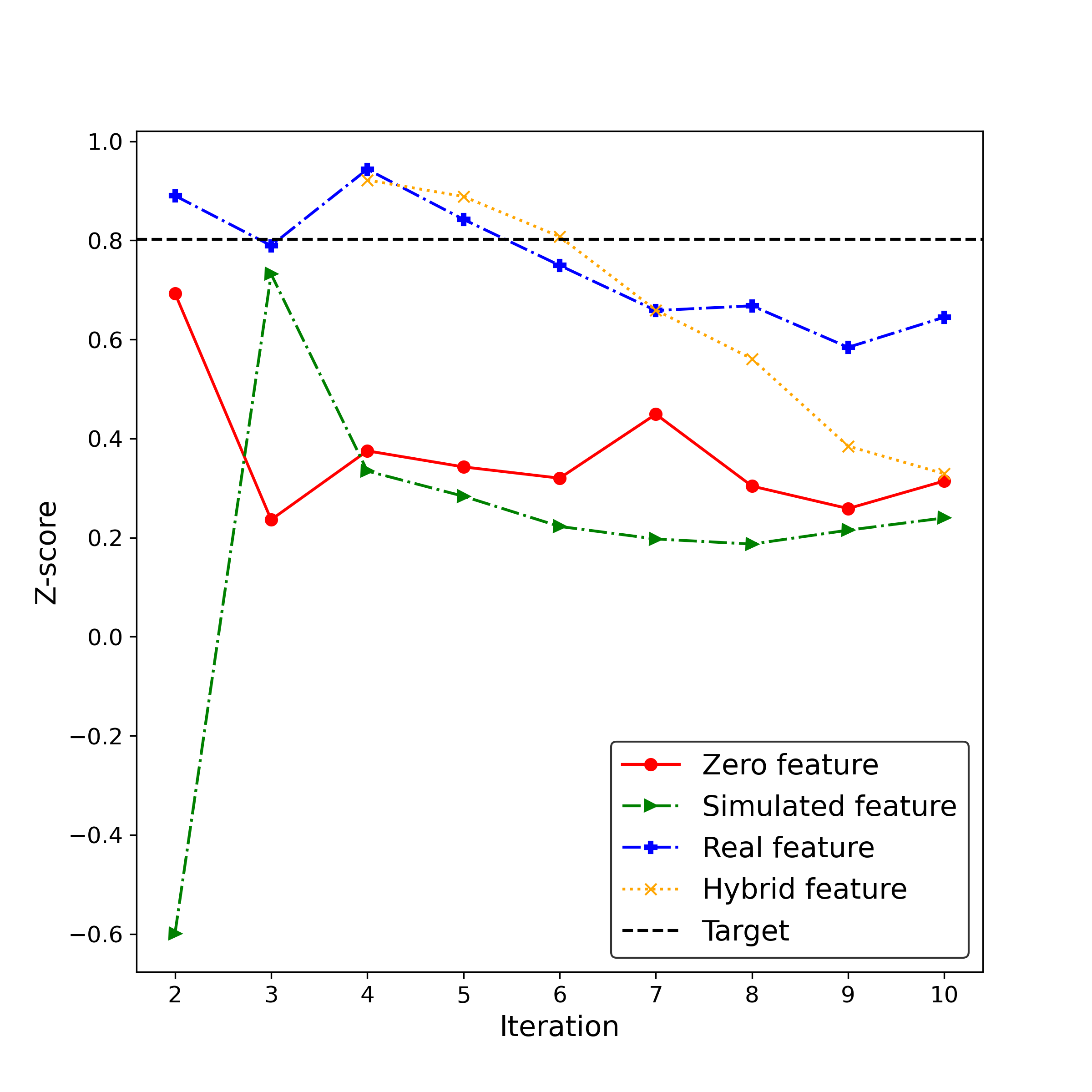}
	\end{minipage}}\\
	\caption{Z--score values of triads in the social network simulations}
\label{SP}
\end{figure*}

Fig.~\ref{trdNum} shows the number of the four triadic closures in the simulated networks, which increases with addition of edges. There are more observations of triadic closure 2, 3 and 4 than triadic closure 1. The real feature-based SNS tends to have more triadic closure 1 simulated and hits the target number in the second iteration within the shortest time (see Fig.~\ref{trdNum}(a)). All the SNSs share a similar increasing trend for triadic closure 2 and 3 over iterations, with the target number reached in around the second iteration (see Fig.~\ref{trdNum}(b) and Fig.~\ref{trdNum}(c)). The real feature--based and the hybrid feature--based SNSs firstly reach the target occurrences of triadic closure 4 in the second iteration and then simulate more of them, compared with the other SNSs (see Fig.~\ref{trdNum}(d)). To better understand the significance of these triadic closures, we calculate their Z--scores (see Fig.~\ref{SP}). 

As shown in Fig.~\ref{SP}, the positive/negative Z--score indicates a more/less frequent occurrence of triadic closures in comparison to random networks. There are missing Z--scores for the networks simulated by the hybrid feature-based SNS in the first two iterations. This is caused by the unavailability of triadic closures when only small number of edges are added.

The zero feature--based, the real feature--based and the simulated feature-based SNSs achieve a generally positive Z--score for triadic closure 1 over the iterations, while the hybrid feature--based SNS produces negative ones (close to zero) (see Fig.~\ref{SP}(a)). The zero feature--based, the simulated feature--based and the hybrid feature--based SNSs achieve a positive Z--score for triadic closure 2, deviating from the negative target, while the real feature--based SNS fluctuates around the target over the first five iterations (see Fig.~\ref{SP}(b)). All the SNSs generally enable a positive Z--score for triadic closure 2 and 4, except for the simulated feature--based SNS, which produces negative values in the first iteration (see Fig.~\ref{SP}(c) and Fig.~\ref{SP}(d)). 

The real feature--based SNS, compared with the other SNSs, attains a higher value of  Z--score for the triadic closure 1 and the triadic closure 4, and a generally lower Z--score for the triadic closure 2 and the triadic closure 3. The real feature--based SNS gets closest to the target Z--scores, which, for triadic closure 1, 2, 3 and 4, is reached in the seventh, third, first and fifth iteration respectively (See Fig.~\ref{SP}). 

\subsubsection{A brief summary}
To conclude the analysis, from a global perspective, the density, the modularity and the average node degrees increase with more edges added over the iterations for each SNS. This indicates a denser network and a stronger community structure and it is the expected result as the number of nodes remains static and only number of edges increases over the iterations. The average shortest path lengths decrease when more nodes get connected, and stabilise at a value that is smaller than $3$, showing small-world properties. From a local perspective, the average clustering coefficient and number of triadic closures increase with edge addition over iterations, indicating more interpersonal connections within the subgraphs of the simulated networks.

Among all the SNSs and across iterations, the real feature--based SNS gets closest to the target when considering modularity, assortativity, degree distribution, and subgraph significance profiles. The simulated feature--based SNS gets closest to the target clustering coefficient distribution and the shortest path length distribution. Generally all the network measures for all the social network simulators approach their target values and then deviate from them as the number of edges exceeds the number of edges of the target network. The SNS that achieves the closest distance to the target network measure (as measured through KL divergence) varies from case to case and requires different numbers of iterations. Real feature-based SNS can generally achieve a higher level of similarity with the target value of each measure through a smaller number of iterations within a shorter time and this indicates that including different features in the simulation process leads to different network patterns, which may respectively deviate from that of the real networks to a different extent. The trade-offs between the similarity levels considering multiple network measures pose a challenge to the performance evaluation of SNSs and their extension towards a DT. Current studies on network simulators generally employ a single network measure or combine a few to evaluate the similarity between the simulated and real social networks. The selection and combination of various features in the SNS extension and the selection and combination of multiple network measures for SNS performance evaluation in a systemic way remains a research gap and requires further study.

\section{Conclusions}
\label{section5}
In this paper, we firstly review the current state-of-the-art of the Social Network Simulators under the modelling and assessment framework proposed in our previous study \cite{IEEEexample:wen2022towards} and identify research gaps in the space of social network modelling. To progress the field further and achieve a simulator that can better model real networks, we extend one of the promising SNS proposed by \cite{IEEEexample:ashraf2019simulation} towards a Digital Twin Oriented SNS by proposing one possible pathway of increasing structural complexity, which is to include informative features that are simulated or observed in the real world.
With this SNS, each node attribute is composed of a node feature and a social DNA (the node's preference for a similar feature and the corresponding weight of preference).

We test different settings of sDNA in DT Oriented SNS to see which one results in social network that is closest to the target network. We propose to assess the similarity between target and simulated networks using metrics calculated at both global and local levels while taking the runtime of SNS as an indicator of efficiency. This DT Oriented SNS also serves as a tool to analyse the complexity of social network simulations built up over iterations. As an illustrative example, we conduct experiments and assessments of this DT Oriented SNS with the Karate Club network, which results show the possibility of 
optimisation and assessment of DT Oriented SNS towards a Digital Twin.

Literature review revealed that the  majority of existing SNSs focus on generating purely simulated networks, while only a small proportion of SNSs aim to model real nodes and node attributes with/without observable information about edges. Under this constraint, we review and discuss the modelling social networks through generations towards the ultimate goal of a DT. Current SNSs generally focus on static networks or dynamic networks. Few SNSs consider networks, the dynamic process on the network and their interrelations. The interrelated social networks and dynamic processes, with the continuous acquisition of real-time information and feedback, also remain a research gap and require further study. 
We also discuss the complexity of social network simulations from the perspective of four dimensions and review the current ways of assessing SNSs. Most SNSs simulate social networks composed of fixed nodes and edges, while some SNSs incorporate network attributes and slow changes in network topologies. Current SNSs consider the global level similarity between the social network simulation and the target network as an indicator of its performance, where neither the local level similarity nor the efficiency is incorporated.

In the experiment, we extend an SNS towards a DT Oriented SNS to help illustrate the possible pathways and the challenges of developing a DT Oriented SNS.
The extension of an SNS towards a DT Oriented SNS incorporates different levels of structural complexity and involves experiments on the benchmark, zero feature-based SNS, and its extensions, including simulated feature-based SNS, real feature-based SNS and the hybrid feature-based SNS concerned about both the real and the simulated features. We calculate the composite performance index of different networks by combining the results of various network measures. For a deeper understanding of the composite performance index that changes over iterations, we also conduct a comparative analysis of these SNSs to see the complexity of their social network simulations over iterations. The experimental results show that among all the SNSs, the real feature-based SNS, with an appropriately increased structural complexity, has the best performance in the efficiency and similarity from both the local and the global perspective. To be more specific with each measure, real feature-based SNS can also get closer to the target value through a smaller number of iterations. However, this conclusion holds for the experiments presented in this study, considering the pathway of SNS extension, the information employed in network simulation and the way we calculate the composite performance index. We need to develop a more systematic approach to developing DT Oriented SNS and propose referable SNS evaluation metrics for further study.
And as SNSs add more edges over the iterations, the runtime of SNSs increases, and a stronger community structure and an assortative mixing pattern emerge, with more triadic closures as more neighbouring nodes get connected. Overall, all the measures involved in the composite performance index approach their target values and then deviate from them as the simulated network exceeds the target network's density. However, different similarity levels achieved by SNSs considering different network measures reveal the challenge of an accurate SNS performance evaluation given the specific requirements of social network simulations. This is a research gap to be addressed in our future study.

The DT Oriented SNSs can be extended with structural variations and temporal changes to approach a DT of the real systems. There is a requirement for a future study on the sDNA that varies across groups or individuals. Further research is also required on the network topology that changes over time and the process dimension that simultaneously interacts with the network dimension. Specific SNS performance evaluation criteria, considering various SNS complexity levels, also require further study. More generally, this study serves as a starting point of our future work on exploring the complexities of the real systems.

\section*{Acknowledgements}\label{section-acknowledge}

This work was supported by the Australian Research Council, Dynamics and Control of Complex Social Networks under Grant DP190101087.

\section*{Appendix}
\label{appendix}
In this appendix, we list some extensible SNSs found through the review of literature and websites (\url{http://caagt.ugent.be/CaGe/index.html}; \url{https://hog.grinvin.org/}). The Table~\ref{appendixTable} below includes the names, references and links to the code for these SNSs while briefly describing their functions.

\begin{table}[h]
\centering
\scriptsize
\caption{SNSs in the current studies}
\label{appendixTable}
\setlength{\tabcolsep}{3pt}
\renewcommand{\arraystretch}{1.5}
\begin{tabular}{|l|c|p{160pt}|p{160pt}|}
\hline
SNS & Study & Link to the code& Description\\
\hline
VirtualSoc & \cite{IEEEexample:ashraf2019simulation} & \url{https://github.com/AkandaAshraf/VirtualSoc} & Dynamic Social Network Simulation Data with Ground Truth Labels and Features   \\ 
\hline
Hashkat& &\url{docs.hashkat.org} & A dynamic network simulor designed to model the growth of and information propagation through an online social network. \\ 
 \hline
Minibaum &\cite{IEEEexample:brinkmann1996fast} &\url{http://caagt.ugent.be/minibaum/} &A generator for connected cubic graphs, but can be restricted to generating only graphs that have a fixed minimal girth or are bipartite \\
\hline
snarkhunter &\cite{IEEEexample:brinkmann2017generation} & \url{http://caagt.ugent.be/cubic/}& A generator for connected cubic graphs and snarks.\\
\hline
GenHypohamiltonian &\cite{IEEEexample:zamfirescu2019almost} &\url{http://caagt.ugent.be/hypoham/} & A generator for hypohamiltonian and almost hypohamiltonian graphs\\
\hline
Genreg& \cite{IEEEexample:meringer1999fast}&\url{http://www.mathe2.uni-bayreuth.de/markus/reggraphs.html} & A very fast structure generator for regular graphs\\
\hline
MOLGEN&\cite{IEEEexample:gugisch2015molgen}  &  \url{http://www.molgen.de/?src=documents/molgen4.html} &A structure generator for molecular graphs. \\
\hline
alternating& \cite{IEEEexample:althofer2015alternating}& \url{https://github.com/nvcleemp/alternating} &A generator for alternating planar graphs  \\
\hline
CographGeneration
& \cite{IEEEexample:jones2018cograph}& \url{https://github.com/atilaajones/CographGeneration} & A generator for cographs \\
\hline
CriticalPfreeGraph & \cite{IEEEexample:goedgebeur2018exhaustive} & \url{http://caagt.ugent.be/criticalpfree/} & A generator for k-critical graphs without long induced paths\\
\hline
nauty and Traces & \cite{IEEEexample:mckay2014practical}&\url{https://pallini.di.uniroma1.it/}& A generator for automorphism groups of graphs and digraphs. It can also produce a canonical labelling. \\
\hline
 plantri and fullgen &\cite{IEEEexample:brinkmann2007fast} &\url{https://users.cecs.anu.edu.au/~bdm/plantri/}&A generator for planar graph.\\
\hline
Buckygen& \cite{IEEEexample:goedgebeur2015recursive}& \url{http://caagt.ugent.be/buckygen/}&A generator for all nonisomorphic fullerenes \\
\hline
perihamiltonian&\cite{IEEEexample:fabrici2021non} &\url{https://github.com/nvcleemp/perihamiltonian}& A generator for perihamiltonian graphs with a given connectivity.\\ 
\hline
GenerateUHG  & \cite{IEEEexample:goedgebeur2020graphs} &\url{http://caagt.ugent.be/uhg/} & A generator for graphs with a given number k of hamiltonian cycles (which is especially efficient for small values of k)\\ 
\hline
\end{tabular}
\end{table}

\end{document}